\documentclass[a4paper,11pt]{amsart}
\usepackage{amsmath}
\usepackage{amsfonts}
\usepackage{eucal}
\usepackage{amsthm}

\catcode`\@=11

\newtheorem{thm}{Theorem}[section]
\newtheorem{lem}[thm]{Lemma}
\newtheorem{prop}[thm]{Proposition}

\theoremstyle{definition}

\newtheorem{rem}[thm]{Remark}

\newcommand{\be}{\begin{equation}}
\newcommand{\ee}{\end{equation}}

\newenvironment{pre}{\noindent{Proof.} \hskip 2pt}{{\null\hfill$\Box$}\medskip}

\numberwithin{equation}{section}
\def\R{\mathbb{R}}
\def\C{\mathbb{C}}
\def\Z{\mathbb{Z}}
\def\N{\mathbb{N}}

\font\tenms=msbm10 scaled 1200
\font\sevenms=msbm10
\font\fivems=msbm8
\newfam\bbfam \textfont\bbfam=\tenms \scriptfont\bbfam=\sevenms
\scriptscriptfont\bbfam=\fivems

\newcommand{\jap}[1]{\langle #1 \rangle}

\def\i{\mbox{\raisebox{.5ex}{$\i$}}}
\def\1{1\!{\rm I}}
\def\t1{\tilde{\1}}

\def\R{\mathbb{R}}
\def\C{\mathbb{C}}
\def\Z{\mathbb{Z}}
\def\N{\mathbb{N}}
\def\re{\mathop{\rm Re}\nolimits}
\def\im{\mathop{\rm Im}\nolimits}
\def\open{{\mathcal {\ddot O}}}
\def\p{{\partial}}
\def\Supp{{\rm Supp}\hskip 1pt}
\def\supp{{\rm Supp}\hskip 1pt}
\def\ord{{\mathcal O}}

\def\adots{\mathinner{\mkern2mu\raise 1pt\hbox{.}\mkern3mu\raise 4pt\hbox
{.}\mkern 1mu\raise 7pt\hbox{.}}}

\def\eq#1{(\ref{#1})}

\def\rf#1#2#3#4{{\parindent 3,5em
\item [{\hbox to \parindent {\enskip [#1]\hfill}}] #2: {\sl
#3}\hfill\break #4
\hfill}}

\author[S. Fujii\'e, A. Lahmar-Benbernou \& A. Martinez]{S. Fujii\'e${}^1$, A. Lahmar-Benbernou${}^2$ and A. Martinez${}^3$}
\title[Shape Resonances for Non Globally Analytic Potentials]{Width of Shape Resonances for Non Globally Analytic Potentials}

\begin{document}

\begin{abstract}
We consider the semiclassical Schr\"odinger operator with a {\it well in an island} potential, on which we assume smoothness only, except near infinity. We give the asymptotic expansion of the imaginary part of the shape resonance at the bottom of the well. This is a generalization of the result by Helffer and Sj\"ostrand \cite{hs1} in the globally analytic case.
We use an almost analytic extension in order to continue the WKB solution coming from the well beyond the caustic set, and, for the justification of the accuracy of this approximation,  we develop some refined microlocal arguments in $h$-dependent small regions.
\end{abstract}

\setcounter{section}{0}
\maketitle

%%%%%%%%%%%%%%%%%%%%%%%%%%%%%%%%%%%%%%
\addtocounter{footnote}{1}
\footnotetext{University of Hyogo, Department of Mathematical Sciences, 2167 Shosha, Himeji, Japan 671-2201. Supported by JSPS Grants in Aid for Scientific Research}
\addtocounter{footnote}{1}
\footnotetext{Universit\'e de Mostaganem, D\'epartement de Math\'ematiques, B.P. 227, 27000 Mostaganem, Algeria}
\addtocounter{footnote}{1}
\footnotetext{Universit\`a di Bologna, Dipartimento di
Matematica, Piazza di Porta San Donato 5, 40127 Bologna,
Italy. Partly supported by Universit\`a di Bologna, Funds
for Selected Research Topics and Founds for Agreements with
Foreign Universities}

\tableofcontents
%%%%%%%%%%%%%%%%%%%%%%%%%%%%%%%%%%%%%%%%

\section{Introduction}
This paper is concerned with the quantum resonances of the semiclassical Schr\"odinger operator in $\R^n$,
\begin{equation}
\label{schop}
P=-h^2\Delta+V(x).
\end{equation}
From the physical point of view, such resonances are associated to metastable states, that is, states with a finite life-time, and the  life-time is given by the inverse of the absolute value of the imaginary part (width) of the corresponding resonance.
\vskip 0.2cm

In the literature, many geometrical situations have been studied where  the location of the resonances of $P$ has been determined up to errors of order ${\mathcal O}(h^\infty)$: see, e.g.,  \cite{BaMn, BCD, GeSj, HiSi, Sj2}. In particular, when the approximated location is at a distance $\sim h^{N_0}$ of the real line with some fixed $N_0 >0$, then such results automatically produce good estimates on the widths of such resonances, too. However, in the physically most interesting situations where the resonances are exponentially close to the real axis, only very few results give a lower bound on the width. To our knowledge, and apart from the one-dimensional case (see, e.g., \cite{Ha, FeLa}), the only available results are \cite{Bu, hs1}. In \cite{Bu}, a very nice and general estimate is obtained, independently of the geometrical situation presented by the potential $V$. In particular, because of its wide generality, such a result does not give any precise asymptotic of the width. On the contrary, the result of \cite{hs1} gives a full asymptotic of the width when
the (globally analytic) potential $V$ presents the geometric situation of a {\it well in an island}, at the origin of the creation of the so-called {\it shape resonances}.

\medskip
The purpose of this article is to extend the results of \cite{hs1} to smooth potentials that are not assumed to be globally analytic, but only near infinity (allowing  the definition of resonances by analytic distorsion).
In that case, we obtain a classical expansion of the first resonance $\rho$ just as in the globally analytic case of \cite{hs1}.

\medskip
Roughly speaking, a quantum resonance is a complex energie $\rho$ for which the Schr\"odinger equation
$$
Pu=\rho \,u
$$
admits a non-trivial solution called {\it resonant state}, which is {\it outgoing} for large $x$ (see \S 2 for a rigorous definition by analytic distorsion).
In the case of shape resonances, the resonant state describes a  quantum particle, concentrated in the well for a long period, but then escaping to the {\it sea} (classically allowed region outside the island) by tunneling effect.
This effect is reflected by the width of the resonance $\rho$.

\medskip
More precisely, let $\Gamma$ be the set of points on the boundary of the island $\open$ where the Agmon distance from the bottom of the well $x_0$ reaches its minimum $S$.
In the semiclassical regime, it happens that almost all the tunneling occurs along a small neighborhood of the geodesic curves from $x_0$ to $\Gamma$. Then,
the width of the resonance $\rho$
is determined by the amplitude of $u$ near $\Gamma$, and the result is,
$$
\im \rho(h)\sim -h^{\frac{1-n_\Gamma}2}\big(\sum_{j\geq 0}h^jf_j\big)e^{-2S/h} \quad \mbox{mod }{\mathcal O}(h^\infty)e^{-2S/h},
$$
where $n_\Gamma$ is some geometrical constant, and $f_0>0$ (see Theorem \ref{main}).

\medskip
 In order to prove this result for non globally analytic potentials, we mainly follow the strategy of \cite{hs1}, with some additional technical difficulties that we explain now.
 
 \medskip
Let $W$ be a bounded domain and $(\cdot,\cdot)_W,
||\cdot||_W$ the scalar product and the norm in $L^2(W)$. Then,
$\im \rho$ can be
represented in terms of the corresponding resonant
state $u$ on $W$, by applying Green's formula to the identity
$((P-\rho)u,u)_W=0$. This leads to,
\begin{equation}
\label{Green}
\im
\rho=-\frac{h^2}{||u||_W^2}\im\int_{\partial
W}\frac{\partial u}{\partial n}\bar udS,
\end{equation}
where $\partial/\partial n$ denote the exterior normal derivative on
$\partial W$. 
The point is that if we take for $W$ a small neighborhood of
$\open$, then the contribution from
the integral of the RHS of \eq{Green} is concentrated near $\Gamma$, and the problem is mainly reduced
to the study of the asymptotic behavior of
$u$ near $\Gamma$. 

\medskip
In the globally analytic case studied by Helffer and Sj\"ostrand in \cite{hs1}, the first step in order to obtain this asymptotic behavior consisted in extending to the complex domain the WKB solutions coming from the well, in such a way that one could go round the caustic set (see \cite{hs1} Proposition 10.9). In our case, such an analytic extension is no longer possible but, by means of almost-analytic extensions, it is still possible to go round the caustic set on the condition of staying close enough to the real domain, namely, at a distance ${\mathcal O}((h\ln h^{-1})^{2/3})$ of the caustic set. In this way, one can recover an (outgoing) WKB expression out of the island, but still at a distance of order $(h\ln h^{-1})^{2/3}$ from the boundary (see Proposition \ref{summary}).

\medskip
The next step in the proof of \cite{hs1} was to use an argument of microlocal analytic propagation that, thanks  to a suitable  a priori estimate, permitted to compare the solution $u$ with the previous WKB constructions  near $\Gamma$. In our case, the analogous a priori estimate can be obtained only at a distance ${\mathcal O}((h\ln h^{-1})^{2/3})$  of $\Gamma$ (see Proposition \ref{prop7}). For this reason, the usual results of propagation do not apply, and we need to refine them in order to be able to work in $h$-dependent open sets (see Lemma \ref{propagation1}). Then, after a suitable change of scale, and still using almost-analytic extensions, the comparison between $u$ and the WKB constructions is obtained up to a distance of order $(h\ln h^{-1})^{2/3}$ of $\Gamma$ (see Proposition \ref{propaganalyt}, and Proposition \ref{comparison}).

\medskip
The final step in \cite{hs1} consisted in replacing $u$ by its WKB approximation into (\ref{Green}) (with $W$ a fixed small enough neighborhood of $\open$), and in using a stationary phase argument in order to obtain the asymptotic of $\im\rho$. In our case, a similar argument can be used with $W=\{ d(x,\open)<|h\ln h|^{2/3}$\} , but this makes appear  terms in $h^j(\ln h)^k$ in the integrand of the right-hand side of (\ref{Green}). However, by slightly changing the choice of $W$, and by observing that the left-hand side of (\ref{Green}) does not depend on this choice, we can prove that, indeed, the final expansion does not contain terms in $(\ln h)^k$, $k\not= 0$ (see (\ref{last}) and Lemma \ref{lastlemma}).

\medskip
Summing up, the method consists in the three following parts: 
\begin{enumerate}
\item Extension of the WKB solution constructed at the bottom of the well $x_0$ up to a neighborhood of
$\Gamma$.
\item Estimate on the difference between the WKB solution $w$ and the
resonant state $u$.
\item
Computation of the asymptotics of the integral \eq{Green} using $w$
instead of $u$.
\end{enumerate}

\medskip
Point 1 is performed by solving transport equations  along the minimal geodesics, and then,  by using an Airy integral representation of the solution near $\Gamma$. In order to recover a WKB asymptotic expansion outside $\open$, almost-analytic extensions (or, more precisely, holomorphic approximations) are used as well as a stationary-phase expansion at a distance of order $|h\ln h|^{2/3}$ from $\Gamma$.

\medskip 
Point 2 is obtained by using three propagation arguments. At first, a standard $C^\infty$-propagation, exploiting the fact that $u$ is outgoing, gives a microlocal information on $u$ in the incoming region up to a small distance of $\Gamma$. Then, a refined $C^\infty$ propagation result permits to extend this microlocal information up to a distance of order $|h\ln h|^{2/3}$ of $\Gamma$. Finally, performing a change of scale both in the variables and in the semiclassical parameter, an analytic-type propagation argument gives the required estimate in a full neighborhood of $\Gamma$ of size $|h\ln h|^{2/3}$.

\medskip For Point 3, we take $W=\{ d(x,\open)<|h\ln h|^{2/3}$\} and we use Green's formula as we explained before. Then, by a deformation argument, we show that the final expansion does not contain terms in $\ln h$, and actually coincides with the expansion obtained in \cite{hs1}.

\medskip
 The article is organized as follows:
 In \S 2, we assume conditions on the potential $V$ and state the main results. 
 In \S 3, we prove Theorem \ref{th3}, especially the global a priori estimate \eq{compdiric3} of $u$.
\S 4 is devoted to the above point 1, and
 \S 5, \S 6 and \S 7 are devoted to the point 2.
 In \S 8, we study the point 3, and prove the main Theorem \ref{main}
 
 \medskip
 We also refer to \cite{LFM} for a shorter version of some aspects of this work.

\section{Assumptions and Main Results}
Let us state precisely the assumptions on the potential.
\begin{itemize}
\item[{\bf (A1)}]
$V\in C^\infty$ is real-valued and there exists some compact set
$K_0\subset \R^n$ such that $V$ is analytic on
$K_0^C=\R^n\backslash K_0$ and can be extended as a holomorphic
function on a sector
$$
D_0=\{x\in\C^n;|\im x|<\sigma_0|\re x|, \re x\in K_0^C\}
$$
for some constant $\sigma_0>0$. Moreover, there exists $\sigma_1>0$ such
that
$$
|V(x)|={\mathcal O}(|x|^{-\sigma_1})\quad\mbox{as}\quad |\re x|\to\infty, \quad x\in
D.
$$
\end{itemize}

This assumption enables us to define resonances near the real axis
as complex eigenvalues of the distorted operator $P_\theta$ of $P$.
Let $F(x)\in C^\infty(\R^n,\R^n)$ such that $F(x)=0$ on the compact set $K_0$,
$F(x)=x$ for large enough $|x|$, and $|F(x)|\leq |x|$ everywhere. For small $\nu>0$, we define a unitary
operator on $L^2(\R^n)$ by
\begin{equation}
\label{Unu}
U_\nu\phi(x)=\det (1+\nu dF(x))^{-1/2}\phi(x+\nu F(x)).
\end{equation}
The conjugate operator $\tilde P_\nu=U_\nu PU_{-\nu}$ is analytic with
respect to
$\nu$ and we can define the distorted operator by $P_\theta=\tilde
P_{i\theta}$ for positive small $\theta$.
Then the essential spectrum is given by
$\sigma_{ess}(P_\theta)=e^{-2i\theta}\R_+$ (Weyl's perturbation theorem)
and the spectrum $\sigma(P_\theta)$ in the sector
$S_\theta=\{E\in\C;-2\theta<\arg E <0\}$ is discrete (see \cite{hu}).
The elements of $\sigma(P_\theta)\cap S_\theta$ are called {\it resonances}.
This definition is independent of $\theta$ in the sense
that $\sigma(P_{\theta'})\cap S_\theta=\sigma(P_\theta)\cap S_\theta$
if $\theta'>\theta$, and also of the function $F(x)$. Moreover, if $u_\theta$ is an eigenfunction of $P_\theta$, it can be proved (see, e.g., \cite{hm}) that there exists $u\in C^\infty (\R^n)$, holomorphic in $D_0$, such that $u_\theta = U_{i\theta}u$. Such functions $u$
are called  {\it resonant states} of $P$.

The next assumption describes the shape of $V(x)$ in the {\it island}:
\begin{itemize}
\item[{\bf (A2)}]
There exist a bounded open domain $\open$ with smooth boundary, a point
$x_0$ in $\open$ and a positive number $E_0$ such that
$$
V(x_0)=E_0, \quad \frac{\p V}{\p x}(x_0)=0,\quad \frac{\p ^2V}{\p x^2}(x_0)>0,
$$
and 
$$
V(x)>E_0\quad {\rm in}\quad \open\backslash\{x_0\},\qquad
V(x)=E_0\quad {\rm on}\quad 
\partial \open
$$
\end{itemize}

To the {\it well} $\{ x_0\}$ of the potential, we can associate a Dirichlet problem.
Let us denote by $d(x,y)$  the Agmon distance  associated with the pseudo-metric $ds^2=\max
(V(x),0)dx^2$, $S=d(x_0,\partial \open)$ the minimal
distance from $x_0$ to the boundary of $\open$, and
$B_d(x_0,S):=\{ x\,;\, d(x,x_0) < S\}$ the open  ball centered
at $x_0$ of radius $S$ with respect to the distance $d$.
We consider a Dirichlet realization $P_D$ of the operator
$P$ on the domain $\overline{B_d(x_0,S-\eta)}$ for sufficiently small
$\eta$. The following result is due to
Helffer and Sj\"ostrand \cite{hs2} (see also Simon \cite{si} for a partial version):

\begin{thm} \sl {\bf (Helffer-Sj\"ostrand)}
Let $\lambda_D(h)$ be the first eigenvalue of
$P_D$ and $u_D(x,h)$ the corresponding normalized eigenfunction.
Then $\lambda_D(h)$ has a complete
classical asymptotic expansion with respect to
$h$:
$$
\lambda_D(h)=E_0+E_1h+E_2h^2+\cdots,
$$
where $E_1={\rm tr}\sqrt{\frac 12\frac{\p ^2V}{\p x^2}(x_0)}$ is the first eigenvalue of the
corresponding harmonic oscillator $-\Delta+\frac 12\left<\frac{\p ^2V}{\p x^2}(x_0)x,x\right>$. Moreover,  in a neighborhood $\omega$ of $x_0$, $u_D(x,h)$ can be written in the
WKB form:
\begin{equation}
\label{wkb}
u_D(x,h)=h^{-n/4}a(x,h)e^{-d(x_0,x)/h},
\end{equation}
where $a$ is a realization of a classical symbol:
\begin{equation}
\label{symbol}
a(x,h)\sim \sum_{j=0}^\infty a_j(x)h^j, \quad a_0(0)>0.
\end{equation}
\end{thm}

In the {\it sea} $\open^C$, on the other hand, we assume the {\it non-trapping
condition} on the energy surface $p^{-1}(E_0)=\{(x,\xi);p(x,\xi)=E_0\}$:
\begin{itemize}
\item[{\bf (A3)}]
For any $(x,\xi)\in p^{-1}(E_0)$ with $x\in\open^C$, the quantity
$|\exp tH_p(x,\xi)|$ tends to infinity as $|t|$ tends to $\infty$. 
\end{itemize}

Here $H_p$ is the Hamilton vector field
$$
H_p=\frac{\partial p}{\partial\xi}\cdot\frac{\partial}{\partial x}
-\frac{\partial p}{\partial x}\cdot\frac{\partial}{\partial\xi}
=2\xi\cdot\frac{\partial}{\partial x}-\frac{\p V}{\p x}\cdot\frac{\partial}{\partial\xi}.
$$

If $x\in\partial\open$, in particular, the only $\xi\in\R^n$ such that $p(x,\xi)=E_0$ is 0,
and $H_p=-\nabla V(x)\cdot\frac{\partial}{\partial\xi}$. Hence (A3) also implies,
\begin{equation}
\label{nabla}
\nabla V(x)\ne 0\quad{\rm on}\quad \partial\open.
\end{equation}

Under the conditions (A1), (A2), (A3), we have the following theorem (it is an analog to our situation of a result due to Heffer and Sj\"ostrand in the globally analytic case):

\begin{thm}\sl
\label{th3}
Assume (A1)--(A3).
Then, there exists a unique resonance $\rho(h)$ of $P$ such that $h^{-1}|\rho (h)- \lambda_D(h)| \rightarrow 0$ as $h\rightarrow 0_+$, and it verifies,
\begin{equation}
\label{compdiric1}
|\lambda_D(h)-\rho(h)|={\mathcal O}(e^{-(2S-\epsilon(\eta))/h}).
\end{equation}
Moreover, denoting by 
$u(x,h)$ the  corresponding (conveniently normalized) resonant state, one has,
\begin{equation}
\label{compdiric2}
|u_D(x,h)-u(x,h)|={\mathcal O}(e^{-(2S-d(x_0,x)-\epsilon(\eta))/h}),
\end{equation}
uniformly in $\overline{B_d(x_0,S-\eta)}$,
where $\epsilon(\eta)\to 0$ as $\eta\to 0$, and, for any $K\subset \R^n$ compact,
there exists $N_K\in \N$ such that,
\begin{equation}
\label{compdiric3}
\Vert e^{s(x)/h}u(x,h)\Vert_{H^1(K)}={\mathcal O}(h^{-N_K}),
\end{equation}
uniformly as $h\to 0$, where $s(x)=d(x_0,x)$ if $x\in
B_d(x_0,S)$ and $s(x)=S$ otherwise.
\end{thm}

Finally we assume some conditions on the set $\partial \open\cap
\overline{B_d(x_0,S)}$ and on the {\it caustic set} ${\mathcal
C}=\overline{\{x\in\open;d(x_0,x)=d(x,\partial\open)+S\}}$.

The points of the set $\partial \open\cap
\overline{B_d(x_0,S)}$ are called {\it points of type 1} in \cite{hs1}.
Since they mainly concentrate the interactions between the well and the sea,
here we prefer to call them {\it points of interaction}. 

Our additional assumption is,

\begin{itemize}
\item[{\bf (A4)}]
$\partial \open\cap \overline{B_d(x_0,S)}$ is a submanifold $\Gamma$ of $\partial\open$, and $\mathcal C$ has a contact of order exactly 2
with $\partial\open$ along $\Gamma$. 
\end{itemize}
We denote by $n_\Gamma$ ($\leq n-1$) the dimension of $\Gamma$.
Then, our main result is,

\begin{thm}\sl
\label{main}
Under Assumptions  (A1)--(A4),
there exists a classical symbol 
$$
f(h)\sim\sum_{j\geq 0}h^jf_j,
$$
with $f_0>0$, such that
$$
\im \rho(h)=-h^{\frac{1-n_\Gamma}2}f(h)e^{-2S/h}.
$$
\end{thm}

\begin{rem}
In the globally analytic case, it has been shown in \cite{hs1} that $f(h)$ is  an analytic
classical symbol. In the general case, however, such a property is no longer satisfied.
\end{rem}
\begin{rem}
Our arguments can also be applied to resonances of $P$ with a larger real part, and give, as in \cite{hs1} Theorem 10.14, an upper bound on their width when the corresponding eigenvalue of $P_D$ is asymptotically simple.
\end{rem}

\section{Proof of Theorem \ref{th3}}

Proceeding as in \cite{cmr} and \cite{ma2}, we consider the distorded operator $P_\theta$ with  $\theta = h\ln (1/h)$, constructed with the unitary operator $U_\nu$ as in (\ref{Unu}), and with $F(x)=0$ on some arbitrary large compact set $K\subset \R^n$, that we can assume to contain $\open$. Moreover, following the same idea as in \cite{hs1} Section 9, we also consider the corresponding operator $\tilde P_\theta$ where the well has been filled-up, that is,
$$
\tilde P_\theta = P_\theta + W(x),
$$
where $W\in C_0^\infty (\open )$, $W(x_0)>0$, $W\geq 0$ everywhere, ${\rm Supp}\hskip 1pt W$ arbitrarily small around $x_0$ (in particular, the Hamilton flow of $p+W$ has no trapped trajectories with energy $E_0$).

Then, proceeding as in \cite{cmr} Section 7 (or \cite{ma2} Section 4), one can construct a function $\psi_0\in C_0^\infty (\R^{2n}\backslash \Supp W)$ such that, 
\begin{equation}
\label{ellptilde}
-\im \tilde p_\theta (x-t \partial_x\psi_0-it\partial_\xi\psi_0, \xi -t\partial_\xi\psi_0+it\partial_x\psi_0)\geq \frac1{C_0}h\ln (1/h),
\end{equation}
for some constant $C_0>0$ large enough, where $t:=2C_0\theta$, $\tilde p_\theta$ is the semiclassical  principal symbol of $\tilde P_\theta$, and the inequality holds for  $(x,\xi)$ such that $|\re \tilde p_\theta (x,\xi) - E_0|\leq \jap{\xi}^2/C_0$.

In particular, by \cite{cmr} Proposition 7.2, we easily obtain,
\begin{equation}
\label{Fuj1}
\Vert e^{t\psi_0/h}Tv\Vert_{L^2(\R^{2n})} ={\mathcal O}(|h\ln h|^{-1})\Vert e^{t\psi_0/h}T(\tilde P_\theta - z)v\Vert_{L^2(\R^{2n})},
\end{equation}
uniformly for $h>0$ small enough, $v\in L^2(\R^n)$,  and $|z-E_0| < <h\ln(1/h)$. 
Here $T=T_1$ is the so-called FBI-Bargmann transform defined by \eq{defT}.

This means that the norm of $(\tilde P_\theta - z)^{-1}$ is uniformly ${\mathcal O}(|h\ln h|^{-1})$ when we consider $\tilde P_\theta$ as acting on the space $H_t := L^2(\R^n)$ endowed with the norm,
\begin{equation}
\label{Ht}
\Vert v\Vert_t := \Vert e^{t\psi_0/h}Tv\Vert_{L^2(\R^{2n})} .
\end{equation}
From this point, one can proceed exactly as in \cite{hs1}, Proof of Proposition 9.6 (that is, by considering a Grushin problem for $P_\theta$, the inverse of which is obtained by using  the corresponding Grushin problem for $P_D$, and by using Agmon-type estimates inside the island $\open$), and, working with $u_\theta:= U_{i\theta}u$ ($=u$ on $K$), one concludes (\ref{compdiric1}). (Indeed, this proof never uses the analyticity of $V$, but just the fact that one has a good enough control on the resolvent of the  ``filled-up well''-operator.)
\vskip 0.2cm
Similarly, (\ref{compdiric2}) is obtained as in the proof of \cite{hs1} Theorem 9.9, mainly by considering the spectral projector of $P_\theta$ as in \cite{hs1} Formula (9.37) (in that case, $u$ must be normalized, e.g., by requiring  that $\Vert u_\theta\Vert_t =1$).
\vskip 0.2cm
Moreover, the same arguments also show that,
$$
\Vert e^{t\psi_0/h}Tu_\theta\Vert_{L^2((\R^{n}\backslash B_d(x_0, S-\eta))\times \R^n)} = {\mathcal O}(e^{-(S-\epsilon (\eta ))/h}),
$$
and thus, also,
$$
\Vert u_\theta\Vert_{L^2(\R^{n}\backslash B_d(x_0, S-\eta))} = {\mathcal O}(e^{-(S-\epsilon' (\eta ))/h}).
$$
As a consequence, up to some constant factor of the type $1+{\mathcal O}(e^{-\delta/h})$ ($\delta >0$), the normalization of $u$ does not depend on the particular choices of $K$, $F$, and $\psi_0$.
\vskip 0.2cm
Now, we come to the proof of (\ref{compdiric3}).

Let $\chi_1\in C_0^\infty (B_d(x_0,S-\eta))$, such that $\chi_1 =1$ in $B_d(x_0,S-2\eta)$ ($\eta >0$ fixed arbitrarily small), and let $\chi_2\in C_0^\infty (B_d(x_0,\frac12(S+\eta))$, such that $\chi_2 =1$ in $B_d(x_0,\frac12(S-\eta))$. Setting,
\begin{equation}
\label{Fuj2}
R_\theta(z):= \chi_1 (P_D-z)^{-1}\chi_2 + (\tilde P_\theta -z)^{-1}(1-\chi_2),
\end{equation}
 one has (see \cite{hs1} Formula (9.39)),
\begin{equation}
\label{resolPtheta}
(P_\theta - z)R_\theta (z) = I + K_\theta (z),
\end{equation}
with,
\begin{eqnarray*}
K_\theta (z):&=& [P_\theta ,\chi_1](P_D-z)^{-1}\chi_2 - W(\tilde P_\theta -z)^{-1}(1-\chi_2)\\
&=& [-h^2\Delta ,\chi_1](P_D-z)^{-1}\chi_2 - W(\tilde P_\theta -z)^{-1}(1-\chi_2).
\end{eqnarray*}
Now, if $W$ is taken in such a way that ${\rm Supp}\hskip 1pt W \subset B_d(x_0,\eta )$, then, as in \cite{hs1} (in particular the proof of Lemma 9.4),  Agmon estimates show that,
$$
\Vert K_\theta\Vert ={\mathcal O}(e^{-(S-\varepsilon (\eta ))/2h}),
$$
uniformly with respect to $h>0$ small enough and $z\in\gamma$, and where $\varepsilon (\eta )\to 0$ as $\eta\to 0_+$ (here, $K_\theta$ is considered as an operator acting on $H_t$). In particular, we deduce from (\ref{resolPtheta}),
$$
(P_\theta -z)^{-1} = R_\theta (z)(I  + {\mathcal O}(e^{-(S-\varepsilon (\eta ))/2h})).
$$
Moreover, setting
\be
k:=h\ln\frac1{h},
\ee
by (\ref{Fuj1}) and (\ref{Fuj2}), for $z\in\gamma$, we have, 
$$
\Vert R_\theta (z)\Vert_{{\mathcal L}(H_t)} ={\mathcal O}(h^{-2} + k^{-1})={\mathcal O}(h^{-2}),
$$
and thus, we obtain,
\begin{equation}
\label{estresolv}
\Vert (P_\theta -z)^{-1}\Vert_{{\mathcal L}(H_t)} ={\mathcal O}(h^{-2}),
\end{equation}
uniformly for $z\in\gamma$ and $h>0$ small enough.

Now, we consider the Dirichlet realization $P_h$ of $P$ on the $h$-dependent domain,
$$
M_h:= \{ x\in \open\, ;\, {\rm dist}(x,\partial\open )\geq k^{2/3}\}.
$$
We denote by $\lambda_h$ its first eigenvalue, and by $v_h$ the corresponding normalized eigenstate. We also denote by $d_h$ the Agmon distance on $M_h$, associated with the pseudo-metric $(V-\lambda_h)_+dx^2$ (so that,  in particular, $d_h$ depends on $h$, too), and we set,
$$
\varphi_h(x):= d_h(x,x_0).
$$ 

At first, with $\chi_1$ as before, we observe that $(P_h-\lambda_D)\chi_1u_D = (P_D-\lambda_D)\chi_1u_D={\mathcal O}(h^\infty)$, and thus, by the Min-max principle (and since $V\geq E_0$ on $M_h$), we have $E_0\leq \lambda_h\leq \lambda_D +{\mathcal O}(h^\infty)$ (in particular, $\lambda_h =E_0+{\mathcal O}(h)$).
Moreover, since $x_0$ is the only point of $M_h$ where $V$ reaches its (non-degenerate) minimum $E_0$, and since $V-E_0\geq \delta k^{2/3} >> h$ near $\partial M_h$, standard techniques (see, e.g., \cite{hs2, si}) show that, actually, $\lambda_h = \lambda_D +{\mathcal O}(h^\infty )$, and the gap between $\lambda_h$ and the second eigenvalue of $P_h$ behaves like $h$. 

Since  $V-\lambda_h \leq V-E_0$, we also have, 
\begin{equation}
\label{compdist}
\varphi_h(x)\leq d(x_0,x).
\end{equation}
\begin{lem}\sl
\label{minorphi0}
There exists a constant $C_1\geq 0$, such that, for $x\in M_h$ and $h>0$ small enough, one has,
$$
\varphi_h(x)\geq s(x) -C_1k.
$$
where, as before,  $k=h\ln h^{-1}$.
\end{lem}
\begin{pre} We set $U_h^\pm :=\{ x\in M_h\, ;\, \pm V(x)\geq \pm \lambda_h\}$. Then, since $V(x)-E_0\sim k>>\lambda_h-E_0$ on $\partial M_h$, by definition, we have,
$$
\varphi_h(x) =d_h (U_h^-,x)=\inf_{\gamma \in \Gamma_x}\int_\gamma \sqrt{V(y)-\lambda_h}\hskip 1pt |dy|,
$$
where $\Gamma_x$ stands for the set of $C^1$ curves $\gamma \, :\, [0,1] \rightarrow U_h^+$  with $\gamma (0)\in U_h^-$ and $\gamma (1)=x$. Moreover,  since $\lambda_h =E_0+E_1h+{\mathcal O}(h^2) >E_0$, Assumptions (A2)-(A3) imply that $\nabla V \not=0$ on $\{ V=\lambda_h\}$ for $h>0$ small enough. Then, if $\varphi_h(x)< S_h:=d_h(x_0,\open )$, standard arguments of Riemannian geometry (exploiting the fact that, in that case,  $\dot\gamma (t)$ remains colinear to $\nabla \varphi_h (\gamma (t))$ for any minimal geodesic $\gamma$:  see, e.g., \cite{Mi, hs2}, and \cite{ma3} Section 3) show that, for $V(x)>\lambda_h$, $\varphi_h(x)$  is reached at some minimal geodesic $\gamma$ that can be re-parametrized in such a way that the map $t\mapsto (\gamma (t), \frac12\dot\gamma (t))$ becomes a nul-bicharacteristic of $q_h(x,\xi):= \xi^2 -(V(x)-\lambda_h)$. In particular,  such a $\gamma$ verifies $|\dot\gamma (t)| = 2\sqrt{V(\gamma (t))-\lambda_h}$ , and thus, we obtain,
$$
\varphi_h (x) =2\int_0^{T_x} (V(\gamma(t)) -\lambda_h)dt,
$$
where $T_x=T_x(h) >0$ represents the time employed to go from $U_h^-$ to $x$ in the new parametrization. Writing $\lambda_h = E_0 +h\mu_h$, where $\mu_h =E_1+{\mathcal O}(h)$, this gives,
\begin{eqnarray}
\varphi_h (x) &=&2\int_0^{T_x} (V(\gamma(t)) -E_0)dt -2T_xh\mu_h\nonumber\\
&=& \int_0^{T_x} \sqrt{V(\gamma(t)) -E_0}\sqrt{|\dot\gamma (t)|^2 + 4h\mu_h}\hskip 2pt dt -2T_xh\mu_h\nonumber\\
&\geq&  \int_\gamma \sqrt{V(y) -E_0}\hskip 2pt d|y| -2T_xh\mu_h\nonumber\\
\label{compphid}
&\geq& d(U_h^-,x) -2T_xh\mu_h.
\end{eqnarray}
Moreover, since, for $x$ close to $x_0$, we have $d(x_0,x)={\mathcal O}(|x-x_0|^2)$, we see that  $d(U_h^-,x_0)={\mathcal O}(h)$, and thus, by the triangular inequality, $d(x_0,x)\leq d(U_h^-,x) +Ch$ for some constant $C>0$. Combining with (\ref{compphid}), we obtain,
\begin{equation}
\label{estphi1}
\varphi_h (x)\geq d(x_0,x)-Ch-2T_xh\mu_h\geq s(x)-Ch-2T_xh\mu_h,
\end{equation}
still for $x$ verifying $\varphi_h(x) <S_h$. 
Thus, in  that case, it remains only to prove that $T_x={\mathcal O}(\ln\frac1{h})$. To do so, we set $(x(t),\xi(t)):=(\gamma (t), \frac12\dot\gamma (t))=\exp tH_q(x_h, 0)$, where $H_q(x,\xi):=(2\xi , \nabla V(x))$ is the Hamilton field of $q_h$, and $x_h:=\gamma (0)\in U_h^-$. If $\varepsilon >0$ is any arbitrarily small fixed number, we see that  $H_q(x,\xi)$ remains outside some fixed neighborhood of $0$ on $\{ (x,\xi)\, ; |x-x_0|\geq \varepsilon,\, q_h(x,\xi)=0\}$. As a consequence, if $|x-x_0|\geq\varepsilon$, the time employed  by $\gamma$ to go from $x$ to the set $\{ y\,; |y-x_0|=\varepsilon \}$ is bounded, uniformly with respect to $h$. Therefore, it remains to estimate the time $\tilde T$ employed by $\gamma$ to go from $\gamma (0)\in U_h^-$ to $\{ y\,; |y-x_0|=\varepsilon \}$. Since we stay in an arbitrarily small neighborhood of $x_0$, we can assume $x(t)\cdot \nabla V(x(t)) \geq 4\delta^2 |x(t)|^2$ for $t\in [0,\tilde T]$ and  with some $\delta >0$ constant. Then, setting $f(t):= x(t)\cdot\xi(t)/|x(t)|^2$, we compute,
$$
\dot f(t)  +4(f(t))^2 = \frac{2|\xi(t)|^2+x(t)\cdot \nabla V(x(t)}{|x(t)|^2} \geq 4\delta^2.
$$
Therefore, on its domain of definition, the function  $g(t):=(\delta -f(t))^{-1}$ verifies,
$$
\dot g\geq 4\frac{\delta^2-f^2}{(\delta -f)^2}=4\frac{\delta +f}{\delta -f}= 8\delta g -4,
$$
and thus, since $g(0)=\delta^{-1}$, we easily deduce,
$$
g(t)\geq \frac1{2\delta}(1+e^{8\delta t}),
$$
as long as $f(t)<\delta$, and thus,
$$
f(t)\geq \delta -\frac{2\delta}{1+e^{8\delta t}},
$$
on the same interval.
Now, if $f(t_1)=\delta$ for some $t_1\in[0,\tilde T)$, we fix $\delta_1<\delta$ arbitrary, and, for $t$ close to $t_1$, we set $g_1(t):=(\delta_1-f)^{-1}$. Using that $\dot f(t)  +4(f(t))^2\geq 4\delta_1^2$, the same procedure gives,
$$
g_1(t)\geq \frac1{2\delta_1} - (\frac1{2\delta_1} +\frac1{\delta -\delta_1}) e^{8\delta_1 (t-t_1)},
$$ 
as long as $f(t)>\delta_1$, that is,
$$
\frac1{f(t)-\delta_1}\leq (\frac1{2\delta_1} +\frac1{\delta -\delta_1}) e^{8\delta_1 (t-t_1)}-\frac1{2\delta_1}.
$$
In particular, $(f(t)-\delta_1)^{-1}$ remains bounded on any finite interval $[t_1, T_1]$ where $f(t)>\delta_1$, and this means that $f(t)$ cannot take the value $\delta_1$ on $[t_1, \tilde T]$. Thus, in this case, $f(t)$ necessarily remains $\geq \delta$ on $[t_1, \tilde T]$. Summing up, we have proved,
\begin{equation}
\label{estsurf}
f(t)\geq \delta -\frac{2\delta}{1+e^{8\delta t}},
\end{equation}
on the whole interval $[0,\tilde T]$. Since,
$$
\frac{d}{dt}\ln |x(t)| = \frac{x(t)\cdot \dot x(t)}{|x(t)|^2}=2f(t),
$$
and $|x(0)|\geq \delta'\sqrt h$ for some $\delta'>0$ constant, we deduce from (\ref{estsurf}),
$$
\ln |x(t)|\geq \ln (\delta'\sqrt h) + 2\delta t -\int_0^t \frac{2\delta ds}{1+e^{8\delta s}}\geq \ln (\delta'\sqrt h) + 2\delta t -\int_0^{+\infty} \frac{2\delta ds}{1+e^{8\delta s}},
$$
and thus, on $[0,\tilde T]$,
\begin{equation}
\label{estnormx}
|x(t)| \geq \delta''\sqrt he^{2\delta t},
\end{equation}
with $\delta'' = \delta'e^{-C_2}$, $C_2:=\int_0^{+\infty} \frac{2\delta ds}{1+e^{8\delta s}}$. Since $\delta''\sqrt he^{2\delta t}=\varepsilon$ when $t=(2\delta )^{-1}\ln (\varepsilon /\delta''\sqrt h)$, we deduce from (\ref{estnormx}) that, necessarily, one has $\tilde T\leq (2\delta )^{-1}\ln (\varepsilon /\delta''\sqrt h)$, and, by (\ref{estphi1}), it follows that,
\begin{equation}
\label{estphi2}
\varphi_h (x)\geq  s(x)-C_1'k,
\end{equation}
for $x$ verifying $\varphi_h(x) <S_h$, and some constant $C_1'>0$, independent of $x$. On the other hand, $\varphi_h(x)$ reaches $S_h$ at some point $x_h$ verifying $V(x_h)=\lambda_0=E_0+{\mathcal O}(h)$ (and $x_h$ away from some fix neighborhood of $x_0$), and thus ${\rm dist}(x_h,\open) ={\mathcal O}(h)$. As a consequence, $d(x_h,\open)={\mathcal O}(h)$, too, and thus, $d(x_0,x_h)\geq S-Ch$ for some constant $C>0$. Therefore, by continuity, 
(\ref{estphi2}) also proves that $S_h\geq S-C_1k$ with $C_1=C'_1+C$, and it follows that (\ref{estphi2}) is still valid if $\varphi_h(x)\geq S_h$ (since, by definition, $s(x)\leq S$ everywhere).
\end{pre} 
\begin{lem}\sl
\label{estuh}
There exists a constant $N_0\geq 0$, such that,
$$
\Vert e^{\varphi_h/h}v_h\Vert_{H^1(M_h)} ={\mathcal O}(h^{-N_0}),
$$
uniformly for $h>0$ small enough.
\end{lem}
\begin{pre} Following \cite{hs2}, we set,
$$
\phi  (x) := \left\{
\begin{array}{l}
\varphi_h(x) -Ch\ln [\varphi_h(x)/h] \, \mbox{ if } \varphi_h(x)\geq Ch;\\
\varphi_h(x)-Ch\ln C\,  \mbox{ if } \varphi_h(x)\leq Ch,
\end{array}
\right.
$$
where $C\geq 1$ is some constant that will be fixed large enough later on.
Then, we use the following identity (that is at the origin of Agmon estimates: see, e.g., \cite{hs2} Theorem 1.1),
\begin{eqnarray}
&&\re \langle e^{\phi  /h}(P-\lambda_h)v_h, e^{\phi  /h}v_h\rangle\nonumber\\ 
\label{Agmon}
&& = h^2\Vert \nabla(e^{\phi /h}v_h)\Vert^2  + \langle \left(V-\lambda_h-(\nabla\phi )^2\right)e^{\phi /h}v_h,e^{\phi /h}v_h\rangle.
\end{eqnarray}
Now, if $C$ is taken large enough, by  (\ref{compdist}) we see that $M_h\cap \{ \varphi_h(x)\geq Ch\} \subset \{ V\geq \lambda_h\}$. Moreover, on this set we have,
$$
\nabla\phi  = (1-Ch/\varphi_h)\nabla\varphi_h,
$$
and thus, using that $(\nabla\varphi_h)^2\leq (V-\lambda_h)_+$,
$$
V-\lambda_h-(\nabla\phi )^2\geq V-\lambda_h- (1-Ch/\varphi_h)^2(V-\lambda_h)\geq Ch\frac{V-\lambda_h}{\varphi_h}.
$$
Writing again $\lambda_h =E_0 +\mu_hh$ with $\mu_h=E_1+{\mathcal O}(h)$, and using again (\ref{compdist}) and the fact that $d(x_0,x)\leq C_0(V(x)-E_0$ near $x_0$ (for some $C_0>0$ constant), we deduce,
$$
V(x)-\lambda_h-(\nabla\phi (x))^2 \geq Ch\frac{V(x)-E_0}{d(x_0,x)} -Ch\frac{\mu_hh}{\varphi_h}\geq \frac{Ch}{C_0}-\mu_hh,
$$
on $M_h\cap \{ \varphi_h(x)\geq Ch\}$, and thus, taking $C$ large enough,
$$
V(x)-\lambda_h-(\nabla\phi (x))^2 \geq \frac{Ch}{2C_0},
$$
on the same set. Inserting this estimate into (\ref{Agmon}), and using that the left-hand side  is 0, we obtain,
$$
h^2\Vert \nabla(e^{\phi /h}v_h)\Vert^2+h\Vert e^{\phi /h}v_h\Vert_{\{\varphi_h\geq Ch\}}^2={\mathcal O}(\Vert e^{\phi /h}v_h\Vert_{\{\varphi_h\leq Ch\}}^2),
$$
and thus, since $\phi \leq Ch(1-\ln C)\leq 0$ on $\{\varphi_h\leq Ch\}$, we conclude,
$$
h^2\Vert \nabla(e^{\phi /h}v_h)\Vert^2+h\Vert e^{\phi /h}v_h\Vert^2={\mathcal O}(1).
$$
Then, observing that $e^{\phi   /h}\geq (h/M)^C e^{\varphi_h /h}$ with $M:=\sup\varphi_h={\mathcal O}(1)$, the result follows.
\end{pre}

Now, let $\chi_h\in C_0^\infty (M_h)$, such that $\chi_h=1$ on $\{ x\in\open\, ;\, {\rm dist}(x,\partial\open )\geq 2k^{2/3}\}$ and, for all $\alpha$, $\partial^\alpha \chi_h ={\mathcal O}(k^{-2|\alpha|/3})$ (such a $\chi_h$ exists because $\partial\open$ is a hypersurface of $\R^n$). In particular, $\chi_h v_h$ is in the domain of $P_\theta$, and one has,
\begin{lem}\sl
\label{chihuh}
There exists a constant $N_1\geq 0$, such that,
$$
\left\Vert\frac1{2i\pi}\int_{\gamma} (z-P_\theta)^{-1}\chi_h v_h dz - \chi_h v_h\right\Vert_t ={\mathcal O}(h^{-N_1}e^{-S/h}),
$$
uniformly for $h>0$ small enough.
\end{lem}
\begin{pre} Setting $w_h:=\chi_h v_h$, we have,
\begin{equation}
\label{Pv0}
P_\theta w_h = P_h w_h= \lambda_h w_h + [P, \chi_h]v_h = \lambda_h w_h -2h^2(\nabla \chi_h)(\nabla v_h) -h^2(\Delta \chi_h)v_h.
\end{equation}
Moreover, on the support of $\nabla\chi_h$, we have ${\rm dist}(x, \partial\open) \leq 2k^{2/3}$, and thus, by standard properties of the Agmon distance near $\partial\open$ (see, e.g., below Lemma \ref{criticalpoint}, and \cite{hs1} Section 10), we see that $s(x)\geq S-Ck$ for some constant $C>0$. Therefore, by Lemma \ref{minorphi0}, we also have $\varphi_h(x)\geq S-C'k$ ($C'=C+C_0$), and by Lemma \ref{estuh}, we deduce from (\ref{Pv0}),
$$
\Vert  (P_\theta -\lambda_h)w_h\Vert_{L^2} ={\mathcal O}(h^{-N}e^{-S/h}),
$$
with some constant $N\geq 0$. As a consequence, since $\psi_0$ is bounded and $T$ is an isometry, $r_h:=(P_\theta -\lambda_h)w_h$ verifies,
\begin{equation}
\label{estrh}
\Vert r_h\Vert_t =\Vert h^{-2C_0\psi_0}Tr_h\Vert ={\mathcal O}(h^{-M})\Vert r_h\Vert_{L^2} ={\mathcal O}(h^{-M'}e^{-S/h}),
\end{equation}
with $M,M'>0$ constant. 
Then, writing,
\begin{eqnarray}
\frac1{2i\pi}\int_{\gamma} (z-P_\theta)^{-1}w_h dz - w_h&=& \frac1{2i\pi}\int_{\gamma} [(z-P_\theta)^{-1}-(z-\lambda_h)^{-1}]w_h dz\nonumber\\
\label{estdiff}
&=& \frac1{2i\pi}\int_{\gamma} \frac1{z-\lambda_h}(z-P_\theta)^{-1}r_h dz,
\end{eqnarray}
and using (\ref{estresolv}) and (\ref{estrh}), the result immediately follows.
\end{pre}

Using the equation $(P_h-\lambda_h)v_h=0$ and the ellipticity (in the standard sense) of $P_h$, it is not difficult to deduce from (\ref{estrh}) and Lemma \ref{estuh}, that, for all $\ell\geq 0$, there exists $M_\ell\geq 0$ such that,
\begin{equation}
\label{estrh1}
\Vert P_\theta^\ell r_h\Vert_t ={\mathcal O}(h^{-M_\ell}e^{-S/h}).
\end{equation}
As a consequence, applying $P_\theta^\ell$ to (\ref{estdiff}),  we deduce from (\ref{estrh1}),
\begin{equation}
\label{iterP}
\left\Vert P_\theta^\ell \left(\frac1{2i\pi}\int_{\gamma} (z-P_\theta)^{-1}\chi_h v_h dz - \chi_h v_h\right) \right\Vert_t ={\mathcal O}(h^{-N_\ell}e^{-S/h}),
\end{equation}
for all $\ell \geq 0$ and some $N_\ell\geq 0$ constant. Then, using the ellipticity of $P_\theta$ and the fact that, for all $u$,   $\Vert u\Vert_{L^2} =\Vert Tu\Vert_{L^2}={\mathcal O}(h^{-M}\Vert u\Vert_t)$, we deduce from (\ref{iterP}),
\begin{equation}
\label{estHs1}
\left\Vert \frac1{2i\pi}\int_{\gamma} (z-P_\theta)^{-1}\chi_h v_h dz - \chi_h v_h\right\Vert_{H^s(\R^n)} ={\mathcal O}(h^{-N_s}e^{-S/h}),
\end{equation}
for all $s\geq 0$ and some constant $N_s\geq 0$.

It also follows from Lemmas \ref{estuh} and \ref{chihuh}, that,
$$
\Vert \frac1{2i\pi}\int_{\gamma} (z-P_\theta)^{-1}\chi_h v_h dz\Vert_t = 1+{\mathcal O}(e^{-\delta /h}),
$$
 for some $\delta >0$ constant, and therefore, we necessarily have, 
$$
u_\theta =\frac{\alpha'}{2i\pi}\int_{\gamma} (z-P_\theta)^{-1}\chi_h v_h dz,
$$
with $|\alpha'|=1+{\mathcal O}(e^{-\delta /h})$.  In particular, by (\ref{estHs1}), 
\begin{equation}
\label{estHs2}
\Vert u_\theta -\alpha'\chi_h v_h\Vert_{H^s} = {\mathcal O}(h^{-N_s}e^{-S/h}).
\end{equation}
Then, since $u_\theta = u$ on $K$, (\ref{compdiric3}) easily follows from Lemma \ref{minorphi0}, Lemma \ref{estuh}, and (\ref{estHs2}).

\section{Extension of the WKB Solution}\label{sec-ExtWB}
Let $x^1\in \Gamma$.
In this section, we will extend to a neighborhood of $x^1$ the WKB solution,
\begin{equation}
\label{wkbw}
w(x,h)\approx h^{-n/4}a(x,h)e^{-d(x_0,x)/h},
\end{equation}
that approximates both the eigenfunction $u_D(x,h)$ of $P_D$ near $x_0$ (see
\eq{wkb}) and the resonant state $u(x,h)$ (Theorem \ref{th3}).

\subsection{Extension up to the Caustic Set}
\label{coord}
The extension in the island will be done along the geodesic with respect to
the Agmon distance from $x_0$ to $x^1$.

Let $q(x,\xi)=\xi^2-V(x)$ and
$H_q=2\xi\cdot\partial/\partial x+\partial_xV(x)\cdot\partial/\partial\xi$
its Hamilton vector field. Then, one can see as in \cite{hs1} that the integral curve
$\tilde\gamma(t)=(\tilde x(t),\tilde \xi(t))$ of $H_q$ starting at $(\tilde x(0),\tilde \xi(0))=(x^1,0)$ verifies,
$$
(x(-\infty),\xi(-\infty))=(x_0,0).
$$
Moreover, its projection on the $x$-space is the unique minimal geodesic between $x_0$ and $x^1$ staying in $\open\cup\{ x^1\}$ (see \cite{hs1} Section 10 and \cite{hs2}).

If $\phi(x):=d(x_0,x)-S$, we learn from \cite{hs2} that $\phi$ is $C^\infty$ in a neighborhood $\Omega$ of 
$\tilde x ([-\infty ,0))$, and  the Lagrangian manifold,
$$
\Lambda=\left\{(x, \nabla\phi (x)) \, ;\, x\in\Omega\right\},
$$
is  the outgoing stable manifold of dimension $n$ associated to the fixed point $(x_0,0)$ of $H_q$.

In particular, $\Lambda$ is $H_q$-invariant and contains $\tilde\gamma([-\infty,0))$. Moreover, since $H_q$ does not vanish at $\tilde\gamma(0)=(x^1,0)$ by
(A3),  $\Lambda$ can be extended by the flow of  $H_q$ to a larger Lagrangian manifold (that we still denote by $\Lambda$)  that  contains 
$\tilde\gamma([-\infty,0])$.
The natural projection $\Pi : \Lambda\to \R_x^n$ is
singular at
$\tilde\gamma(0)$, and, as  shown in \cite{hs1} Lemma 10.1,
the kernel of $d\Pi(x^1,0)$ is a one-dimensional vector space
generated by $H_q(x^1,0)$.

Let us choose Euclidian coordinates $x$ centered at $x^1$ such
that $T_{x^1}(\partial \open)$ is given by $x_n=0$, and
$\partial/\partial x_n$ is the exterior normal of $\open$ at
this point.
Then by Assumption (A3),
\begin{equation}
\label{local}
V(x)-E_0=-C_0x_n+W(x),
\end{equation}
where $C_0>0$ is a constant and $W(x)={\mathcal O}(|x|^2)$.
In a neighborhood of the point $\tilde\gamma(0)$, the
Lagrangian manifold $\Lambda$ is defined by a real-valued $C^\infty$ function
$g(x',\xi_n)$ with $g(0)=0$, $dg(0)=0$, that is, 
\begin{equation}
\label{lag}
\Lambda=\left\{(x,\xi);\xi'=\frac{\p g}{\p x'}(x',\xi_n),\,\,x_n=-\frac{\p g}{\p \xi_n}(x',\xi_n)\right\},
\end{equation}
Moreover, there exist real-valued smooth functions $\xi^c_n(x')$, $a(x')$, $b(x')$ and
$\nu_0(x',\xi_n)$, $\nu_1(x',\xi_n)$,
such that,
\begin{eqnarray}
\label{odex'2}
 &&|\xi^c_n(x')|+ |a(x')| + |b(x')|={\mathcal O}(|x'|^2) \mbox{ as } |x'|\to 0;\\
\label{odex'22}
 &&\nu_0 = \nu_1+{\mathcal O}(|\xi_n-\xi_n^c(x')|) = \frac1{C_0} + {\mathcal O}(|x'| + |\xi_n|);\\
 \label{caustics0}
&&g(x',\xi_n)=a(x')+b(x')(\xi_n-\xi_n^c(x'))+\frac 13\nu_0(x',\xi_n)
(\xi_n-\xi_n^c(x'))^3;\\
 \label{caustics1}
&&\frac{\p g}{\p\xi_n}(x',\xi_n)=b(x')+\nu_1(x',\xi_n)
(\xi_n-\xi_n^c(x'))^2.
\end{eqnarray}
All these properties are proved in \cite{hs1} (pages 136-148) and do not require the analyticity of the potential.

Then, near $(x^1, 0)$, the caustic set ${\mathcal C}$ (that, by definition, is the set where $\phi$ fails to be smooth, and thus, the set of $x$ for which the roots $\xi_n(x)$ of the equation $x_n=-\partial_{\xi_n}g(x',\xi_n)$ are not smooth) is  given by,
$$
{\mathcal C}=\{x\, ;\, x_n+b(x')=0\}.
$$
It is shown in \cite{hs1} Lemma 10.2, that there exists a positive constant $C$ such that
\begin{equation}
\label{lemma}
\phi(x)|_{\mathcal C}\ge C(V(x)-E_0).
\end{equation}
(The proof of (\ref{lemma}) does not use the analyticity, but rather the fact that, since $V|_{\mathcal C}\geq 0$, one has  $|\nabla (V\left\vert_{\mathcal C})\right.|={\mathcal O}(\sqrt{V|_{\mathcal C}})$.)

This estimate together with  the assumptions (A3) and (A4) mean that
$\phi(x)|_{\mathcal C}$ is non-negative and quadratic along $\Gamma$, with $\phi\left\vert_\Gamma\right. =0$.

Let 
$\tilde\Omega$ be a small neighborhood of
$\gamma([-\infty,0])$ and let
\begin{equation}
\label{omega}
\tilde\Omega_+=\{x\in\tilde\Omega;x_n+b(x')>0\},\quad
\tilde\Omega_-=\tilde\Omega\backslash (\tilde\Omega_+\cup {\mathcal C}).
\end{equation}

Then the phase function $-d(x_0,x)=-\phi(x)-S$ of \eq{wkbw} is a
$C^\infty$ function defined in $\tilde\Omega_-$.
The symbol
$a(x,h)$ can also be extended to $\tilde\Omega_-$ by solving successively
for $a_j(x)$ in \eq{symbol} the transport equations (that are first-order ordinary
differential equations along the integral curves of
$H_q$), and by re-summing the series $\sum_{j\le 0}h^ja_j$.

\medskip
More generally, the previous arguments also permit us to extend $w$ in the open set defined as the union of all smooth minimal geodesics included in $\open$ and starting from $x_0$. We denote  this set by $\bf\Omega$.

\subsection{Extension Beyond the Caustic Set}
In order to extend the WKB solution $w$ beyond the
caustic set, we follow the idea of \cite{hs1} and represent
$h^{n/4}e^{S/h}w$ in the integral form,
\begin{equation}
\label{airy}
I[c](x,h)=h^{-1/2}\int_{\gamma(x)}e^{-(x_n\xi_n+
g(x',\xi_n))/h} c(x',\xi_n,h)d\xi_n.
\end{equation}
For $x$ in $\tilde\Omega_-$ close to $x^1$, the phase function
$x_n\xi_n+g(x',\xi_n)$ has two real critical points (see \eq{caustics1}).
The steepest descent method at one of these points gives us the asymptotic
expansion of
$I[c]$. Comparing this with the symbol $a$, we can determine 
$c(x',\xi_n;h)$ so that the asymptotic expansion of
$e^{S/h}w$ coincides with that of $I[c]$ in $\tilde\Omega_-$.
 
Hence, in order to determine a possible asymptotic expansion of $e^{S/h}w$ in
$\tilde\Omega_+$, it is enough to compute it for $I[c]$.

If $g$ was analytic with respect to $\xi_n$, we would find two complex
critical points for $x\in \tilde\Omega_+$, one of them corresponding to an
outgoing solution (i.e. resonant state). 
In our
$C^\infty$ case, however, $g$ is only defined for real $\xi_n$.
So we will extend $g(x',\cdot)$ {\it almost-analytically}  (see Appendix for the
definition) to a ($h$-dependent) small complex
neighborhood of $\xi_n=0$. Then, we will apply the 
steepest descent method for $x\in\tilde\Omega_+$ sufficiently close to
$x^1$ so that the imaginary part of the critical point $\xi_n$ remains sufficiently
small.

In the following, we carry out the above procedure in several steps.

\subsubsection{Integral Representation in $\tilde\Omega_-$}
We first determine the
$C^\infty$ classical symbol $c(x',\xi_n,h)\sim\sum_{j=0}^\infty
c_j(x',\xi_n)h^j$ and the integration contour $\gamma(x)$ for
$x\in\tilde\Omega_-$.

Let $x^3$ be in $\tilde\Omega_-$ close to $x^1$ and $\tilde U$ a small
neighborhood of $x^3$. The critical points of the phase function
$x_n\xi_n+g(x',\xi_n)$ are the zeros of the function
$x_n+\frac{\p g}{\p \xi_n}(x',\xi_n)$. From
\eq{caustics1}, we see that for $x\in\tilde U$, there are two real critical
points
$\xi_n^+(x)$, $\xi_n^-(x)$, and they verify,
$$
\xi_n^\pm(x)\sim
\xi_n^c(x')\pm\sqrt{\frac{-(x_n+b(x'))}{\nu_1(x',\xi_n^c(x'))}},
$$
as $|x_n+b(x')|\to 0$.
We define a sufficiently small real open interval $\gamma(x)$ so that it
contains $\xi_n^+(x)$ inside as the only non-degenerate minimal point of
$x_n\xi_n+g(x',\xi_n)$. The minimal value is
\begin{equation}
\label{repphi}
\phi(x)=x_n\xi_n^+(x)+g(x',\xi_n^+(x)).
\end{equation}

Then by the steepest descent method, we obtain the asymptotic expansion as
$h\to 0$ of $I[c]$:
$$
I[c](x,h)\sim e^{-\phi(x)/h}\sum_{j=0}^\infty b_j(x)h^j,
$$
for some $C^\infty$ functions $b_j(x)$ defined on $\tilde U$.
In particular,
$$
b_0(x)=\sqrt{\frac\pi{r(x)}}c_0(x',\xi_n^+(x)),
$$
where
$$
r(x)=\frac 12\frac{\p^2 g}{\p\xi_n^2}(x',\xi_n^+(x)).
$$
Moreover, the map that associates a sequence of functions $\{b_j(x)\}_{j=0}^\infty$
on $\tilde U$ to a sequence of functions $\{c_j(x',\xi_n)\}_{j=0}^\infty$
on $U=\{(x',\xi_n^+(x));x\in\tilde U\}$ is
bijective, and we define the function $c(x',\xi_n,h)$ as a realization of
the inverse image of 
$\{a_j(x)\}_{j=0}^\infty$ by this map. In particular,
\begin{equation}
\label{c0}
c_0(x',\xi_n^+(x))=\sqrt{\frac{r(x)}\pi} a_0(x).
\end{equation}

\subsubsection{Extension of $c(x',\xi_n,h)$ to a Neighborhood of
$(x',\xi_n)=(0,0)$}

The symbol $c(x',\xi_n;h)$, previously defined in $U$, formally verifies,
\begin{equation}
\label{tildep}
e^{g/h}(\hat P-\rho(h))(e^{-g/h}c)\sim 0.
\end{equation}
Here $\hat P=-h^2\Delta_{x'}-\xi_n^2+V(x',h\frac{\p}{\p\xi_n})$, where
$V(x',h\frac{\p}{\p\xi_n})$ is considered as a pseudodifferential operator
whose action on $e^{-g/h}c$ is defined by the standard asymptotic
expansion,
\begin{eqnarray*}
V(x',h\frac{\p}{\p\xi_n})(e^{-g/h}c)&:=& e^{-g/h}\sum_{\ell\geq 0}\frac{h^\ell}{\ell !}\p_{x_n}^\ell V(-\p_{\xi_n}g)\\
&& \hskip 2cm \times \p_{\eta}^\ell \left( c(x',\eta)e^{-\kappa (x',\xi_n,\eta)/h}\right)\left\vert_{\eta=\xi_n}\right.,
\end{eqnarray*}
where $\kappa (x',\xi_n,\eta):=g(x',\eta)-g(x',\xi_n)-(\eta-\xi_n)\p_{\xi_n}g(x',\xi_n)$.

\eq{tildep} leads us to transport equations, which are also differential
equations along the integral curves of $H_q$ on $\Lambda$. The flows emanating from $U$ covers a full neighborhood
of $(x',\xi_n)=(0,0)$, and thus we have extended $c$ there.

\subsubsection{Critical Points and the Extension of $\phi$}

Let $N\geq 1$, $k=h\ln\frac1{h}$, and let $\tilde \nu_0$ be a holomorphic $(Nk)^{1/3}$-approximation of $\nu_0$ with respect to $\xi_n$ (in the sense of Lemma \ref{almost}), where $\nu_0$ is the function appearing in (\ref{caustics0}). Then, setting,
\be
\label{holapproxg}
\tilde g(x',\xi_n):=a(x')+b(x')(\xi_n-\xi_n^c(x'))+\frac 13\tilde \nu_0(x',\xi_n)
(\xi_n-\xi_n^c(x'))^3
\ee
we see that
$\tilde g$ be a holomorphic $(Nk)^{1/3}$-approximation of $g$ with respect to $\xi_n$, and
we look for the critical points of 
$\xi_n\mapsto x_n\xi_n+\tilde g(x',\xi_n)$, that is,  the roots of the equation with respect
to $\xi_n$,
\begin{equation}
\label{critical}
x_n+\frac{\p\tilde g}{\p\xi_n}(x',\xi_n)=0.
\end{equation}
Recalling the definition of $\nu_1(x',\xi_n)$ in (\ref{caustics1}), we choose a constant $c_1>0$ such that $c_1 < \inf_{\Omega'}\nu_1 $, where $\Omega'$ is some fixed small enough neighborhood of $(x',\xi_n)=(0,0)$.
\begin{lem}\sl
\label{criticalpoint}
Let $x\in\tilde\Omega_+\cap\{x_n+b(x')\leq c_1(Nk)^{2/3}\}$.
Then, the equation \eq{critical} has
two complex roots
$\xi_n^{-i}(x)$, $\xi_n^{+i}(x)$ satisfying
$$
\xi_n^{\pm i}(x)\sim
\xi_n^c(x')\pm i\sqrt{\frac{x_n+b(x')}{\tilde\nu_1(x',\xi_n^c(x'))}}
$$
as $x_n+b(x')$ tends to 0, where $\tilde\nu_1$ is a holomorphic $(Nk)^{1/3}$-approximations in the $\xi_n$-variable of 
 $\nu_1(x',\xi_n)$.
Moreover, setting,
\begin{equation}
\label{tildephi}
\tilde\phi(x)=x_n\xi_n^{-i}(x)+\tilde g(x',\xi_n^{-i}(x)),
\end{equation}
one has,
\begin{equation}
\label{imphi}
\im \nabla_x\tilde\phi(x)=-\frac 1{\sqrt{\tilde \nu_1(x', \xi_n^c(x'))}}(x_n+b(x'))^{1/2}\nabla(x_n+b(x'))
+{\mathcal O}(x_n+b(x')),
\end{equation}
and there exists $\varepsilon (h)={\mathcal O}(h^\infty )$ real, such that,
\begin{equation}
\label{reimphi}
\re\tilde\phi (x)\geq \varepsilon (h),
\end{equation}
for all $x\in\tilde\Omega_+\cap\{x_n+b(x')\leq c_1(Nk)^{2/3}\}$ and $\alpha\in\N^n$.
\end{lem}

\begin{pre}
From \eq{holapproxg}, we have,
\begin{equation}
\label{caustics00}
\begin{array}{rl}
x_n\xi_n+\tilde g(x',\xi_n)=&a(x')+x_n\xi_n^c(x')\\[10pt]
+&(x_n+b(x'))(\xi_n-\xi_n^c(x'))+\frac 13\tilde\nu_0(x',\xi_n)
(\xi_n-\xi_n^c(x'))^3,
\end{array}
\end{equation}
\begin{equation}
\label{caustics10}
x_n+\frac{\p \tilde g}{\p\xi_n}(x',\xi_n)=x_n+b(x')+\tilde\nu_1(x',\xi_n)
(\xi_n-\xi_n^c(x'))^2,
\end{equation}
where, actually,  $\tilde\nu_1(x',\xi_n)$ is a holomorphic $(Nk)^{1/3}$-approximations in the $\xi_n$-variable of 
 $\nu_1(x',\xi_n)$.

We set,
$$
x_n+b(x')=-z^2.
$$ 

If $\xi_n$ is a critical point of the phase, the left-hand side of \eq{caustics10} vanishes, and one has,
$$
z=\sqrt{\tilde\nu_1(x',\xi_n)}(\xi_n-\xi_n^c(x')).
$$
Since $\tilde\nu_1(x',\xi_n^c(x'))= 1/C_0 + {\mathcal O}(|x'|)$, and $\tilde\nu_1(x',\xi_n)$ is holomorphic with respect to $\xi_n$ in $\{| \im \xi_n |\leq (Nk)^{1/3}\}$, for $z$ and $x'$ small enough this equation 
is solvable with respect to $\xi_n$, and the solution is given by the Lagrange inversion formula,
$$
\xi_n= \xi_n^c(x')+Y(x', z),
$$
with,
\begin{equation}
\label{defY}
Y(x', z):=\sum_{k=1}^\infty \frac{d^{k-1}}{d\xi_n^{k-1}}\left({\tilde\nu_1(x',\xi_n)}\right)^{-k/2}\left|_{\xi_n=\xi_n^c(x')}\right. \frac{z^k}{k!},
\end{equation}
that  is holomorphic
with respect to
$z$ in $\{ |\im z|\leq \sqrt{ c_1}(Nk)^{1/3}\}$. Then, taking the sign into account, we have,
$$
\xi_n^\pm(x)=\xi_n^c(x')+Y(x', \pm\sqrt{-x_n-b(x')}),
$$
for $x\in\tilde\Omega_-$, and, 
$$
\xi_n^{\pm i}(x)=\xi_n^c(x')+Y(x', \pm i\sqrt{x_n+b(x')}),
$$
for $x\in\tilde\Omega_+\cap\{x_n+b(x')\leq c_1(Nk)^{2/3}\}$. 

We again suppose $\xi_n$ is a critical point. Then $x_n\xi_n+\tilde g$ can be represented in 
terms of $x'$ and $z$, as, 
\begin{equation}
\label{criticalvalue}
x_n\xi_n+\tilde g(x',\xi_n)=a(x')-b(x')\xi_n^c(x')-\xi_n^c(x')z^2-\tilde\nu(x', z)z^3
\end{equation}
where $\tilde\nu(x',z)$ is smooth in $x'$, holomorphic in $z$ for $\{| \im z |\leq \sqrt{c_1}(Nk)^{1/3}\}$, and,
\begin{equation}
\label{tildedelta}
\tilde\nu(x', z)=\frac 2{3\sqrt{\tilde\nu_1(x', \xi_n^c(x'))}}+{\mathcal O}(z)
\end{equation}
as $z\to 0$. Let $\Phi(x', z)$ be the right hand side of \eq{criticalvalue}.
Then, for $x\in\tilde\Omega_-$, the critical value is,
\begin{equation}
\label{phi-}
\tilde\phi(x)=\Phi(x', \sqrt{-x_n-b(x')})= \phi (x)+{\mathcal O}(h^\infty),
\end{equation}
and, for $x\in\tilde\Omega_+\cap\{x_n+b(x')\leq c_1(Nk)^{2/3}\}$,
\begin{equation}
\label{phi+}
\tilde\phi(x)=\Phi(x', -i\sqrt{x_n+b(x')}).
\end{equation}
In particular, since the functions $a$, $b$ and $\xi_n^c$ are all real valued, we have,
$$
\im \tilde \phi (x) = -(x_n + b(x'))^{3/2}\re\tilde\nu (x', -i\sqrt{x_n+b(x')}),
$$
for $x\in\tilde\Omega_+\cap\{x_n+b(x')\leq c_1(Nk)^{2/3}\}$, and thus, in view of (\ref{odex'2}) and (\ref{tildedelta}), the estimate 
 \eq{imphi} easily follows.

In order to prove \eq{reimphi}, recall that $\phi$ is solution of the eikonal equation,
$$
q\left (x,\frac{\p\phi}{\p x}\right )+E_0=\left (\frac{\p\phi}{\p x}\right
)^2-V(x)+E_0=0.
$$
By (\ref{phi-}), this implies that $\Phi$ verifies,
$$
\left(\frac{\p\Phi}{\p x'}-\frac{\p b}{\p x'}\frac 1{2z}\frac{\p\Phi}{\p z}\right )^2
+\left (\frac 1{2z}\frac{\p\Phi}{\p z}\right )^2
-V(x', -z^2-b(x'))+E_0={\mathcal O}(h^\infty),
$$
for $z$ real close enough to 0. Since in addition $\partial\Phi /\partial z={\mathcal O}(|z|)$ by (\ref{criticalvalue}), we see that  the left-hand side is holomorphic  in $z$ for $\{| \im z |\leq \sqrt{c_1}(Nk)^{1/3}\}$, and thus,  returning to the $x$ variable, we easily deduce that $\tilde\phi$ verifies,
$$
\left (\frac{\p\tilde\phi}{\p x}\right
)^2-V(x)+E_0={\mathcal O}(h^\infty),
$$
uniformly for $x\in\tilde\Omega_+\cap\{x_n+b(x')\leq c_1(Nk)^{2/3}\}$ and $h>0$ small enough. In particular, taking the imaginary part, we obtain,
\begin{equation}
\label{imeiko}
\nabla\im\tilde\phi (x)\cdot \nabla\re\tilde\phi (x)={\mathcal O}(h^\infty).
\end{equation}
Then, following \cite{hs1}, we take local coordinates $(y',y_n)$ such that $\{ x_n + b= 0\}=\{ y_n= 0\}$ and $\nabla(x_n +b) \cdot \nabla= \partial /\partial y_n$.  In view of (\ref{imphi}), we obtain,
$$
\nabla\im\tilde\phi (x)\cdot \nabla = -\frac 1{\sqrt{\nu_1(x', \xi_n^c(x'))}}y_n^{1/2}\frac{\partial}{\partial y_n}+\sum_{j=1}^n{\mathcal O}(y_n)\frac{\partial}{\partial y_j},
$$
and  the vector field $\nabla\im\tilde\phi (x)\cdot \nabla$ can be desingularized at $y_n=0$ by setting $(z',z_n):=(y', y_n^{1/2})$, leading to,
 $$
\nabla\im\tilde\phi (x)\cdot \nabla = \left (-\frac 1{2\sqrt{\nu_1(x', \xi_n^c(x'))}}+{\mathcal O}(z_n)\right )\frac{\partial}{\partial z_n}+\sum_{j=1}^{n-1}{\mathcal O}(z_n^2)\frac{\partial}{\partial z_j}.
$$
Therefore, using  (\ref{lemma}) and (\ref{imeiko}), we immediately deduce (\ref{reimphi}).
\end{pre}

\subsubsection{Modification of $I[c]$}
Let us introduce the notation:
\begin{equation}
\label{omegaepsilon}
\Omega(\varepsilon_1,\varepsilon_2)=\{x\in\tilde\Omega;\varepsilon_1<x_n+b(x')<\varepsilon_2\}
\end{equation}
for two real small numbers $\varepsilon_1<\varepsilon_2$.
We want to extend $I[c]$, which is so far defined in $\Omega(-\delta,0)$ for some
$\delta>0$, to
$\Omega(-\delta, c_1(Nk)^{2/3})$.

If $x\in \Omega(0,c_1(Nk)^{2/3})$, then
by Lemma \ref{criticalpoint}, $|\im\xi_n^{-i}(x)|\leq (Nk)^{1/3}$.
We modify the integration contour $\gamma(x)$ in \eq{airy} within this complex strip so 
that it remains 
to be a steepest descent curve passing by $\xi_n^{-i}(x)$ for $x\in
\Omega(0,c_1(Nk)^{2/3})$. A careful observation of the real part of the phase as in
\cite{hs1} gives the following lemma:

\begin{lem}\sl
\label{contour}
Let $\delta >0$ be small enough. 
Then, for $x\in \Omega(-\delta, c_1(Nk)^{2/3})$, there exists a piecewise smooth curve $\gamma_N(x,h)$
in a small complex neighborhood of 
$\xi_n=\xi_n^c(x')$ satisfying the following properties:
\begin{description}
\item[(i)]
$\gamma_N(x,h)$ is included in a band
$\{\xi_n\in\C;|\im\xi_n|\le (Nk)^{1/3}\}$  and the extremities are
independent of $x$ (i.e. fixed when $h$ is fixed).

\item[(ii)]
If $x\in \Omega(-\delta,0)$, $\gamma_N(x,h)$ contains $\xi_n^+(x)$, and along $\gamma_N(x,h)$, one has, 
$$
\re (x_n\xi_n+\tilde g(x',\xi_n)) -\phi(x)\geq \delta_1 (|x_n+b(x')|^{\frac12} +|\xi_n -\xi_n^+(x)|) |\xi_n -\xi_n^+(x)|^2
$$
 for some constant $\delta_1>0$. Moreover, $|\xi_n -\xi_n^+(x)|\geq \delta_1(Nk)^{1/3}$ at the extremities of $\gamma_N (x,h)$.
\item[(iii)]
If $x\in \Omega(0,c_1(Nk)^{2/3})$, $\gamma_N(x,h)$ contains $\xi_n^{-i}(x)$, and along the contour $\gamma_N(x,h)$, one has, 
$$
\re (x_n\xi_n+\tilde g(x',\xi_n)-\tilde\phi(x))\geq \delta_1 (|x_n+b(x')|^{\frac12} +|\xi_n -\xi_n^{-i}(x)|) |\xi_n -\xi_n^{-i}(x)|^2
$$
 for some constant $\delta_1>0$. Moreover, $|\xi_n -\xi_n^{-i}(x)|\geq \delta_1(Nk)^{1/3}$ at the extremities of $\gamma_N (x,h)$.
\end{description}
\end{lem}

Now, we also define a holomorphic $(Nk)^{1/3}$-approximation $\tilde
c(x',\cdot;h)$ of the symbol$c(x',\cdot;h)$ by writing $c\sim \sum_{j\geq 0}h^jc_j$, by taking a holomorphic $(Nk)^{1/3}$-approximation $\tilde
c_j$ of $c_j$, and by re-summing the formal symbol $\sum_{j\geq 0}h^j\tilde c_j$ (note that here, each $\tilde c_j$ depends on $h$, but in a very well-controlled way).

With these $\tilde
c(x',\cdot;h)$ and $\gamma_N(x,h)$, we define the modified integral
representation of $e^{S/h}w$ by the formula,
\begin{equation}
\label{airy2}
\tilde I_N[\tilde c](x,h)=h^{-1/2}\int_{\gamma_N(x,h)}e^{-(x_n\xi_n+ \tilde g(x',\xi_n))/h}
\tilde c(x',\xi_n;h)d\xi_n.
\end{equation}

\begin{lem}\sl
There exists a constant $\delta_2>0$ such that, for $x\in \Omega(-\delta,0)$, and for all $N\geq 1$, one has,
$$
\partial^\alpha\left (I[c](x,h)-\tilde I_N[\tilde c](x,h)\right )={\mathcal O}(h^{\delta_2N-1/2-|\alpha|} e^{-\phi(x)/h}).
$$
\end{lem}

\begin{pre}
By definition $(c,g)$ and $(\tilde c, \tilde g)$ coincide on the real, up to ${\mathcal O}(h^\infty)$. Therefore, substituting the real contour $\gamma (x)$ to $\gamma_N (x,h)$ in the expression of $\tilde I_N[\tilde c](x,h)$, we obtain an integral $J_N(x)$ that coincides with $I[c](x,h)$ up to ${\mathcal O}(h^\infty e^{-\phi(x)/h})$. Then, modifying continuously $\gamma (x)$ into $\gamma_N (x,h)$ in $J_N(x)$, we recover $\tilde I_N[\tilde c](x,h)$ up to error terms coming from the fact that $\gamma (x)$ and $\gamma_N (x,h)$ do not have the same extremities. However, in view of Lemma \ref{contour} (ii) and the fact that, along $\gamma (x)$, the minimum of $\re (x_n\xi_n+\tilde g(x',\xi_n)) -\phi(x)$ is non-degenerate, we see that the deformation can be done in such a way that these error terms are ${\mathcal O}(e^{-(\phi(x)+\delta_2Nk)/h}) = {\mathcal O}(h^{\delta_2 N}e^{-\phi (x)/h})$, with $\delta_2=\delta_1^4$.
\end{pre}

\begin{lem}\sl
As $h\to 0$, one has
$$
(P-\rho(h))\tilde I_N[\tilde c]={\mathcal O}(h^{\delta_2N} e^{-\re \tilde\phi(x)/h}),
$$
uniformly in $ \Omega(-\delta,c_1(Nk)^{2/3})$.
\end{lem}

\begin{pre}
In view of \eq{tildep}, it is enough to check,
$$
P(\tilde I_N[\tilde c])=h^{-1/2}\int_{\gamma_N(x,h)}e^{-x_n\xi_n/h}\hat
P(e^{-\tilde g/h}\tilde c)d\xi_n+ {\mathcal O}(h^{\delta_2N} e^{-\re \tilde\phi(x)/h}).
$$
This can be done exactly in the same way as Proposition 10.5 in \cite{hs1}, with the only difference that, in our case, the values of $\re (x_n\xi_n + g)$ at the extremities of $\gamma_N(x,h)$ are greater that $\re\tilde\phi(x) + \delta_2Nk$.
 \end{pre}

\subsubsection{Asymptotic Expansion of $\tilde I_N[\tilde c]$}

 Here, we fix $c_2\in (0,c_1)$, and we study the asymptotic behavior of $\tilde I_N[\tilde c](x,h)$ as $h$ tends to 0 , for $x$ in $\Omega(c_2(Nk)^{2/3}, c_1(Nk)^{2/3})$.
 Setting,
 $$
 \tilde r(x):= \frac12\frac{\partial^2\tilde g}{\partial\xi_n^2}(x', \xi_n^{-i}(x))
 $$
 $$
  (=-i\sqrt{\tilde\nu_1(x',\xi_n^c(x'))(x_n+b(x'))} +{\mathcal O}(|x_n+b(x')|)),
 $$
 we have,
\begin{prop}\sl
\label{col}
 For all integers $L$, 
$M$ and $N$ large enough, and for 
$x$ in $ \Omega(c_2(Nk)^{2/3},c_1(Nk)^{2/3})$,   one has,
\begin{equation}
\label{asymptotique}
\tilde I_N[\tilde c](x,h)= \frac{e^{-\tilde\phi(x)/h}}{\sqrt{\tilde r(x)}}
\left\{\sum_{m=0}^{L+[M/2]} \beta_m(x)\left\{\frac
{h}{\tilde r(x)^3}\right\}^m+R_{L,M,N}(x,h) \right\},
\end{equation}
 with, for any $\alpha\in\N^n$,
 \begin{eqnarray}
|\partial^{\alpha}_xR_{L,M,N}(x,h)|& \leq& C_{N,\alpha}h^{\delta_\alpha N-\frac12}\sum_{m=0}^{M}C_{L,\alpha}^{m+1}\left(\frac{hm}{Nk}\right)^{\frac{m}2} + C_{L,\alpha}h^{L+\frac12}\nonumber\\
 \label{restelaplace}
 && \hskip 2cm+ h^{-\frac12}C_{L,\alpha}^{M+1}\left(\frac{hM}{Nk}\right)^{\frac{M}2},
\end{eqnarray}
 where the positive constant $\delta_\alpha$ does not depend on $(L,M,N)$, while $C_{N,\alpha}$ does not depend on $(L,M)$, and $C_{L,\alpha}$ does not depend on $(M,N)$.
Moreover,  the coefficients of the symbol verify,
\begin{equation}
\label{beta}
\beta_0(x)=\sqrt\pi\tilde c_0(x',\xi_n^{-i}(x))=\sqrt\pi\tilde
c_0(x',\xi_n^c(x'))+{\mathcal O}(\sqrt{x_n+b(x')}),
\end{equation}
$$
\beta_m(x)={\mathcal O}(1)\,\,\,(m=0,1,\ldots)\quad \mbox{as}\quad x_n+b(x')\to 0.
$$
In particular, taking $M=2[Nk/C'_Lh]$ with $C'_L>0$ large enough depending on $L$ only, one obtains,
\begin{equation}
\label{asymptotique1}
\tilde I_N[\tilde c](x,h)= \frac{e^{-\tilde\phi(x)/h}}{\sqrt{\tilde r(x)}}
\left\{\sum_{m=0}^{L+[Nk/C'_Lh]} \beta_m(x)\left\{\frac
{h}{\tilde r(x)^3}\right\}^m+{\mathcal O}(h^{\delta_L N} + h^{L}) \right\},
\end{equation}
uniformly for $x\in\Omega(c_2(Nk)^{2/3},c_1(Nk)^{2/3})$ and $h>0$ small enough, and where the positive constant $\delta_L$ does not depend on $N$ large enough.
\end{prop}

\begin{pre}              
For $x\in \Omega(c_2(Nk)^{2/3},c_1(Nk)^{2/3})$, setting
$\eta=\xi_n-\xi_n^{-i}(x)$, we can write,
$$
x_n\xi_n+\tilde g(x',\xi_n)=\tilde\phi(x)+\tilde
r(x)\eta^2+G(x,\eta)\eta^3
$$
where $G(x,\eta):=\int_0^1\frac{(1-t)^3}2\partial_{\xi_n}^3\tilde g(x', \xi_n^{-i}(x) + t\eta )dt$ is holomorphic with respect to $\eta$ in $\{|\re\eta|<\delta_1, \, |\im\eta|<\delta_1(Nk)^{1/3}\}$, with $\delta_1>0$ small enough (independent of $N$).
Then, we set,
$$
\tilde r(x)\eta^2+G(x,\eta)\eta^3=\tilde r(x)\zeta^2,
$$
so that $\hat\eta=\eta/\tilde r$, $\hat\zeta=\zeta/\tilde r$ verify,
$$
\hat\eta\sqrt{1+G(x,\tilde r(x)\hat\eta)\hat\eta}=\hat\zeta,
$$
where the square root is 1 for $\hat\eta =0$.
This equation is solvable with respect to $\hat\eta$, and  gives
$\hat\eta=\hat\eta(x,\hat\zeta)$ where the function $\hat\eta(x,\hat\zeta)$ is smooth with respect to $x\in \Omega(c_2(Nk)^{2/3},c_1(Nk)^{2/3})$ and holomorphic with respect to $\hat\zeta$ in some fixed neighborhood of 0, and  $\frac{\p \hat
\eta}{\p\hat\zeta}|_{\hat\zeta=0}=1$.

Changing the variables from $\xi_n$ to $\zeta$ in \eq{airy2}, we obtain,
$$
\tilde I_N[\tilde c](x,h)=h^{-1/2}e^{-\tilde\phi(x)/h}\int_{\Gamma_N (x,h)}e^{-\tilde r(x)\zeta^2/h}F\left(x,\frac\zeta {\tilde r};h\right)d\zeta,
$$
where the contour $\Gamma_N (x,h)$ is such that,
\begin{eqnarray}
\label{GammaN1}
0\in \Gamma_N(x,h); \, \re(\tilde r(x)\zeta^2)\geq \delta_3|\tilde r(x)|\cdot|\zeta |^2\, \mbox{ along } \Gamma_N (x,h);\\
\label{GammaN2}
|\zeta|\geq \delta_3|\tilde r(x)| \mbox{ at the extremities of } \Gamma_N (x,h),
\end{eqnarray}
for some positive constant $\delta_3$,
and where the symbol,
$$
F(x,\hat\zeta,h):=\tilde c(x',\xi_n^{-i}(x)+\tilde r(x)\hat\eta(x,\hat\zeta);h)\frac{\p
\hat\eta}{\p\hat\zeta}(x,\hat\zeta)
$$
can be developed asymptotically into,
$$
F(x,\hat\zeta,h)\sim\sum_{\ell\ge 0}F_{\ell}(x, \hat\zeta)h^\ell,
$$
with $F_{\ell}$ holomorphic with respect to $\hat\zeta$ in a fixed neighborhood of 0, and $F_{0}(x,0)=\tilde c_0(x', \xi_n^{-i}(x))$. Actually, $F_{\ell}$ also depends on $N$, but using Lemma \ref{almost} (ii)  and the fact that $|\tilde r(x)|\sim (Nk)^{1/3}$, we easily obtain that the derivatives of $F_\ell$ verify,
$$
|\partial_{\hat\zeta}^\beta F_\ell |\leq C_\ell^{1+|\beta|}\beta!,
$$
for some constant $C_\ell >0$ independent of $N$.

Now we set $y=\tilde r(x)^{1/2}\zeta$, where $\tilde r(x)^{1/2}$ is the branch such that $\tilde r(x)^{1/2}\sim e^{-i\pi /4}(\nu_1(x_n+b))^{1/4}$,
and, using (\ref{GammaN1})-(\ref{GammaN2}), we see that  a new slight modification of the contour of integration gives,
$$
\tilde I_N[\tilde c](x,h) = \frac {e^{-\tilde\phi(x)/h}}{\sqrt{
h\tilde r(x)}}\left(\int_{-\delta_4\tilde r^{3/2}}^{\delta_4\tilde r^{3/2}}e^{-y^2/h}F\left(x,\frac y {\tilde r^{3/2}};h\right)dy +{\mathcal O}(e^{-\delta_5\tilde r(x)^3/h})\right)
$$
for some constants $\delta_4,\delta_5 >0$. As a consequence, using again that $|\tilde r(x)|\sim (Nk)^{1/3}$ and writing $F=\sum_{\ell =0}^LF_\ell h^\ell+ {\mathcal O}(h^{L+1})$, we obtain,
\begin{eqnarray*}
&& \tilde I_N[\tilde c](x,h) \\
&&= \frac {e^{-\tilde\phi(x)/h}}{\sqrt{
h\tilde r(x)}}\left(\sum_{\ell =0}^Lh^\ell\int_{-\delta_4\tilde r^{3/2}}^{\delta_4\tilde r^{3/2}}e^{-y^2/h}F_\ell\left(x,\frac y {\tilde r^{3/2}}\right)dy +{\mathcal O}(h^{\delta_6N}+ h^{L+1})\right)
\end{eqnarray*}
with some new positive constant $\delta_6$.

At this point, we can proceed with the usual Laplace method in order to estimate each term of the previous sum. Writing,
$$
F_\ell\left(x,\frac y {\tilde r^{3/2}}\right)=\sum_{m=0}^\infty F_{\ell,m}(x)\frac{y^m}{\tilde r^{3m/2}}=\sum_{m=0}^M F_{\ell,m}(x)\frac{y^m}{\tilde r^{3m/2}} + S_{M,\ell},
$$
with $|F_{\ell,m}(x)|\leq C_\ell^{m+1}$ and $|S_{M,\ell}|\leq 2C_\ell^{M+2}|y/\tilde r^{3/2}|^{M+1}$, we obtain,
\begin{eqnarray}
\tilde I_N[\tilde c](x,h) &=& \frac {e^{-\tilde\phi(x)/h}}{\sqrt{
h\tilde r(x)}}\Big( \sum_{\ell=0}^L\sum_{m=0}^MF_{\ell,m}(x)h^\ell
\int_{-\delta_4\tilde r^{3/2}}^{\delta_4\tilde r^{3/2}}e^{-y^2/h}\left
(\frac y{\tilde r^{3/2}}\right )^mdy \nonumber\\
&& \hskip 4cm+ R^{(1)}_{L,M,N}(x) \Big)\nonumber\\
&=& \frac {e^{-\tilde\phi(x)/h}}{\sqrt{
h\tilde r(x)}}\Big(\sum_{\ell=0}^L\sum_{m=0}^MF_{\ell,m}(x)h^\ell
\int_{-\infty}^{+\infty}e^{-y^2/h}\left
(\frac y{\tilde r^{3/2}}\right )^mdy \nonumber\\
&& \hskip 4cm+ R^{(1)}_{L,M,N}(x) + R^{(2)}_{L,M,N}(x)\Big) \nonumber\\
 \label{int}
&=& \frac {e^{-\tilde\phi(x)/h}}{\sqrt{h\tilde r(x)}}\Big(\sum_{\ell=0}^L\sum_{m=0}^{[M/2]}\Gamma\big( m+\frac 12\big)F_{\ell,2m}(x)\frac{h^{m+\ell+\frac12}}{\tilde r^{3m}}\nonumber\\
&& \hskip 3.5cm+ R^{(1)}_{L,M,N}(x) + R^{(2)}_{L,M,N}(x)\Big) ,
\end{eqnarray}
with,
\begin{eqnarray*}
&& |R^{(1)}_{L,M,N}(x)|\leq C_Nh^{\delta_6N}+C_Lh^{L+1}+ C_L^{M+1}(h/Nk)^{M/2} M^{M/2};\\
&& |R^{(2)}_{L,M,N}(x)|\leq C_Nh^{\delta_6N}\sum_{m=0}^{M}C_L^{m+1}(h/Nk)^{m/2}m^{m/2} ,
\end{eqnarray*}
for some positive constant $\delta_6$ independent of $(L,M,N)$,  some positive constant $C_N$ independent of $(L,M)$,  and some positive constant $C_L$ independent of $(M,N)$. Similar estimates hold true for all the derivatives with respect to $x$ of $R^{(1)}_{L,M,N}$ and $R^{(2)}_{L,M,N}$.

Hence, we obtain \eq{asymptotique} if we set,
$$
\beta_m(x):=\sum_{j+\ell=m}\Gamma\big(j+\frac 12\big) F_{\ell,2j}(x)\tilde r(x)^{3\ell},
$$
and,
\begin{eqnarray*}
R_{L,M,N}(x)&:=&h^{-1/2}\Big(R^{(1)}_{L,M,N}(x) + R^{(2)}_{L,M,N}(x)\\
&& -\sum_{\genfrac{}{}{0cm}{}{\ell + m\leq L+[M/2]}{\ell > L \,\rm{or}\, m > [M/2]}} \Gamma\big( m+\frac 12\big)F_{\ell,2m}(x)\frac{h^{m+\ell+\frac12}}{\tilde r^{3m}}\Big).
\end{eqnarray*}
In particular, (\ref{restelaplace}) with $\alpha =0$ is verified, as well as the estimates on $\beta_m(x)$ and, moreover,
$$
\beta_0(x)=\Gamma\left (\frac 12\right)F_{0}(x,0)
=\sqrt\pi\tilde c_0(x',\xi_n^{-i}(x)).
$$
The estimate (\ref{restelaplace}) for all $\alpha$ is obtained in the same way. 

Finally, substituting $M=2[Nk/C'_Lh]$ into (\ref{restelaplace}) with $\alpha=0$, \eq{asymptotique1} follows by taking $C'_L>4C_{L,0}^2$, since in that case, we have,
\begin{eqnarray*}
C_{L,0}^{M+1}\left (\frac{hM}{Nk}\right )^{\frac{M}2}&\leq& C_{L,0}\left(2C_{L,0}^2/C'_L\right)^{\frac{M}2}
\leq C_{L,0}2^{-[Nk/C'_Lh]}\\
&\leq&2C_{L,0}2^{-Nk/C'_Lh} =2C_{L,0}h^{\delta_LN},
\end{eqnarray*}
with $\delta_L:= (\ln 2)/C'_L$,
and,
$$
\sum_{m=0}^{M}C_{L,0}^{m+1}\left (\frac{hm}{Nk}\right )^{\frac{m}2}\leq \sum_{m=0}^{M}C_{L,0}^{m+1}\left(\frac2{C'_L}\right)^{\frac{m}2}\leq C_{L,0}\sum_{m=0}^{\infty}2^{-\frac{m}2}=\frac{\sqrt 2}{\sqrt{2}-1}C_{L,0}.
$$
\end{pre}

Let us observe that the principal symbol (denoted by $\tilde a_0(x)$) of the asymptotic expansion of
$\tilde I_N[\tilde c]$ is,
$$
\tilde a_0(x)=\frac{\beta_0(x)}{\sqrt{\tilde r(x)}}=\sqrt{\frac\pi{\tilde r(x)}}\tilde
c_0(x',\xi_n^{-i}(x)),
$$
and it behaves like,
\begin{equation}
\label{a0}
\tilde a_0(x)=\frac{\sqrt\pi
c_0(x',\xi_n^c(x'))}{\nu_1(x',\xi_n^c(x'))^{1/4}}z^{-1/2}(1+{\mathcal O}(z)),
\end{equation}
as $z=-i\sqrt{x_n+b(x')}\to 0$.
Recall that the principal term of $I[c]$ for $x\in\tilde\Omega_-$ should coincide with
$a_0(x)$ (see \eq{c0}), that is, 
$$
a_0(x)=\sqrt{\frac\pi{r(x)}}c_0(x',\xi_n^+(x)),
$$
and it has the same behavior \eq{a0} as $\tilde a_0(x)$, as $z=\sqrt{-x_n-b(x')}\to 0$.

\subsection{Global WKB Solution near $\partial\ddot O$}

The previous study shows that, for any point $x^1$ of $\Gamma$ and for any $N\geq 1$, the WKB solution $w=h^{-n/4}e^{-S/h}I[c]$ can be extended in a neighborhood of $x^1$ of the form,
$$
\Omega_N (x^1):= \bigcup_{-t_0 <t<(Nk)^{2/3}}\exp t\nabla f (\omega (x_1)),
$$
where $\omega (x_1)$ is a fixed small enough neighborhood of $x^1$ in the caustic set ${\mathcal C}$, and $f$ is such that $f=0$ is an equation of ${\mathcal C}$ near $x^1$,  with  $\{ f>0\}\cap \open^C\not= \emptyset$. Therefore, by using a standard partition of unity in a neighborhood $\omega (\Gamma)$ of $\Gamma$ in ${\mathcal C}$, we obtain an extension $w_N$ of $w$ in an open set of the form,
$$
\Omega_N := \bigcup_{-t_0 <t<(Nk)^{2/3}}\exp tX(\omega (\Gamma)),
$$
where $X$ is any vector-field transverse to ${\mathcal C}$ near $\Gamma$ and directed towards $\open^C$ (for instance, one can take $X=-\nabla V$).

\medskip
Moreover, by Assumption (A4) and (\ref{lemma}) (see also \cite{hs1} Remarque 10.4), we  see that, if $x\in {\mathcal C}$ is such that  ${\rm dist}(\Gamma, x)\sim (Nk)^{1/2}$, then $\re\tilde\phi (x) \sim Nk$, and thus $w_N(x)={\mathcal O}(h^{-n/4}e^{-(S+\re\tilde\phi)/h})={\mathcal O}(h^{\delta N}e^{-S/h})$ for some constant $\delta >0$. We also observe that, on ${\mathcal C}$, we have $d(x_0,x)-S=\re\tilde\phi (x) \sim {\rm dist}(\Gamma, x)^2$. As a consequence (thanks to (\ref{imeiko})), if
we set,
$$
\gamma({\mathcal C}):=\{ x\in {\mathcal C}\, ;\, S+Nk\leq d(x_0,x)\leq S+2Nk\},
$$
then,  we also have,
$$
w_N={\mathcal O}(h^{\delta N}e^{-S/h})\,\, \mbox{ on }\,  \gamma_N^+ ({\mathcal C}):=\bigcup_{0\leq t<(Nk)^{2/3}}\exp tX(\gamma ({\mathcal C})).
$$

Now, for $\varepsilon, t_0 >0$ small enough, we set,
\begin{eqnarray*}
&&\omega(\varepsilon):=  \{ x\in {\mathcal C}\, ;\, d(x_0,x)\leq S+\varepsilon\};\\
&&\omega^+(\varepsilon, t_0):= \bigcup_{0\leq t<t_0}\exp tX(\omega (\varepsilon));\\
&&{\bf \Omega}(\varepsilon, t_0):=\{ x\in {\bf\Omega}\, ;\, d(x_0,x)< S+\varepsilon\}\cup \omega^+(\varepsilon, t_0).
\end{eqnarray*}

Then ${\bf \Omega}(\varepsilon, t_0)$ is an open set, and the previous discussion shows that $w_N$ is well defined and $C^\infty$ on ${\bf \Omega}(2Nk, (Nk)^{2/3})$,
\be
\label{2Nk}
w_N={\mathcal O}(h^{\delta N} e^{-S/h})\,\,\,{\rm on}\,\, {\bf \Omega}(2Nk, (Nk)^{2/3})\backslash {\bf \Omega}(Nk, (Nk)^{2/3}).
\ee
Moreover, by construction, it satisfies $(P-\rho )w_N={\mathcal O}(h^{\delta N} e^{-(S+\re\tilde\phi )/h})$ on ${\bf \Omega}(2Nk, (Nk)^{2/3})$, for some constant $\delta >0$.

\medskip
As a consequence, if we take a cut-off function $\chi_N$ such that $\supp \chi_N\subset {\bf \Omega}(2Nk, (Nk)^{2/3})$, $\chi_N=1$ on ${\bf \Omega}(Nk, \frac12(Nk)^{2/3})$, $\partial^\alpha\chi_N ={\mathcal O}(k^{-N_\alpha})$ (for some $N_\alpha\geq 0$ and all $\alpha\in\N^n$), then, setting,
$$
\tilde w_N:= \chi_N w_N,
$$
we see that $\tilde w_N$ is $C^\infty$ on $\{ {\rm dist} (x,\open)< \delta_0(Nk)^{2/3}\}$ for some fixed $\delta_0>0$, and  it verifies,
$$
(P-\rho )\tilde w_N={\mathcal O}(h^{\delta N} e^{-\re\tilde\phi/h})\,\, \mbox{ in} \,\, \{ {\rm dist} (x,\open)< \delta_0(Nk)^{2/3}\}.
$$
Summing up (and slightly changing the notations by writing $w_N$ instead of $\tilde w_{2^{3/2}N/\delta_0^{3/2}}$), we have proved,

\begin{prop}\sl
\label{summary} 
For any large enough $N$, there exists a smooth function $w_N(x,h)\in C^\infty(\open_N)$, with $\open_N:=\{  {\rm dist} (x,\open)<2(Nk)^{2/3}\}$,  verifying the following properties:
\begin{description}
\item[(i)] 
There exists a constant $\delta>0$, independent of $N$, such that uniformly in $\open_N$, and for all $\alpha\in\Z_+^n$, one has, 
\begin{eqnarray*}
&& \p^\alpha w_N(x,h)={\mathcal O}(h^{-m_\alpha}e^{-(S+\re\tilde\phi(x))/h});\\
&& (P-\rho(h))w_N(x,h)={\mathcal O}(h^{\delta N} e^{-(S+\re\tilde\phi(x))/h}),
\end{eqnarray*}
for some $m_\alpha\geq 0$, and where $\tilde\phi$ is defined by
$\tilde\phi(x)=d(x_0,x)-S$ for
$x\in{\bf \Omega}$, by
\eq{tildephi} for
$x\in\omega^+(2Nk, (Nk)^{2/3})$; 
\item[(ii)]
In any compact subset of ${\bf \Omega}$, for any $M\in\N$, one has, 
$$
w_N(x,h)= h^{-n/4}e^{-(S+\tilde\phi(x))/h}\left (\sum_{j=0}^M a_j(x)h^j+\ord (h^{M+1})\right ),
$$
as $h\to 0$, where $a_j(x)$ are extensions of those given in \eq{symbol}, and $a_0$ is elliptic.
\item[(iii)]
In $\{  (Nk)^{2/3}<{\rm dist} (x,\open)<2(Nk)^{2/3}\}$, for any large enough $L$, there exist $C_L'>0$ and $\delta_L>0$ independent of $N$ such that
\begin{equation}
\label{summarywkb}
w_N(x,h)=
h^{-n/4}e^{-(S+\tilde\phi(x))/h}\left (\sum_{j=0}^{L+[Nk/C_L'h]} \tilde a_j(x)h^j+\ord (h^{\delta_L N}+h^{L})\right ),
\end{equation}
as $h\to 0$,
with $\tilde a_j $ (independent of $h$) of the form,
\be
\label{taj}
\tilde a_j (x) = ({\rm dist}(x,{\mathcal C}))^{-3j/2-1/4}\tilde\beta_j(x,{\rm dist}(x,{\mathcal C})),
\ee 
where $\tilde\beta_j$ is  smooth near 0, $\tilde\beta_0(0)\not=0$. 
\end{description}
\end{prop}

\section{Comparison in the Island}

In this section, we compare the WKB solution $w_N$ with the true
resonant state $u$ inside  $\open$ near  a point of interaction $x^1$. More precisely, we obtain an estimate on the difference up to a distance of order $(Nk)^{2/3}$ of $x^1$.

\medskip
We use the same notations as in Section \ref{sec-ExtWB}.
Let $x\in\tilde\Omega_-$ be a point  sufficiently close to $x^1$. Using the representation formula (\ref{repphi}) for $\phi$ (see also \cite{hs1} Formula (10.22)), we see that,
\begin{equation}
\label{estbelow}
\phi (x) \geq \phi (x', -b(x')) + (x_n+b(x'))\xi_n^c(x')-C_1|x_n+b(x')|^{3/2},
\end{equation}
for some
positive constant $C_1$.
Moreover, thanks to Assumption (A4), we already know that $\phi (x', -b(x')) = \phi\left|_{\mathcal C} \right.(x', -b(x'))\geq \delta |x'|^2$ with $\delta >0$ constant, and thus, using (\ref{odex'2}), we see that $\phi (x', -b(x')) + (x_n+b(x'))\xi_n^c(x')\geq 0$ near $x^1$. As a consequence, we deduce from (\ref{estbelow}),
\begin{equation}
\label{3demi}
d(x_0,x)\ge S-C_1|x_n+b(x')|^{3/2}.
\end{equation}
In particular, if 
$x\in\Omega(-(Nk)^{2/3},0)$,
$k=h\log \frac 1h$,
we have,
$$
e^{-s(x)/h}={\mathcal O}(h^{-C_1N} e^{-S/h}).
$$

The aim of this section is to show  a local a priori estimate near a point of interaction $x^1$
(Proposition \ref{comparison2}), and after then, as a direct consequence, a global a priori estimate in a neighborhood of $\p\open$ (Proposition \ref{prop7}).

\begin{prop}\sl
\label{comparison2}
There exists $N_2\in\Z$ and $C>0$, such that, for any $N>0$, one has,
$$
\Vert u(x,h)-w_{CN}(x,h)\Vert_{H^1(\Omega(-(Nk)^{2/3},0))}={\mathcal O}(h^{-N_2}e^{-S/h}),
$$
uniformly as $h\rightarrow 0$.
\end{prop}

\begin{pre}
Recall (Theorem \ref{th3}) that
there exists $N_0$ such that
$$
\Vert e^{s(x)/h}u(x,h)\Vert_{H^1(\tilde\Omega_-)} ={\mathcal O}(h^{-N_0}).
$$
The WKB solution $w_{CN}$ also satifies the same estimate (see Proposition \ref{summary}), and
hence so does the difference,
$$
\Vert e^{s(x)/h}(u(x,h)-w_{CN}(x,h))\Vert_{H^1(\tilde\Omega_-)} ={\mathcal O}(h^{-N_0}).
$$
In particular,
$$
\Vert u(x,h)-w_{CN}(x,h)\Vert_{H^1(\tilde\Omega_-\cap
\{ d(x_0,x)\geq S-2k\})}={\mathcal O}(h^{-N_0'}e^{-S/h}),
$$
for some other constant $N_0'$.

Now, we set,
$$
\Omega_1 =\Omega_1(h):=B_d(x_0,S-k).
$$
Since every point of $\Omega_1$ can be connected to $x_0$ by a smooth minimal geodesic (with respect to the Agmon distance), the arguments of the previous section show that the WKB solution $w_{CN}(x,h)$ is well defined  in all of $\Omega_1$ (we use its integral representation when $x$ becomes too close to a point of interaction). Moreover, it is not difficult to construct $\chi_h\in C_0^\infty (\Omega_1)$, such that $\chi_h =1$ on $\{ d(x_0,x) \leq S-2k\}$, $0\leq \chi_h\leq 1$ everywhere, and, for all $\alpha\in\N^n$,
$$
 \p^\alpha \chi_h ={\mathcal O}(h^{-N_\alpha}),
$$
for some constant $N_\alpha\geq 0$. Then, we set,
$$
\hat w :=\chi_h (x)w_{CN}(x,h),
$$
and, for  $N\geq 1$ arbitrarily large, 
\begin{eqnarray*}
\phi_N(x)&=&\min\left(
d(x_0,x)+C_1Nk +k(S-d(x_0,x))^{1/3},\right.\\
&& \hskip 4cm \left. S+(1-k^{1/3})(S-d(x_0,x)))
\right).
\end{eqnarray*}
On $\Omega(-(Nk)^{2/3},0)$, by (\ref{3demi}) we have $d(x_0,x)\geq S-C_1Nk$. Therefore,
$\phi_N(x)\ge S$ there, and it suffices to show that there exists $N_0$ such
that, for any $N\geq 1$,
$$
\Vert e^{\phi_N/h}(\chi_h u-\hat w)\Vert_{H^1(\Omega_1)}={\mathcal O}(h^{-N_0}).
$$
We prove it by using Agmon estimates (see lemma \ref{agmon} in the appendix). At first, we observe that, by construction (and since $\phi_N\leq d(x_0,x)+(C_1N+S^{1/3})k$), we have (uniformly in $\Omega_1$),
\begin{eqnarray}
(P-\rho (h))\hat w &=& [P,\chi_h]w_{CN} + {\mathcal O}(h^{\delta CN} e^{-d(x_0,x)/h})  \nonumber\\
\label{estPchiw}
 &=& {\mathcal O}({\bf 1}_{\supp\nabla\chi_h}h^{-M_1}e^{-S/h})+{\mathcal O}(h^{\delta CN-C_1N -S^{1/3}} e^{-\phi_N/h}),
\end{eqnarray}
for some $M_1\geq 0$ constant. Moreover,  using (\ref{compdiric3}),
\begin{equation}
\label{estPchiu}
\Vert e^{\phi_N/h}(P-\rho (h))\chi_h u\Vert_{L^2} = \Vert e^{\phi_N/h}[P,\chi_h]u\Vert_{L^2}={\mathcal O}(h^{-M_1'}e^{(F_N-S)/h}),
\end{equation}
for some other constant $M_1'\geq 0$, and with, 
$$
F_N:=\sup_{{\rm Supp}\nabla\chi_h}\phi_N.
$$
Since  $S-d(x_0,x)\leq 2k$ on  ${\rm Supp}\nabla\chi_h$, we have $F_N\leq S+2(1-k^{1/3})k\leq S+2k$, 
and thus, we deduce from (\ref{estPchiu}),
\begin{equation}
\label{estPchiu2}
\Vert e^{\phi_N/h}(P-\rho (h))\chi_h u\Vert_{L^2}={\mathcal O}(h^{-M_1'-2}).
\end{equation}
Setting,
$$
u'_h:=\chi_h u-\hat w,
$$
and choosing $C$ such that $\delta C\geq C_1$, we obtain from (\ref{estPchiw})-(\ref{estPchiu2}),
\begin{equation}
\label{eqagmon}
\Vert e^{\phi_N/h}(P-\rho (h))u'_h \Vert_{L^2}={\mathcal O}(h^{-M_2}),
\end{equation}
for some constant $M_2\geq 0$, independent of $N$.

We also observe, that, on $\Omega_1^-:=\Omega_1\cap \{ d(x_0,x)+C_1Nk+k(S-d(x_0,x))^{1/3} < S+(1-k^{1/3})(S-d(x_0,x))\}$, we have,
$$
\nabla\phi_N = \left( 1-\frac{k}{3(S-d(x_0,x))^{2/3}}\right)\nabla d(x_0,x),
$$
and, on $\Omega_1^+:=\Omega_1\cap \{ d(x_0,x)+C_1Nk+k(S-d(x_0,x))^{1/3} > S+(1-k^{1/3})(S-d(x_0,x))\}$,
$$
\nabla\phi_N = -( 1-k^{1/3})\nabla d(x_0,x).
$$
Since $k(S-d(x_0,x))^{-2/3}\leq k^{1/3} <<1$,  for $h$ sufficiently small we easily deduce,
$$
V-\re\rho-|\nabla\phi_N|^2\geq \frac{k}{3(S-d(x_0,x))^{2/3}}(V-E_0)-(\re\rho-E_0),\, \mbox{ on } \Omega_1^-;
$$
and,
$$
V-\re\rho-|\nabla\phi_N|^2\geq k^{1/3}(V-E_0)-(\re\rho-E_0),\, \mbox{ on } \Omega_1^+.
$$
Now, since $\nabla V\not= 0$ on $\p\open$, a small examination of the Hamilton curves of $q=\xi^2-V(x)$ starting from $\p\open\times \{0\}$, shows that, for $x\in\open$ close enough to $\p\open$, one has $d(x,\p\open) = {\mathcal O}((V(x)-E_0)^{3/2})$. Therefore, by the triangle inequality, we deduce,
\begin{equation}
\label{3demiV}
d(x_0,x)\geq S-C_2(V(x)-E_0)^{3/2},
\end{equation}
where $C_2>0$ is a constant, and the inequality is actually valid in all of $\open$ except for some  fixed small enough neighborhood $U_0$ of $x_0$ (since $V-E_0>0$ on $\open\backslash\{x_0\}$).

In particular,  $U_0$ can be assumed to be disjoint from $\Omega_1^+$,  and then (\ref{3demiV}) shows that $V(x)-E_0\geq  (k/C_2)^{2/3}$ on $\Omega_1^+$. Therefore, observing also that $|\rho-E_0|\leq C_3h$ with $C_3>0$ constant, on this set, we obtain,
\begin{equation}
\label{ell+}
V-\re\rho-|\nabla\phi_N|^2\geq  \frac{k}{C_2^{2/3}}-C_3h\geq \frac{k}{2C_2^{2/3}},
\end{equation}
for $h>0$ small enough.

Moreover, by (\ref{3demiV}), on $\open\backslash U_0$, we also have,
$$
\frac{V-E_0}{3(S-d(x_0,x))^{2/3}}\geq \frac1{3C_2^{2/3}},
$$
and thus, if $x\in \Omega_1^-\backslash U_0$,
\begin{equation}
\label{ell-}
V-\re\rho-|\nabla\phi_N|^2\geq \frac{k}{3C_2^{2/3}}-C_3h\geq \frac{k}{4C_2^{2/3}},
\end{equation}
for $h>0$ small enough. On the other hand, by (\ref{compdiric1}) and the results of \cite{hs2}, we know that $w_N(x,h)$ is a good approximation of $u(x,h)$ on $U_0$, in the sense that,
$$
\Vert e^{d(x_0,x)/h}u'_h\Vert_{L^2(U_0)} ={\mathcal O}(h^\infty ). 
$$
Since $e^{\phi_N/h} ={\mathcal O}(h^{-C_1N-S^{1/3}}e^{d(x_0,x)/h})$, we deduce,
\begin{equation}
\label{estpuits}
\Vert e^{\phi_N/h}u'_h\Vert_{L^2(U_0)} ={\mathcal O}(h^\infty ).
\end{equation}
Now, we apply the identity (\ref{Agmon}) with $u'_h, \phi_N, \re\rho$ instead of $v_h, \phi, \lambda_h$. Using (\ref{eqagmon}), (\ref{ell+}), (\ref{ell-}), and (\ref{estpuits}), this permits us to obtain,
$$
 h^2\Vert \nabla(e^{\phi_N /h}u'_h)\Vert^2  + k\Vert e^{\phi_N /h}u'_h\Vert^2 ={\mathcal O}(h^\infty + h^{-M_2}\Vert e^{\phi_N /h}u'_h\Vert).
$$
In particular,
$$
\Vert e^{\phi_N /h}u'_h\Vert ={\mathcal O}(h^{-(M_2+1)}),
$$
and thus, also,
$$
\Vert \nabla(e^{\phi_N /h}u'_h)\Vert ={\mathcal O}(h^{-(M_2+3/2)}),
$$
and the result follows.
\end{pre}

\medskip
Now, we estimate $u-w_{CN}$ globally in a $N,h$-dependent small neighborhood $U_N:=\{x;{\rm dist}(x,\partial\open)<2(Nk)^{2/3}\}$ of the boundary of the island.
We show

\begin{prop}
\label{prop7}
There exist  $N_2\in \Z$ and $C>0$ such  that for any $N$ such that, for any $N$, one has as $h\to 0$
\be
\label{glbl}
||u-w_{CN}||_{H^1(U_N)}=\ord (h^{-N_2}e^{-S/h}).
\ee
\end{prop}

\begin{pre}
Let $U_{N,1}$ be the neighborhood of $\Gamma$ in $U_N$ defined by,
\be
\label{defUN1}
U_{N,1}=U_N\cap {\bf \Omega}(Nk, t_0),
\ee
with $t_0>0$ small enough.
We may assume,
$$
%U_{N,2}\subset \bigcup_{x^1\in\Gamma}\omega_{x^1,N}^1(h),\quad
U_{N,1}\subset \bigcup_{x^1\in\Gamma}\Omega_{x^1}^1(-(Nk)^{2/3},(Nk)^{2/3}),
$$
where 
%$\omega_{x^1,N}^1(h)$, 
$\Omega_{x^1}^1(-(Nk)^{2/3},(Nk)^{2/3})$ 
is the neighborhood of each $x^1\in\Gamma$ defined by 
%\eq{omegaN1}, 
\eq{omegaepsilon}.

Then Proposition \ref{comparison2}, \eq{compdiric3} and an estimate of $\tilde I[\tilde c]$ in the sea mean that there exists $N_2$ such that
\be
\label{N2}
||u-w_{CN}||_{H^1(U_{N,2})}=\ord (h^{-N_2}e^{-S/h}).
\ee

%On the other hand, recall (\eq{2Nk} and Proposition \ref{summary} (i))
%that
%\be
%\label{wCNest}
%||w_{CN}||_{H^1(U_N\backslash U_{N,1})}=\ord(h^{\delta N}e^{-S/h}).
%\ee
It follows from  \eq{compdiric3} and the fact  $(U_N\backslash U_{N,2})\cap B_d(x_0,S)=\emptyset$ that
$$
||u||_{H^1(U_{N}\backslash U_{N,2})}=\ord (h^{-N_2}e^{-S/h}),
$$
and since $w_{CN}$ vanishes in $U_N\backslash U_{N,2}$, we obtain \eq{glbl}.
\end{pre}

\section{Comparison in the Sea}

In this section, we give a more precise estimate on $v_N=e^{S/h}(u-w_{CN})$ ($C>0$ being as in Proposition \ref{comparison2}), in a neiborhood $U_N$ of $\p\open$. We will show

\begin{prop}\sl
\label{comparison}
For any $L>0$ and for any $\alpha\in\Z_+^n$, there exists $N_{L,\alpha}\geq 1$ such that, for any $N\geq N_{L,\alpha}$, one has, 
\be
\label{vhl}
\partial_x^\alpha v_N(x,h)={\mathcal O}(h^L)\quad {\rm as}\,\,\,\, h\to 0,
\ee
uniformly in $U_N$.
\end{prop}

Let $\hat x$ be an arbitrary point on $\p\open$. In the sequels, all the estimates we give are locally uniform with respect to $\hat x\in \p\open$ (and thus, indeed, globally uniform since $\p\open$ is compact).

\medskip
Here again, we choose Euclidian coordinates $x$ as in \S \ref{coord} but centered at $\hat x$ such
that $T_{\hat x}(\partial \open)$ is given by $x_n=0$, and
$\partial/\partial x_n$ is the exterior normal of $\open$ at
this point.

Consider  $h$-dependent neighborhoods of $\hat x$, of the form
\be
\label{omegaN1}
\hat\omega_N(h)=\{x; |x_n-\hat x_n|<(Nk)^{2/3}\, ;\, |x'-\hat x'| <(Nk)^{1/2}\},
\ee
 where $k=h\log\frac 1h$.

Let $(x(t),\xi(t))=\exp tH_p(0,0)$ be the Hamilton flow passing
by the origin at time 0 i.e.
\begin{equation}
\label{hamilton}
\left\{
\begin{array}{l}
\frac{dx}{dt}=2\xi, \quad x(0)=0\, ;\\[8pt]
\frac{d\xi}{dt}=-\frac{\p V}{\p x},\quad \xi(0)=0.
\end{array}
\right.
\end{equation}
 Recall that, by the non-trapping condition (A3), there esixts, at any point $\hat x$ on $\p\open$, a positive constant $C_0=C_0(\hat x)$ such that the potential $V$  is written in the form \eq{local}.
Hence $-\frac{\p V}{\p x}=(0,\ldots,0,C_0)+{\mathcal O}(|x|)$, and the flow is tangent
to the
$\xi_n$-axis. As $t\to 0$, one has
\begin{equation}
\label{asymp0}
\begin{array}{ll}
x_n(t)=C_0t^2+{\mathcal O}(t^3),&\xi_n(t)= C_0t+{\mathcal O}(t^2), \\[8pt]
x'(t)={\mathcal O}(t^4),&\xi'(t)={\mathcal O}(t^3).
\end{array}
\end{equation}

When $t\to\pm\infty$, on the other hand, $|x(t)|\to\infty$ by assumption (A3),
and $\xi(t)\to\xi_\infty^\pm$ for some $\xi_\infty^\pm\in\R^n$ satisfying
$|\xi^\pm_\infty|^2=E_0$ by (A1).  That is, as  $t\to\pm\infty$,
\begin{equation}
\label{asymp}
x(t)=2\xi_\infty^\pm t+o(|t|), \quad\xi(t)=\xi_\infty^\pm +o(1),\quad
|\xi_\infty^\pm |=\sqrt{E_0}.
\end{equation}
In particular,
\begin{equation}
\label{psxxi}
\left\{
\begin{array}{l}
x(t)\cdot\xi(t) = 2E_0t + o(|t|)\\
|x(t)|\cdot |\xi(t)| = 2E_0|t| + o(|t|)
\end{array}\right.
\mbox{ as }  t\to\pm\infty.\\
\end{equation}

\subsection{Propagation in the Incoming Region}
Here , we study the microlocal estimate of $e^{S/h}u$ and $e^{S/h}w_{CN}$ independently along the incoming Hamilton flow 
$\bigcup_{t<0}(x(t),\xi(t))$. For the estimate on $e^{S/h}u$, we use the fact that $u$ is {\it outgoing} at infinity and the propagation of {\it frequency set}.
For that of $e^{S/h}w_{CN}$, we use the result of \S 4.

\medskip
\subsubsection{Microlocal Estimate of $e^{S/h}u$}
We first study 
$\tilde u=e^{S/h}u$.  Using the Bargmann-FBI transform $T_\mu$ of (\ref{defT}), we
plan to prove that, for some convenient $\mu >0$, $T_\mu\tilde u(x,\xi,h)$ is exponentially small  for $(x,\xi)$ close enough to $(x(-t), \xi(-t))$,
$t>0$ sufficiently large. Actually, we prove something slightly better, namely,
\begin{lem}
\label{infinity}
For any $S_1>0$, there exist $t_1>0$ and $\mu>0$, such that, for all $t\geq t_1$, one has,
$$
T_\mu u(x,\xi,h)={\mathcal O}(e^{-S_1/h}),
$$
uniformly for $(x,\xi)$ in a neighborhood of $(x(-t), \xi(-t))$.
\end{lem}
\begin{pre}
Let $F(x)$ be the function used to define the distorded operator $P_\theta$ (see (\ref{Unu})), and 
let $\chi\in C^\infty (\R_+)$ verifying $\chi (|x|) =0$ on $\pi_x (\Supp\psi_0)\cup \{ F(x)\not= x\}$, $\chi =1$ on $[R,+\infty )$ for $R>1$ large enough, $\chi'\geq 0$ everywhere.

For $\delta >0$ small enough, we consider the distortion,
$$
G_\delta (x) := x+ i\delta \chi (|x|) x,
$$
and the corresponding distortion operator $\tilde U_\delta$, formally given by,
$$
\tilde U_\delta \phi (x):=(\det G_\delta )^{-1/2}\phi (G_\delta (x)).
$$
Then, the distorded operator $P_{\theta,\delta}:= \tilde U_\delta P_\theta \tilde U_\delta^{-1}$ is well defined, and its principal symbol $p_{\theta,\delta}$ verifies,
\begin{eqnarray*}
&& p_{\theta,\delta}(x,\xi ) = p_\theta (x,\xi ) \mbox{ if } x\chi n \pi_x (\Supp\psi_0)\cup \{ F(x)\not= x\};\\
&& p_{\theta,\delta}(x,\xi ) =(1+i\theta)^{-2} (dG_\delta (x)^{-1}\xi  )^2 +  V((1+i\theta)G_\delta (x))\\
&& \hskip 4cm  \mbox{ if } x\notin \pi_x (\Supp\psi_0)\cup \{ F(x)\not= x\}.
\end{eqnarray*}
Next, we observe that,
$$
dG_\delta (x) = (1+i\delta \chi (|x|))I + i\delta A(x),
$$
with, 
$$
A(x) = \frac{\chi '(|x|)}{|x|} (x_jx_k)_{1\leq j,k\leq n}.
$$
In particular, one has $\langle A(x)y,y\rangle = \frac{\chi'(|x|)}{|x|}\langle x,y\rangle^2\geq 0$ for all $y\in\R^n$, and thus, it is not difficult to deduce that $\im {}^tdG_\delta (x)^{-1}dG_\delta (x)^{-1}\leq 0$ for $\delta >0$ small enough. As a consequence, we see that, for $x\notin \pi_x (\Supp\psi_0)\cup \{ F(x)\not= x\}$, one has,
$$
\im p_{\theta,\delta}(x,\xi )\leq -\theta \xi^2 +{\mathcal O}(\langle x\rangle^{-\delta_1}),
$$
and thus, if $F(x)$ and $\psi_0$ have been conveniently constructed, and using (\ref{ellptilde}), we obtain,
$$
-\im  p_{\theta,\delta} (x-t \partial_x\psi_0-it\partial_\xi\psi_0, \xi -t\partial_\xi\psi_0+it\partial_x\psi_0)\geq \frac{k}{C},
$$
for some constant $C>0$, and for  $(x,\xi)$ such that $|\re  p_{\theta,\delta} (x,\xi)+W(x) - E_0|\leq \jap{\xi}^2/C$. As for $\tilde P_\theta$ (see Section 2), this implies that $(P_{\theta,\delta} + W -\rho )^{-1}$ is well defined and has a norm ${\mathcal O}(k^{-1})$ on $H_t$.

On the other hand, we also know that $u_{\theta,\delta} := \tilde U_\delta u_\theta = \tilde U_\delta U_{i\theta}u$ is well defined, and is in $L^2(\R^n)$ (see, e.g., \cite{hm}). Thus, we can write,
$$
(P_{\theta,\delta} + W -\rho )u_{\theta,\delta} =Wu_{\theta,\delta} = Wu,
$$
that is, $u_{\theta,\delta} = (P_{\theta,\delta} + W -\rho )^{-1}Wu$, and thus,
\begin{equation}
\label{estimudelta}
\Vert u_{\theta,\delta}\Vert_{L^2(\R^n)} = {\mathcal O}(h^{-M})\Vert u_{\theta,\delta} \Vert_t ={\mathcal O}(h^{-M}k^{-1}),
\end{equation}
for some $M>0$ constant, independent of $\delta>0$ small enough.

Making in the expression of $T_\mu u$ the change of contour of integration,
$$
\R^n\ni y\mapsto G_{\theta, \delta}(y):=G_\delta (y)+i\theta F(G_\delta (y)),
$$
we obtain,
$$
T_\mu u(x,\xi ) = c_\mu\int_{\R^n}e^{i(x-G_{\theta,\delta} (y))\xi /h - \mu (x-G_{\theta,\delta} (y))^2/2h}u_{\theta,\delta} (y)\det dG_{\theta,\delta} (y)dy,
$$
and thus, using (\ref{estimudelta}) and the Cauchy-Schwarz inequality,
\begin{equation}
\label{estTusort}
T_\mu u(x,\xi ) ={\mathcal O}(h^{-M_1})\left [\int e^{\left\{2\im G_{\theta,\delta}(y)\xi  +\mu (\im G_{\theta,\delta}(y))^2 - \mu(x-\re G_{\theta,\delta}(y))^2\right\}/h}dy\right ]
\end{equation}
for some $M_1>0$ constant, independent of $\delta$.

Now, let $R_1>>1$ be some fixed number arbitrarily large, and take $t>0$ sufficiently large to have  $|x(-t)|\geq R+2R_1$.
Then, for $(x,\xi)$  close enough to $(x(-t),\xi (-t))$, and setting $\tilde\delta:=\delta + \theta$, we deduce from (\ref{estTusort}),
\begin{eqnarray*}
T_\mu u(x,\xi )= {\mathcal O}(h^{-M_1})\left[ \int_{|y-x|\leq |x|/2}e^{2\tilde\delta y\xi /h +\mu\tilde\delta^2y^2/h - \mu(x-y+\theta\delta y)^2/h}dy\right]^{1/2}\\
 + {\mathcal O}(h^{-M_1}e^{-\mu x^2/16h})\left[ \int_{|y-x|\geq |x|/2}e^{2\tilde\delta y\xi /h +\mu \tilde\delta^2y^2/h - \mu (x-\re G_{\theta,\delta}(y))^2/2h}dy\right]^{1/2},
\end{eqnarray*}
and thus, for $\delta/\mu$ small enough, (and since  $|y-x|\geq |x|/2$ implies $|y-x|\geq |y|/4$,  and we have 
$\re G_{\theta,\delta}(y) =y+{\mathcal O}(\theta\delta |y|)$, and $|\xi|$ remains uniformly bounded),
\begin{eqnarray}
\label{estTusort1}
T_\mu u(x,\xi )= {\mathcal O}(h^{-M_1})\left[ \int_{|y-x|\leq |x|/2}e^{2\tilde\delta y\xi /h +\mu\tilde\delta^2y^2/h }dy\right]^{1/2}\\
 + {\mathcal O}(e^{-\mu R_1^2/4h}).\nonumber
\end{eqnarray}
Moreover, if $|y-x|\leq |x|/2$, we have $y\cdot \xi \leq x\cdot \xi + |x|\cdot |\xi|/2$, and thus, by (\ref{psxxi}), and for  $(x,\xi)$ close enough to $(x(-t),\xi (-t))$, we obtain (possibly by taking $t$ larger),
$$
y\cdot \xi \leq - E_0t/2.
$$
In the same way, using (\ref{asymp}), we also obtain,
$$
|y|\leq 3|x|/2 \leq 4\sqrt{E_0} t,
$$
and thus, inserting  these estimates into (\ref{estTusort1}), we find,
$$
T_\mu u(x,\xi )= {\mathcal O}(h^{-M_1}e^{-E_0\tilde\delta t(1 - 16\mu\tilde\delta t)/h})+ {\mathcal O}(e^{-\mu R_1^2/4h}).
$$
In particular, for any $S_1>0$, if we first fix $\mu>0$ such that $E_0>64\mu S_1$, then $R_1$ such that $\mu R_1^2/4 \geq S_1$, then $t>>1$ such that $|x(-t)|\geq R+2R_1$, and finally $\delta := (32\mu t)^{-1}-\theta =(32\mu t)^{-1}-k$, we obtain,
\begin{eqnarray*}
T_\mu u(x,\xi )&=&{\mathcal O}(h^{-M_1}e^{-E_0\tilde \delta t/2h}+e^{-S_1/h})\\
&=&{\mathcal O}(h^{-M_1}e^{-E_0/64\mu h}+e^{-S_1/h})\\
&=&{\mathcal O}(e^{-S_1/h}),
\end{eqnarray*}
uniformly for $(x,\xi)$ close enough to $(x(-t),\xi (-t))$ and $h>0$ small enough.
{} \hfill \end{pre}
\vskip 0.2cm
In particular, taking $S_1>S$, we obtain, 
$$
T_\mu (e^{S/h}u) ={\mathcal O}(h^\infty ) \mbox{ near } (x(-t_1),\xi (-t_1)),
$$
and therefore,
$$
(x(-t_1),\xi (-t_1))\notin FS(e^{S/h}u),
$$
where $FS(e^{S/h}u)$ stands for the {\it frequency set} of $e^{S/h}u$ (see, e.g., \cite{GuSt, ma1}). Moreover, by (\ref{compdiric3}), we know that, for any  $K\subset \R^n\backslash B_d(x_0,S)$ compact,  $\Vert e^{S/h}u\Vert_{H^1(K)}={\mathcal O}(h^{-N_K})$ for some $N_K>0$ constant. Since $\R^n\backslash  B_d(x_0,S)$ is a neighborhood of $\{ x(-t)\, ;\, t>0\}$, and $(P-\rho)(e^{S/h}u) =0$ in ${\mathcal D}'(\R^n\backslash  \overline{B_d(x_0,S)})$, the standard result of propagation of the frequency set for solutions of real-principal type partial differential equations (see, e.g., \cite{ma1} Chapter 4, exercise 7) can be applied above this set, and  tells us,
\begin{equation}
(x(-t),\xi (-t))\notin FS(e^{S/h}u) \mbox{ for all } t>0.
\end{equation}
In particular, for any $\mu >0$ fixed (indepedent of $h$), for any $t>0$, and for any $K\subset \R^n\backslash  B_d(x_0,S)$ compact containing $(x(-t),\xi (-t))$ in its interior, we have (denoting by ${\bf 1}_{K}$ the characteristic function of $K$),
\begin{equation}
\label{absFS}
T_\mu ({\bf 1}_{K}e^{S/h}u)={\mathcal O}(h^\infty ) \mbox{ uniformly near } (x(-t),\xi (-t)).
\end{equation}
Now, we set,
\be
\label{TN}
{\bf T}_N u(x,\xi ):= \int e^{i(x-y)\xi /h -\mu_n (x_n-y_n)^2/2h -(x'-y')^2/2h}u(y)dy,
\ee
where $\mu_n:= (Nk)^{-1/3}$.

\begin{lem} 
\label{noFS}
For any $t>0$, for any $K\subset \R^n\backslash  B_d(x_0,S)$ compact containing $(x(-t),\xi (-t))$ in its interior, and for any $N\geq 1$, we have,
$$
{\bf T}_N ({\bf 1}_{K}e^{S/h}u)={\mathcal O}(h^\infty ) \mbox{ uniformly near } (x(-t),\xi (-t)).
$$
\end{lem}
\begin{pre} We write,
$$
{\bf T}_N ({\bf 1}_{K}e^{S/h}u) = ({\bf T}_NT_1^*)T_1({\bf 1}_{K}e^{S/h}u),
$$
and a straightforward computation shows that the distribution kernel $K_N$ of ${\bf T}_N T_1^*$ verifies (see, e.g., \cite{ma1} proof of Proposition 3.2.5),
\begin{eqnarray*}
|K_N (x,\xi ; z,\zeta )| =\alpha e^{ -\frac{\mu_n}{1+\mu_n}(x_n-z_n)^2/2h-\frac1{1+\mu_n}(\xi_n -\zeta_n)^2/2h}\\
\times e^{-(x'-z')^2/4h-(\xi' -\zeta')^2/4h},
\end{eqnarray*}
with $\alpha = {\mathcal O}(h^{-n})$. Then, the result easily follows from the obvious observation that, for any fixed $\delta>0$,  one has,
$$
e^{-\frac{\mu_n}{1+\mu_n}\delta/h} + e^{-\frac1{1+\mu_n}\delta/h} = {\mathcal O}(h^\infty ).
$$
\end{pre}
\vskip 0.2cm
Now, we fix once for all a compact set $K_1=K\backslash  B_d(x_0,S)$, where $K\subset\subset\R^n$ is a compact neighborhood of the closure of $\open$.
\begin{lem}
\label{propagation1}
There exists $\delta_0 >0$ such that, for any $\delta\in (0,\delta_0]$, for all $N\geq 1$ large enough, and for $t_N:=\delta^{-1}(Nk)^{1/3}$, one has,
$$
{\bf T}_N ({\bf 1}_{K_1}e^{S/h} u)={\mathcal O}(h^{\delta N}) \mbox{ uniformly in } {\mathcal W}(t_N,h),
$$
where,
\begin{eqnarray*}
{\mathcal W}(t_N,h):= \{  |x_n- x_n(-t_N)|\leq \delta (Nk)^{2/3}\, ,\, |\xi_n -\xi_n (-t_N)| \leq \delta (Nk)^{1/3},\\
|x'- x'(-t_N)|\leq \delta (Nk)^{1/3}\, ,\, |\xi' -\xi' (-t_N)| \leq \delta (Nk)^{1/3}\}.
\end{eqnarray*}
\end{lem}
\begin{pre} At first, we cut off the function $e^{S/h}u$ by setting
$$
u_1:= \chi_+e^{S/h}u,
$$
where $\chi_+\in C^\infty (\R^n)$, $\Supp\chi_+\subset \{ k^{2/3}\leq \mbox{dist} (x, \open)\leq 2\}$, $\chi_+=1$ on $\{ 2k^{2/3}\leq \mbox{dist} (x, \open)\leq 3/2 \}$, and $\partial^\alpha\chi_+={\mathcal O}(k^{-2|\alpha|/3})$ for all $\alpha\in\Z_+^n$. In particular, by (\ref{compdiric3}), we have $\Vert (P-\rho)u_1 \Vert_{L^2}= \Vert [P,\chi_+]u_1\Vert_{L^2} ={\mathcal O}(h^{-N_1})$ for some $N_1\geq 0$ constant, and, if $\psi =\psi (x,\xi) \in C_0^\infty (\R^{2n})$ is such that 
\be
\label{supppsi}
\pi_x\,\Supp\psi\subset \{ 3(Nk)^{2/3}\leq \mbox{dist} (x, \open)\leq 1\},
\quad\sup |\psi|\leq 2, 
\ee
then, for any $M\geq 1$, we have,
$$
h^{-M\psi }{\bf T}_N(P-\rho)u_1(x,\xi)=h^{-M\psi }{\bf T}_N[P,\chi_+]u_1(x,\xi)
$$
and, since $|x_n-y_n| \geq (Nk)^{2/3}$ for $x\in\pi_x\Supp\psi$ and $y\in\Supp [P,\chi_+]u_1$, we easily obtain,
$$
\Vert h^{-M\psi }{\bf T}_N(P-\rho)u_1\Vert_{L^2} ={\mathcal O}(h^{-N_1} + h^{-2M}e^{-\mu_n (Nk)^{4/3}/2h}),
$$
that is, 
$$
\Vert h^{-M\psi }{\bf T}_N(P-\rho)u_1\Vert_{L^2} ={\mathcal O}(h^{-N_1} + h^{-2M+N/2}).
$$
In particular, for $M\leq \frac{N}4$, this gives,
\begin{equation}
\label{estPu1}
\Vert h^{-M\psi }{\bf T}_N(P-\rho)u_1\Vert_{L^2} ={\mathcal O}(h^{-N_1}).
\end{equation}
Now, in order to specify the function $\psi$ we are going to work with, we first make a symplectic change of variables near $(0,0) \in \R^{2n}$:
\begin{equation}
\label{changesymp}
\left\{
\begin{array}{ll}
y'=x',\quad &y_n=x_n-\frac 1{C_0}\xi_n^2, \\
\eta'=\xi',\quad &\eta_n=\xi_n.
\end{array}
\right.
\ee
In this new coordinates, we have
\be
p=\eta'^2-C_0y_n+W(y',y_n+\frac 1{C_0}\eta_n^2),
\ee
\be
\label{hpyeta}
H_p=2\eta'\frac{\partial}{\partial y'}+C_0\frac{\partial}{\partial \eta_n}
-\nabla W\frac{\partial}{\partial \eta}
+\frac 2{C_0}\eta_n\partial_{x_n}W\frac{\partial}{\partial y_n}.
\ee

Now, we fix $t_0>0$ small enough, and we consider the function,
$$
\psi (y,\eta ) := f(\eta_n)\hskip 1pt \chi \left(\frac{|\eta'|}{|\eta_n|}\right)\hskip 1pt \chi\left( \frac{|y_n|}{|\eta_n|^2}\right) \chi\left( \frac{|y'|}{|\eta_n|}\right),
$$
where $\chi \in C_0^\infty (\R_+; [0,1])$ is such that,
$$
\Supp \chi\subset [0,a],
\quad\chi = 1 \mbox{ on } [0, \frac a2],
\quad-\frac 4a\le \chi'\leq 0 \mbox{ on } \R_+,
$$
for some constant $a>0$ small enough,
and $f\in C_0^\infty (\R ; [0,1+t_0+\varepsilon'])$ is defined in the following way,
\begin{equation}
\label{deff}
f(s) =  \chi_0(s)f_1(s):= \chi_0(s)\left( -Cs -2C_1^2 \int_{s(Nk)^{-1/3}}^{-C_1}\frac{dt}{t^{3}}\right),
\end{equation}
where $C, C_1>0$ are large enough constants, and $\chi_0$ is a cut-off function  such that,
\begin{eqnarray*}
&& \chi_0  \in C_0^{\infty} ([-t_0-\varepsilon, -C_1(Nk)^{1/3}] ;[0,1]);\\
&& \chi_0 =1\mbox{ on } [-t_0, -2C_1(Nk)^{1/3}]\\
&& \chi_0'\leq 0 \mbox{ on } [-2C_1(Nk)^{1/3}, -C_1(Nk)^{1/3}];\\
&& \mbox{for all } \ell\geq 0,\, \chi_0^{(\ell)}={\mathcal O}((Nk)^{-|\ell|/3}) \mbox { on } [-2C_1(Nk)^{1/3}, -C_1(Nk)^{1/3}];\\
&& \mbox{for all } \ell\geq 0,\, \chi_0^{(\ell)}={\mathcal O}(1) \mbox { on }[-t_0-\varepsilon, -t_0].
\end{eqnarray*}

This $\psi$ satisfies the condition \eq{supppsi}, since on the support of $\psi$, one has
$$
\begin{array}{rl}
E_0-V(x)=&\eta^2+C_0y_n-W(y',y_n+\frac 1{C_0}\eta_n^2) \\[8pt]
\ge &\eta_n^2-aC_0\eta_n^2+\ord (a^2\eta_n^2+\eta_n^4) \\[8pt]
\ge &\frac 12\eta_n^2 
\ge  \frac{C_1^2}2(Nk)^{2/3}
\end{array}
$$
for sufficiently small $a$.
In particular, (\ref{estPu1}) is valid with such a $\psi$.

\medskip
Moreover, one has
\be
\partial_{y'}^{\alpha'} \partial_{y_n}^{\alpha_n}\partial^\beta_\eta \psi ={\mathcal O}((Nk)^{-(|\alpha'|+2|\alpha_n|+|\beta|)/3}).
 \ee
Therefore, we see that $\psi$ satisfies the conditions (\ref{estpsi})-(\ref{condsura}) in the $(y',\eta')$-coordinates, with $\rho =0$, and the same conditions in the $(y_n,\eta_n)$-coordinates, with $\rho = -1/3$.

Then, by a straightforward generalization of  Proposition \ref{exp}, and by using (\ref{estHp}) and (\ref{estPu1}), we obtain,
\begin{eqnarray}
\label{estapriori}
&&k\langle  (MH_{p}\psi + q_{M\psi} )h^{-M\psi}{\bf T}_N  u_1,
h^{-M\psi} {\bf T}_N  u_1\rangle \\
&& \hskip 2cm ={\mathcal O}(h)||\langle\eta\rangle h^{-M\psi} {\bf T}_N  u_1||^2 +{\mathcal O}(h^{-N_1})||h^{-M\psi} {\bf T}_N  u_1||, \nonumber
\end{eqnarray}
where $M\leq \frac{N}4$ and $N_1\geq 1$ is independent of $N$.

\medskip

In the sequel, we use the notations,
\begin{eqnarray*}
&& I_1 := \Supp\psi \cap\{ \eta_n\in [-t_0-\varepsilon', -t_0+\varepsilon']\};\\
&& I_2:= \Supp\psi \cap\{ \eta_n\in [-t_0+\varepsilon', -2C_1(Nk)^{1/3}]\};\\
&& I_3:= \Supp\psi \cap\{ \eta_n\in [-2C_1(Nk)^{1/3}, -C_1(Nk)^{1/3}]\}.
\end{eqnarray*}

Let us now estimate $|H_p\psi|$ from below on $I_2\cup I_3$.
First, observe that one has
\be
\label{estf}
|f(\eta_n)|\le 2 \chi_0(\eta_n)
\ee
if $t_0\le 1/C$, and in particular on $I_2\cup I_3$, where $\chi_0'\le 0$, one has
\be
\label{estf'}
|f'(\eta_n)|\ge \left (C+\frac{2C_1^2(Nk)^{2/3}}{|\eta_n|^3}\right )\chi_0(\eta_n)\ge C\chi_0(\eta_n).
\ee
Using these estimates, and the expression \eq{hpyeta}, we can easily show 
\begin{lem}
For $t_0<C_0/2$
\be
\label{Hppsi}
|H_p\psi|\ge |f'(\eta_n)|(C_0\chi_1\chi_2\chi_3-\ord (1/C)).
\ee
where
$$
\chi_1=\chi(\frac{|\eta'|}{|\eta_n|}), \quad
\chi_2=\chi(\frac{|y_n|}{|\eta_n|^2}), \quad
\chi_3=\chi(\frac{|y'|}{|\eta_n|}).
$$
\end{lem}

\begin{pre}
One can estimate $|C_0\frac{\partial}{\partial \eta_n}\psi|$ from below by
$$
|C_0\frac{\partial}{\partial \eta_n}\psi|\ge C_0|f'(\eta_n)|\chi_1\chi_2\chi_3.
$$
One can also estimate $2|\eta'\frac{\partial}{\partial y'}\psi|$, $|\nabla W\frac{\partial}{\partial \eta'}\psi|$,  $|\eta_n||\nabla_{x_n} W\frac{\partial}{\partial y_n}\psi|$ from above by $|f'|$ times a constant  of $\ord (1/C)$. For example,
$$
2|\eta'\frac{\partial}{\partial y'}\psi|\le f\chi_1\chi_2\frac{|\eta'|}{|\eta_n|}|\chi'_3|\le 4f\chi_1\chi_3\le \frac 8C\chi_1\chi_2|f'|,
$$
using the facts $\frac{|\eta'|}{|\eta_n|}\le a$, $|\chi_3'|\le 4/a$, $f\le2f'/C$. On the other hand, for 
 $|\nabla W\frac{\partial}{\partial \eta_n}\psi|$, one has
$$
|y||\frac{\partial}{\partial \eta_n}\psi|\le a|\eta_n||\frac{\partial}{\partial \eta_n}\psi|\le |t_0||\frac{\partial}{\partial \eta_n}\psi|,
$$
which is smaller than $\frac{\partial}{\partial \eta_n}\psi|$ for sufficiently small $t_0$.
\end{pre}

\medskip

%Moreover, direct computations show that, on $I_2$ (and since $\chi_1(\eta_n)=1$ there),
%\begin{eqnarray*}
%&& |\nabla_y\nabla_\eta\psi |+ |\nabla_y\psi |\cdot|\nabla_\eta\psi|={\mathcal O}(\nu^{-3} );\\
%&& |\nabla_y^2\psi | + |\nabla_y\psi |^2 ={\mathcal O}(\nu^{-4});\\
%&& |\nabla_\eta^2\psi | +  |\nabla_\eta\psi |^2={\mathcal O}(\nu^{-2} ),
%\end{eqnarray*}
%and therefore, still on $I_2$, we find,
%$$
%q_{M\psi} ={\mathcal O}(\mu kM^2\nu^{-3} +\mu hM\nu^{-3}+ M^3k^2\nu^{-6}+\mu^3M^3k^2\nu^{-3}),
%$$
%that is, since $hM+\mu^{-1}M^3k^2\nu^{-3}+ \mu^2M^3k^2 \leq hM + 2(Nk)^{-2/3}M^3k^2 \leq kM^2$,
%\begin{equation}
%\label{estqI2}
%q_{M\psi} ={\mathcal O}(\mu kM^2\nu^{-3}) \mbox{ on } I_2.
%\end{equation}
%In the same way, using that $|\eta_n|\sim \nu\sim (Nk)^{1/3}$ on $I_3$,  we also find,
%\begin{equation}
%\label{estqI3}
%q_{M\psi} ={\mathcal O}(\mu kM^2\nu^{-3}) \mbox{ on } I_3.
%\end{equation}

\medskip

Now, let $C_2>C_1$ be another large enough constant. We set,
\begin{eqnarray*}
&& \Omega_1:= (I_2\cup I_3)\cap \left\{ \chi_0\geq \frac1{C_2}\, ;\, \chi_1\geq \frac1{C_2}\, ;\,\chi_2\geq \frac1{C_2}\, ;\,\chi_3\geq \frac1{C_2}\, \right\};\\
&& \Omega_2:= (I_2\cup I_3)\backslash \Omega_1.
\end{eqnarray*}
Then, by construction, on $\Omega_2$ we have by \eq{estf},
\begin{equation}
\label{estpsiomeg2}
\psi \leq \frac1{C_2}\sup f\leq \frac{2}{C_2}.
\end{equation}
Moreover, in a neighborhood of $X_N:= (0 ;0, -3C_1(Nk)^{1/3})$, of the form,
$$
{\mathcal W}_N=\{|\eta'|\le\delta |\eta_n|, |y_n|\le\delta |\eta_n|^2, |y'|\le\delta |\eta_n|,  |(Nk)^{-1/3}\eta_n + 3C_1 |<\delta\},
$$
(where $\delta >0$ is a small enough constant), we see that,
\begin{equation}
\label{psiomeg1}
\psi \geq \frac12\psi(X_N) =3CC_1(Nk)^{1/3}-2C_1^2\int_{-3C_1}^{-C_1}\frac{dt}{t^3}\geq \frac 89=:r_0,
\end{equation}
and, by  (\ref{estpsiomeg2}), we can fix $C_2$ large enough, in such a way that,
\begin{equation}
\label{psiomeg2}
\sup_{\Omega_2}\psi \leq \frac12 r_0.
\end{equation}
On the other hand, by \eq{estf'}, (\ref{Hppsi}), on $\Omega_1$, we have,
\begin{equation}
\label{Hpsiell}
|H_p\psi |\ge \frac{C_0}{2C_2^3}|f'|\ge  \frac{C_0}{2C_2^4}
\left (C+\frac{2C_1^2(Nk)^{2/3}}{|\eta_n|^3}\right ).
\end{equation}

\medskip
Using  the expression of $q_{M\psi}$ deduced from Proposition \ref{exp}, and the fact that  here, $p(x,\xi )=\xi^2+V(x)$, we have

\begin{lem}
As $h\to 0$, 
\be
\label{qmps}
q_{M\psi}=-2kM^2\mu_n\partial_{x_n}\psi\partial_{\xi_n}\psi+\ord \left (\frac 1{\ln (1/h)}\right).
\ee
In particular, on $\Omega_1$, for $M/N$ sufficiently small, one has
\be
\label{qmhp}
|q_{M\psi}| \leq \frac M2 |H_p\psi|.
\ee
\end{lem}

\begin{pre}
Taking into account that
$$
\begin{array}{ll}
\partial_{x'}\psi=\ord (k^{-1/3}),\quad&\partial_{x_n}\psi=\ord (k^{-2/3}), \\
\partial_{\xi'}\psi=\ord (k^{-1/3}),\quad&\partial_{\xi_n}\psi=\ord (k^{-1/3}),
\end{array}
$$
and hence that
$$
\partial_{z_\mu'}\psi=\ord (k^{-1/3}),\quad\mu_n^{-1}\partial_{z_{\mu_n}}\psi=\ord (k^{-1/3}), 
$$
we see that for $p=\xi^2+V(x)$,
$$
\im p(x-2k\mu^{-1}\partial_{z_\mu}\psi, \xi+ik\partial_{z_\mu}\psi)
=kH_p\psi-2k^2\mu_n\partial_{x_n}\psi\partial_{\xi_n}\psi+\ord (k^{4/3}),
$$
and that
$$
\begin{array}{l}
h\partial_{z_\mu}\left [\frac 1\mu\frac{\partial p}{\partial \re x}-i\frac{\partial p}{\partial \re \xi}\right ](x-2k\mu^{-1}\partial_{z_\mu}\psi, \xi+ik\partial_{z_\mu}\psi) \\
= h\partial_{z_\mu}\left [\frac 1\mu\frac{\partial p}{\partial \re x}-i\frac{\partial p}{\partial \re \xi}\right ](x,\xi)+\ord (h).
\end{array}
$$
Since $\partial_{z_\mu}\left [\frac 1\mu\frac{\partial p}{\partial \re x}-i\frac{\partial p}{\partial \re \xi}\right ](x,\xi)$ is real, we obtain \eq{qmps}.

The estimate \eq{qmhp} follows from \eq{qmps} and \eq{Hpsiell},
because using the estimate
$$
 \partial_{\eta_n}\psi=\ord (|\eta_n|^{-1}),
$$
one sees that
$$
kM^2\mu_n\partial_{x_n}\psi\partial_{\xi_n}\psi=\tilde C\frac MN \cdot M
\frac{(Nk)^{2/3}}{|\eta_n|^3},
$$
for some constant $\tilde C$ independent of $M, N$.
\end{pre}

\medskip
Thus, still by (\ref{Hpsiell}),
\begin{equation}
\label{ellip1}
|MH_p\psi + q_{M\psi}|\geq \frac{MCC_0}{4C_2^4}  \mbox{ on } \Omega_1.
\end{equation}

\medskip

Now, we turn back to (\ref{estapriori}), that we rewrite as,
\begin{eqnarray*}
&&\langle  (MH_{p}\psi + q_{M\psi} )h^{-M\psi}{\bf T}_N  u_1,
h^{-M\psi} {\bf T}_N  u_1\rangle_{\Omega_1} \\
 &&=-\langle  (MH_{p}\psi + q_{M\psi} )h^{-M\psi}{\bf T}_N  u_1,
h^{-M\psi} {\bf T}_N  u_1\rangle_{I_1\cup \Omega_2} \\
&& \hskip 2cm +{\mathcal O}(hk^{-1}||\langle\eta\rangle h^{-M\psi} {\bf T}_N  u_1||^2 +h^{-N_1}k^{-1}||h^{-M\psi} {\bf T}_N  u_1||), 
\end{eqnarray*}
and thus, since $MH_{p}\psi + q_{M\psi} ={\mathcal O}(h^{-N_3}) $ for some $N_3\geq 1$ constant, and $\eta$ is bounded on $\Supp\psi$,
\begin{eqnarray*}
&&\langle  (MH_{p}\psi + q_{M\psi} )h^{-M\psi}{\bf T}_N  u_1,
h^{-M\psi} {\bf T}_N  u_1\rangle_{\Omega_1} \\
 &&={\mathcal O}(h^{-N_3}\Vert  h^{-M\psi}{\bf T}_N  u_1\Vert^2_{I_1\cup \Omega_2}+hk^{-1}||h^{-M\psi} {\bf T}_N  u_1||^2_{\Omega_1}) \\
&& \hskip 3cm +{\mathcal O}( hk^{-1}||\langle\eta\rangle  {\bf T}_N  u_1||^2 +h^{-N_1}k^{-1}||h^{-M\psi} {\bf T}_N  u_1||), 
\end{eqnarray*}
Using (\ref{ellip1}), we deduce,
$$
\Vert h^{-M\psi}{\bf T}_N  u_1\Vert^2_{\Omega_1} = {\mathcal O}(h^{-N_3}\Vert  h^{-M\psi}{\bf T}_N  u_1\Vert^2_{I_1\cup \Omega_2}+hk^{-1}||h^{-M\psi} {\bf T}_N  u_1||^2_{\Omega_1}) $$
$$
\hspace{2cm}+{\mathcal O}( hk^{-1}||\langle\eta\rangle  {\bf T}_N  u_1||^2 +h^{-N_1}k^{-1}||h^{-M\psi} {\bf T}_N  u_1||), 
$$
and  thus, since $hk^{-1} =|\ln h|^{-1}\rightarrow 0$ as $h\rightarrow 0_+$,
\begin{eqnarray*}
\Vert h^{-M\psi}{\bf T}_N  u_1\Vert^2_{\Omega_1} &=& {\mathcal O}(h^{-N_3}\Vert  h^{-M\psi}{\bf T}_N  u_1\Vert^2_{I_1\cup \Omega_2}) \\
&&+{\mathcal O}( hk^{-1}||\langle\eta\rangle  {\bf T}_N u_1||^2 +h^{-N_1}k^{-1}||h^{-M\psi} {\bf T}_N  u_1||), 
\end{eqnarray*}
uniformly for $h>0$ small enough. Therefore, setting $N_4=\max (N_3,  N_1+1)$, we obtain,
\begin{eqnarray*}
&& \Vert h^{-M\psi}{\bf T}_N  u_1\Vert^2_{\Omega_1} - \tilde Ch^{-N_4}||h^{-M\psi} {\bf T}_N  u_1||_{\Omega_1} \\
&&\hskip 3cm\leq \tilde Ch^{-N_4}(\Vert  h^{-M\psi}{\bf T}_N  u_1\Vert^2_{I_1\cup \Omega_2}+ ||\langle\xi\rangle  {\bf T}_N u_1||^2),
\end{eqnarray*}
for some positive constant $\tilde C$, and thus,
$$
\Vert h^{-M\psi}{\bf T}_N  u_1\Vert_{\Omega_1}\leq C'h^{-N_4}(1+\Vert  h^{-M\psi}{\bf T}_N  u_1\Vert_{I_1\cup \Omega_2}+ ||\langle\xi\rangle  {\bf T}_N u_1||),
$$
for some other constant $C'>0$.

In particular, since  ${\mathcal W}_N\subset \Omega_1$ and $||\langle\xi\rangle  {\bf T}_N  u_1||=||  u_1||_{H^1} ={\mathcal O}(1)$,
\begin{eqnarray*}
\Vert h^{-M\psi}{\bf T}_N  u_1\Vert_{{\mathcal W}_N} &=&{\mathcal O}(h^{-N_4})(1+\Vert  h^{-M\psi}{\bf T}_N u_1\Vert_{I_1\cup \Omega_2})\\
&=&{\mathcal O}(h^{-N_4})(1+\Vert  h^{-M\psi}{\bf T}_N  u_1\Vert_{ \Omega_2}+ h^{-2M}\Vert {\bf T}_N u_1\Vert_{I_1} )
\end{eqnarray*}
Then, using Lemma \ref{noFS}, we see that $\Vert {\bf T}_N  u_1\Vert_{I_1} ={\mathcal O}(h^\infty )$ if $a$ has been taken sufficiently small, and thus, using 
(\ref{psiomeg1})-(\ref{psiomeg2}), we obtain,
$$
h^{-Mr_0}\Vert {\bf T}_N  u_1\Vert_{{\mathcal W}_N}={\mathcal O}(h^{-N_4-\frac{M}2r_0}),
$$
that is,
\begin{equation}
\label{estfinale}
\Vert {\bf T}_N  u_1\Vert_{{\mathcal W}_N}={\mathcal O}(h^{-N_4+\frac{M}2r_0}),
\end{equation}
where the estimate is valid for $N$ large enough, $M/N$ small enough, and is uniform with respect to $h>0$ small enough. This completes the proof.
\end{pre}

\medskip
\subsubsection{Microlocal Estimate of $e^{S/h}w_{CN}$}
Now, we study the microlocal behavior of the WKB solution $w_{CN}$ in  ${\mathcal W}(t_N,h)$. 
For $N\geq 1$ arbitrary, we denote by $\chi_N$ a cut-off function of the type,
\begin{equation}
\label{defchiN}
\chi_N (x):=\chi_0\left(\frac{|x_n-\hat x_n|}{(Nk)^{2/3}}\right)\chi_0\left(\frac{|x'-\hat x'|}{(Nk)^{1/2}}\right),
\end{equation}
where the function $\chi_0\in C_0^\infty (\R_+ ; [0,1])$ verifies $\chi_0 =1$ in a sufficiently large neighborhood of $0$, and is fixed  in such a way that $\chi_N(x) =1$ in $\{|x_n-\hat x_n| \leq |x_n(-t_N)-\hat x_n| + 2\delta (Nk)^{2/3}\,;\, |x'-\hat x'|\leq |x'(-t_N)-\hat x'|+2\delta (Nk)^{1/2}\}$ (here, $t_N$ and $\delta$ are those of Lemma \ref{propagation1}). Then, setting,
$$
\tilde w_N:=1_{B_d(x_0,S)^C}\chi_Nw_{CN},
$$
(with $C>0$ fixed large enough, as in Proposition \ref{comparison2}),we have,
\begin{lem} 
\label{wsort}
For any $L\in \N$ large enough, there exists $\delta_L >0$ such that, for any $\delta\in (0,\delta_L]$, for all $N\geq 1$ large enough, and for
$t_N:=\delta^{-1}(Nk)^{1/3}$, one has,
$$
{\bf T}_N (e^{S/h} \tilde w_N)={\mathcal O}(h^{\delta N}+h^L) \mbox{ uniformly in } {\mathcal W}(t_N,h).
$$
\end{lem}
\begin{pre}
Let $\chi^1(r)\in C_0^\infty(\R^+;[0,1])$ be a cut-off function such that
$\chi^1=1$ for $0\le r\le 2\delta$ and $\chi^1=0$ for $3\delta\le r$ and set,
$$
\chi_N^1(x):=\chi_1\left(\frac{|x_n-x_n(-t_N)|}{(Nk)^{2/3}}\right)\chi_1\left(\frac{|x'-x'(-t_N)|}{(Nk)^{1/2}}\right).
$$
We write,
$$
{\bf T}_N (e^{S/h}\tilde w_N)={\bf T}_N (e^{S/h}\chi_N^1\tilde w_N)+{\bf T}_N (e^{S/h}(1-\chi_N^1)\tilde w_N)=:I_1+I_2
$$

First we study $I_2$.
We have $|x_n-y_n|\geq\delta (Nk)^{2/3}$ or $|x'-y'|\geq \delta (Nk)^{1/2}$ if $|x_n-x_n(-t_N)|\le\delta (Nk)^{2/3}$, and $|x'-x'(-t_N)|\le\delta (Nk)^{1/2}$, and $y\in {\rm supp} (1-\chi_N^1)$. Hence, there we have
$$
e^{-\mu_n (x_n-y_n)^2/2h-(x'-y')^2/2h}\le h^{\delta^2 N/2}.
$$
With the estimate of  $e^{S/h}w_N$ in Proposition \ref{summary} (i), we deduce,
$$
|I_2(x,\xi;h)|= {\mathcal O}(h^{\delta^2 N/4}),
$$
uniformly for $(x,\xi)\in{\mathcal W}(t_N,h)$.

\medskip
Next we study $I_1$. Since ${\rm Supp}\, \chi^1_N\subset \Omega(c_2(Nk)^{2/3}, c_1(Nk)^{2/3})$ for some $c_1 >c_2>0$, we can use the WKB expansion \eq{summarywkb}, that we prefer to write in the  coordinates $z'=y'$, $z_n=y_n+b(y')$ as in (\ref{asymptotique1}). Using also (\ref{criticalvalue}) and (\ref{phi+}), we obtain,
$$
e^{S/h}w_N(y;h)=h^{-n/4}e^{-\tilde\phi(z',z_n)/h}A(z',z_n;h)+\ord (h^{\delta_LN}+h^L),
$$
where
$$
\tilde\phi(z',z_n)=a(z')-b(z')\xi_n^c(z')+\xi_n^c(z')z_n-i\tilde\nu(z', -iz_n^{1/2})z_n^{3/2},
$$
$$
A(z',z_n;h)=\sum_{j=0}^{L+[Nk/c_L'h]}\frac{\tilde f_j(z', -iz_n^{1/2})}{(-iz_n^{1/2})^{1/2+3j}}h^j,
$$
with $\tilde\nu(z', -iz_n^{1/2})$ and  $\tilde f_j(z', -iz_n^{1/2})$  holomorphic with respect to $z_n^{1/2}$ for $|z_n|<c_1(Nk)^{2/3}$, and $\tilde\nu(0,0)=\frac 23\sqrt{C_0}$ (see \eq{criticalvalue}, \eq{tildedelta}, \eq{odex'22}). In particular, on $\supp \chi^1_N$, we have, 
\begin{equation}
\label{checka}
|A(z',z_n;h)|={\mathcal O}((Nk)^{-1/6}).
\end{equation}

Now, $I_1$ is written as,
$$
I_1(x,\xi;h)= h^{-n/4}\int_{\R^n}e^{i\psi(x,z,\xi)/h}d(x,z;h)dz+\ord (h^{\delta_LN}+h^L),
$$
with
$$\begin{array}{rl}
\psi(x,z,\xi)=&(x'-z')\cdot\xi'+(x_n-z_n+b(z'))\xi_n+i\tilde\phi(z',z_n)\\[8pt]
&+i(x'-z')^2/2+i\mu_n(x_n-z_n+b(z'))^2/2;
\end{array}
$$
$$
d(x,z;h)=\chi^1_N(z',z_n-b(z'))A(z',z_n;h).
$$
By the change of scale,
\begin{eqnarray}
&& x'=(Nk)^{1/2}\tilde x'\, ; \quad x_n=(Nk)^{2/3}\tilde x_n;\nonumber\\ 
&& z'=(Nk)^{1/2}\tilde z'\, ; \quad z_n=(Nk)^{2/3}\tilde z_n;\\ 
&&\xi'=(Nk)^{1/2}\tilde \xi'\, ;\quad \xi_n=(Nk)^{1/3}\tilde \xi_n,\nonumber
\end{eqnarray}
and setting
\be
\tilde h:=(Nk)^{-1}h \quad (<<1),
\ee
$I_1$ becomes,
\begin{eqnarray*}
I_1(x,\xi;h)= (Nk)^{\frac{n+1}3 + (n-1)\varepsilon} h^{-n/4}\int e^{i\tilde\psi(\tilde x, \tilde z, \tilde \xi)/\tilde h}
\tilde d(\tilde x,\tilde z;\tilde h)d\tilde z\\
+\ord (h^{\delta_LN}+h^L),
\end{eqnarray*}
where
$$
\tilde\psi=
(\tilde x-\tilde z)\cdot\tilde\xi+\frac 23\sqrt{C_0}\tilde z_n^{3/2}+i(\tilde x-\tilde z)^2/2+\tilde a(\tilde z')+\ord ((Nk)^{1/3}),
$$
with $\tilde a(\tilde z')={\mathcal O}(|\tilde z'|^2)$ real-valued,
and $\tilde d(\tilde x,\tilde z;\tilde h)$ is a smooth function in $\tilde z$ supported in
\begin{equation}
\label{support}
\{|\tilde z_n-C_0\delta^{-2}+\ord ((Nk)^{1/3})|<3\delta\}\cap \{ |\tilde z'+{\mathcal O}((Nk)^{5/6}|\leq 3\delta\},
\end{equation}
(recall from \eq{asymp0} and  \eq{odex'2} that
$x(-t_N)=(\ord ((Nk)^{4/3}),C_0\delta^{-2}(Nk)^{2/3}+\ord (Nk))$ and $a(z'), b(z'), \xi_n^c(z')$ are real-valued functions that are $\ord (|z'|^2)$). Moreover,  $\tilde d$ satisfies the same estimate as \eq{checka}, and it is holomorphic 
with respect to  $\tilde z_n$ in a ($h$-independent) small neighborhood of $\tilde z_n=\tilde x_n$.

\medskip
Then it suffices to show, 
\begin{equation}
\label{below}
\re\frac{\partial\tilde\psi}{\partial\tilde z_n}\ge \tilde\delta,
\end{equation}
for some positive constant $\tilde\delta$ independent of $h$ and $N$. Indeed, in that case, a standard modification of the integration path with respect to $\tilde z_n$ around $\tilde z_n=\tilde x_n$ to the upper complex plane with small enough radius, shows that  $I_1=\ord (e^{-\delta/\tilde h})$ as $\tilde h\to 0$, with another constant $\delta >0$, and this means that $I_1=\ord (h^{\delta N})$ as $h\to 0$.

\medskip
The fact that \eq{below} holds for $\tilde z$ in the support of $\tilde d$ (where \eq{support} holds) and $(x,\xi)\in{\mathcal W}(t_N,h)$ follows from,
$$\re\frac{\partial\tilde\psi}{\partial\tilde z_n}=-\tilde\xi_n+\sqrt{C_0}\tilde z_n^{1/2}+\ord ((Nk)^{1/3}),
$$
and the estimates \eq{support} and
$
-\tilde\xi_n\ge C_0\delta^{-1}-\delta
$
implies 
$$-\tilde\xi_n+\sqrt{C_0}\tilde z_n^{1/2}\ge 2C_0\delta^{-1}-5\delta.
$$
This completes the proof.
\end{pre}

\subsection{Propagation up to the Outgoing Region}
Now, for $N\geq 1$ large enough, we set,
$$
v_N:=\chi_Nv= \chi_Ne^{S/h}(u-w_{CN}),
$$
where $\chi_N$ is as in \eq{defchiN}, and $C>0$ must be fixed large enough as in Proposition \ref{comparison2}. Lemmas \ref{propagation1} and \ref{wsort} imply that for any $L$ large enough, there exists $\delta_L>0$ such that for any $N\geq L/\delta_L$, 
\begin{equation}
\label{vNsort}
{\bf T}_N (v_N)={\mathcal O}(h^{\delta N}+h^L) \mbox{ uniformly in } {\mathcal W}(t_N,h),
\end{equation}
where $\delta>0$ is a fixed small enough constant independent of $N,L$, and $t_N=\delta^{-1}(Nk)^{1/3}$. Moreover, since $\re\tilde\phi \geq -C_1Nk$ on $\supp\chi_N$ (for some $C_1>0$ constant), we deduce from Proposition \ref{summary} that, if $C>0$ has been chosen sufficiently large, then,
$$
(P-\rho (h))v_N = [P,\chi_N]e^{S/h}(u-w_{CN}) + {\mathcal O}(h^{\delta N}).
$$
Now, we introduce the  $(N,h)$-dependent distance $\tilde d_N$, associated with the metric,
$$
\frac{|dx'|^2}{Nk} + \frac{dx_n^2}{(Nk)^{4/3}}.
$$
Then, 
using Proposition \ref{comparison2}, 
we see that if 
\be
\label{dsupp}
\tilde d_N(x,\supp\nabla\chi_N)\geq \varepsilon
\ee
 for some fixed $\varepsilon >0$,  thanks to the Gaussian factor in the definition of ${\bf T}_N$ (see \eq{TN} for the definition), we have,
$$
{\bf T}_N (P-\rho (h))v_N (x,\xi)={\mathcal O}(h^{-N_5+\varepsilon^2N/2}),
$$
for some $N_5>0$ independent of $N$, and thus,
\begin{equation}
\label{eqvN}
{\bf T}_N(P-\rho (h))v_N (x,\xi)={\mathcal O}(h^{\varepsilon^2N/4}),
\end{equation}
for all $N$ large enough, and uniformly  with respect to $h>0$ small enough and $(x,\xi)\in\R^{2n}$ verifying \eq{dsupp}.

\medskip
Still working in the same coordinates (for which $\hat x=0$),
we consider the $(h,N)$-dependent change of variables,
$$
x=(x',x_n) \mapsto \tilde x =(\tilde x', \tilde x_n):= ((Nk)^{-1/2}x', (Nk)^{-2/3}x_n),
$$
and the corresponding unitary operator $U_N$ on $L^2(\R^n)$.
Under this change, the function $v_N$ is transformed into,
$$
\tilde v_N (\tilde x) := U_Nv_N (\tilde x)=(Nk)^{n/4 +1/12} v_N((Nk)^{1/2}\tilde x', (Nk)^{2/3}\tilde x_n),
$$
and one can check,
\begin{equation}
\label{newT}
T\tilde v_N (\tilde x,\tilde\xi ;\tilde h) = c_1(Nk)^{-n/4 -1/12} {\bf T}_N v_N ( A_N (\tilde x, \tilde \xi ); h),
\end{equation}
where $T=T_1$ is the standard FBI transform defined in (\ref{defT}) with $c_1=2^{-n/2}(\pi h)^{-3n/4}$, and,
\begin{eqnarray*}
&& A_N(\tilde x,\tilde\xi):= ((Nk)^{1/2}\tilde x', (Nk)^{2/3}\tilde x_n ; (Nk)^{1/2}\tilde \xi',(Nk)^{1/3}\tilde\xi_n);\\
&& \tilde h := h/(Nk)= (N\ln\frac1h )^{-1}.
\end{eqnarray*}
Then, setting,
$$
(Nk)^{-2/3}(P-\rho)=:\tilde P = -(Nk)^{1/3}\tilde h^2\Delta_{\tilde x'}-\tilde h^2\partial_{\tilde x_n}^2 + \tilde V(\tilde x),
$$
with,
\begin{eqnarray*}
\tilde V(\tilde x,\tilde h):&=&  (Nk)^{-2/3}V((Nk)^{1/2}\tilde x', (Nk)^{2/3}\tilde x_n)-(Nk)^{-2/3}\rho \\
&=& -C_0\tilde x_n+(Nk)^{-2/3}\left[E_0-\rho(h)+W ((Nk)^{1/2}\tilde x', (Nk)^{2/3}\tilde x_n)\right] ,
\end{eqnarray*}
we deduce from (\ref{newT}) that
(\ref{eqvN}) becomes,
\begin{equation}
\label{eqvNtilde}
T\tilde P\tilde v_N (\tilde x,\tilde \xi ;\tilde h)={\mathcal O}(c_1(Nk)^{-\frac{n+3}4}e^{-\varepsilon^2/4\tilde h}) ={\mathcal O}( e^{-\varepsilon^2/6\tilde h}),
\end{equation}
for any $N\geq 1$ large enough, and 
uniformly with respect to $h>0$ small enough and  $(\tilde x, \tilde\xi)\in\R^{2n}$ verifying \eq{dsupp}. (Here, we have used the fact that $Nk=h/\tilde h = \tilde h^{-1}e^{-1/(N\tilde h)}$.)
\vskip 0.2cm

Moreover, setting,
$$
\tilde p(\tilde x,\tilde\xi ) :=(Nk)^{1/3}|\tilde\xi'|^2 +\tilde \xi_n^2+ \tilde V(\tilde x, \tilde h)=(Nk)^{-2/3}p\circ A_N(\tilde x,\tilde \xi),
$$
a direct computation shows that, for all $\tilde t\in\R$, one has,
$$
\exp \tilde tH_{\tilde p} = A_N^{-1}\circ (\exp (Nk)^{1/3}\tilde t H_p)\circ A_N.
$$

As a consequence, still using (\ref{newT}), we see that (\ref{vNsort}) can be rewritten as,
\begin{equation}
\label{estT1v}
\begin{array}{rl}
T\tilde v_N (\tilde x,\tilde\xi ;\tilde h) &= c_1(Nk)^{-n/4 -1/12} {\mathcal O}(e^{-\delta /\tilde h}+e^{-\delta_L/\tilde h}) \\
&= {\mathcal O}(e^{-\delta'_L/2\tilde h}),\quad \delta_L'=\min(\delta,\delta_L)
\end{array}
\end{equation}
uniformly in the tubular domain
\be
\begin{array}{rll}
\tilde {\mathcal W}(\tilde h):= \{  &|\tilde x_n- \tilde x_n(-\delta^{-1})|\leq \delta \, ,\, &|\tilde \xi_n -\tilde \xi_n (-\delta^{-1})| \leq \delta,\\
&|\tilde x'- \tilde x'(-\delta^{-1})|\leq \delta (Nk)^{-\frac 16}\, ,\, &|\tilde \xi' -\tilde \xi' (-\delta^{-1})| \leq \delta (Nk)^{-\frac 16}\},
\end{array}
\ee
where $(\tilde x(\tilde t), \tilde\xi(\tilde t))=\exp\tilde t H_{\tilde p}(0,0)$.

Moreover,  using (\ref{compdiric3}), Proposition \ref{comparison2}, and the properties of $w_{CN}$ in the sea, we see that there exists $N_1\geq 0$, such that,
\begin{equation}
\label{vn}
\Vert \tilde v_N\Vert_{H^1}={\mathcal O}(h^{-N_1}) ={\mathcal O}(e^{N_1/(N\tilde h)}).
\end{equation}
In particular, for any $\varepsilon >0$, one has $\Vert \tilde v_N\Vert_{H^1}={\mathcal O}(e^{\varepsilon /\tilde h})$ when $N$ is large enough.

Now, we are in a situation very similar to that of the propagation of analytic singularities, except for the fact that the symbol of $\tilde P$ is not analytic. However, denoting by $W_N$ a holomorphic $C(Nk)^{2/3}$-approximation of $W$ near $0$ (in the sense of Lemma \ref{almost}, and with $C>0$ sufficiently large), and setting,
\begin{eqnarray*}
&& \tilde W_N(\tilde x):= W_N((Nk)^{1/2}\tilde x', (Nk)^{2/3}\tilde x_n);\\
&& \tilde V_N(\tilde x) := -C_0\tilde x_n +(Nk)^{-2/3}\left[ E_0-\rho(h) +\tilde W_N(\tilde x)\right];\\
&& \tilde P_N:= -(Nk)^{1/3}\tilde h^2\Delta_{\tilde x'}-\tilde h^2\partial_{\tilde x_n}^2 +\tilde V_N(\tilde x),
\end{eqnarray*}
we deduce from  (\ref{eqvNtilde}), \eq{vn},  (and, e.g.,  the fact that $(Nk)^N =\tilde h^{-N}e^{-1/\tilde h}={\mathcal O}_N(e^{-1/2\tilde h})$), that, for any $\varepsilon >0$ fixed small enough and for any $N\geq 1$ large enough,  we have,
\begin{equation}
\label{equanalyt}
T\tilde P_N\tilde v_N(\tilde x,\tilde\xi ;\tilde h) ={\mathcal O}_N( e^{-\varepsilon^2/6\tilde h}),
\end{equation}
uniformly with respect to $\tilde h>0$ small enough and  $(\tilde x, \tilde\xi)\in\R^{2n}$ verifying \eq{dsupp}, which can be expresses as
\be
\label{deltasupp}
|\tilde x'|<|\tilde x'(-\delta^{-1})|+2\delta-\varepsilon,\quad
|\tilde x_n|<|\tilde x_n(-\delta^{-1})|+2\delta-\varepsilon.
\ee

\vskip 0.2cm
Now, by construction, the symbol of $\tilde P_N$ is holomorphic in a (arbitrarily large) complex neighborhood of $(0,0)$, and since $E_0-\rho (h)={\mathcal O}(h)$ and $\partial^\alpha W(x) ={\mathcal O}(|x|^{(2-|\alpha|)_+})$, we see that the total $\tilde h$-semiclassical symbol  $\tilde p_N$ of $\tilde P_N$ verifies,
$$
\tilde p_N (\tilde x,\tilde\xi) = \tilde \xi_n ^2 -C_0\tilde x_n+(Nk)^{1/3}(\tilde \xi')^2+(Nk)^{1/3}{\mathcal O}(|\tilde x|^2+(N\ln 1/h)^{-1}).
$$
that tends to,
$$
\tilde p_0(\tilde x,\tilde \xi) := \tilde \xi_n ^2 -C_0\tilde x_n,
$$
as $\tilde h$ tends to 0. In particular, $\tilde p_0$ does not depend on $N$, and the point $\exp(- \delta^{-1}H_{\tilde p}(0,0))$ tends to $\exp (-\delta^{-1}H_{\tilde p_0}(0,0))$ as $\tilde h\rightarrow 0_+$.

\medskip
From now on, we fix $\varepsilon >0$ small enough and the cut-off function $\chi_0$ in such a way that ${\rm dist}(\pi_x(\exp \tilde tH_{\tilde p_0}(0,0)), \supp\nabla\chi_0)\geq 2\varepsilon$ for all $\tilde t\in [-\delta^{-1}, \delta^{-1}]$, where $\delta$ is the same as in (\ref{estT1v}).

Then, modifying the proof of the theorem of the propagation of analytic singularities (see, e.g., \cite{sj1} Theorem 9.1, or \cite{ma1} Theorem 4.3.7), we can show that, in our case, the estimates \eq{vn}, (\ref{estT1v}) and (\ref{equanalyt}) imply 

\begin{prop}
\label{propaganalyt}
There exists a constant $\delta_1>0$ independent of $L$, such that, for all $L$ large enough (and $N=L/\delta_L$), one has, for $\tilde h>0$ small enough,
\begin{equation}
\label{0notinMS}
T\tilde v_N (\tilde x,\tilde\xi ;\tilde h) ={\mathcal O}(e^{-\delta_1\delta_L /\tilde h})
\end{equation}
uniformly in $V(\delta_1)=\left\{\tilde x;|\tilde x|\leq \delta_1\right\}\times \left\{\tilde \xi;\,\,(Nk)^{\frac 16}|\tilde \xi'|+|\tilde \xi_n|\leq \delta_1\right\}$.
\end{prop}

\begin{pre}
As in \eq{changesymp}, we make a symplectic change of coordinates  
\begin{equation}
\left\{
\begin{array}{ll}
\tilde y'=\tilde x',\quad &\tilde y_n=\tilde x_n-\frac 1{C_0}\tilde \xi_n^2, \\
\tilde \eta'=\tilde \xi',\quad &\tilde \eta_n=\tilde \xi_n,
\end{array}
\right.
\ee
which leads to
\be
\tilde p_0=-C_0\tilde y_n,
\quad H_{p_0}=C_0\frac{\partial}{\partial \tilde \eta_n}.
\ee

For positive constants $a,b, c,d$ with $b<a$ and $\alpha,\beta$ with $\alpha<2\beta d$, we take $f\in C_0^\infty(]-a,d[;[0,\alpha])$ and $\chi\in C_0^\infty(]-c,c[;[0,1])$ such that
$$
f'\leq -\beta\,\,{\rm on}\,\, [-b,\frac d2],\,\, f(0)=\frac \alpha 2,
$$
$$
\,\chi=1\,\,{\rm on}\,\, [-\frac c4,\frac c4],\,\, \chi\geq\frac 14\,\,{\rm on}\,\, [-\frac c2,\frac c2],\,\, 
\chi\leq \frac 14\,\,{\rm outside}\,\, [-\frac c2,\frac c2].
$$
Then the weight function
$$
\psi(\tilde y,\tilde \eta)=f(\tilde\eta_n)\chi(|\tilde y_n|)\chi((Nk)^{\frac 16}|\tilde y'|)\chi ((Nk)^{\frac 16}|\tilde \eta'|)
$$
satisfies
\be
\label{hppb}
|H_{\tilde p_0}\psi|\geq \frac{C_0\beta}{64}\quad{\rm in}\,\, \tilde \Omega_1,
\ee
\be
\label{psal}
|\psi|\leq \max (\frac{\alpha}4,\frac{\alpha-\beta d}2)=\frac{\alpha}4\quad{\rm in}\,\, \tilde \Omega_2,
\ee
where $\supp \psi\subset \tilde\Omega_1\cup  \tilde\Omega_2 \cup\tilde\Omega_3$, with
$$
\begin{array}{l}
\tilde \Omega_1={\mathcal V}_{c/2}\times [-b,\frac d2], \\  
\tilde \Omega_2=({\mathcal V}_{c}\times [-b,d])\backslash\tilde \Omega_1, \\
\tilde\Omega_3={\mathcal V}_{c}\times [-a,-b]
\end{array}
$$
and
$$
{\mathcal V_{c}}:=\left\{(\tilde y,\tilde \eta');\,\,|(\tilde y',\tilde\eta')|\leq c(Nk)^{\frac 16},\,\,\,|\tilde y_n|\leq c\right\}.
$$
Remark that $\tilde\Omega_3\subset\tilde{\mathcal W}(\tilde h)$ if $a,b$ are suitably chosen. 

\medskip
A microlocal exponential estimate leads us, in our case, to
$$
\theta^2||(H_{\tilde p_0}\psi)e^{\theta\psi/\tilde h}T\tilde v_N||^2\leq C(k^{1/3}+\theta^3)||e^{\theta\psi/\tilde h}T\tilde v_N||^2+||e^{\theta\psi/\tilde h}T\tilde P_N\tilde v_N||^2,
$$
for a small parameter $\theta$ and for each fixed $L$, $N=L/\delta_L$.
Let $\theta_L$ be such small number that $C(\frac{k^{1/3}}{\theta^2}+\theta)\times \frac{C_0\beta}{64}<\frac 12$ holds for sufficiently small $h$, and let denote again by $\delta_L$ the minimum of $\delta'_L$ and $\theta_L$. Then we have, by \eq{hppb},
\be
\label{edlp}
||e^{\delta_L\psi/\tilde h}T\tilde v_N||^2_{L^2(\tilde\Omega_1)} 
\leq ||e^{\delta_L\psi/\tilde h}T\tilde v_N||^2_{L^2(\tilde\Omega_1\cup\tilde\Omega_3)}+C_L||e^{\delta_L\psi/\tilde h}T
\tilde P_N\tilde v_N||^2.
\ee
for some constant $C_L>0$. First using \eq{psal} and $|f|\leq \alpha$, one can estimate the RHS of \eq{edlp}  by.
$$
2e^{\alpha\delta_L\psi/4\tilde h}||T\tilde v_N||^2_{L^2(\tilde \Omega_2)}
+2e^{\alpha\delta_L\psi/\tilde h}||T\tilde v_N||^2_{L^2(\tilde \Omega_3)}
+C_Le^{\alpha\delta_L\psi/\tilde h}||T\tilde v_N||^2,
$$
and next, by using  \eq{vn}, (\ref{estT1v}) and (\ref{equanalyt}) (observing that \eq{deltasupp} is satisfied with $\tilde x=\pi_x\exp \tilde t H_{\tilde p_0}$ for all $-\delta^{-1}\leq \tilde t\leq 0$), by,
$$
C_L'\left (
e^{(\frac\alpha 4+\frac{N_1}L)\delta_L/\tilde h}+e^{(\alpha-\frac 12)\delta_L/\tilde h}+e^{(\alpha\delta_L-\frac {\varepsilon^2}6)/\tilde h}
\right ).
$$
The LHS of \eq{edlp} is estimated from below by
$$
||e^{\delta_L\psi/\tilde h}T\tilde v_N||^2_{L^2(\tilde\Omega_1)} \geq
||e^{\delta_L\psi/\tilde h}T\tilde v_N||^2_{L^2(V(\delta_1))} \geq
e^{\frac 38\alpha\delta_L/\tilde h}||T\tilde v_N||^2_{L^2(V(\delta_1))} ,
$$
if $\delta_1$ is so small that $\psi\geq\frac 38$ on $V(\delta_1)$. Thus we obtain
$$
||T\tilde v_N||^2_{L^2(V(\delta_1))} \leq
C_L'\left (
e^{(-\frac\alpha 8+\frac{N_1}L)\delta_L/\tilde h}+e^{(\frac 58\alpha-\frac 12)\delta_L/\tilde h}+e^{(\frac 58\alpha\delta_L-\frac {\varepsilon^2}6)/\tilde h}
\right ).
$$
This implies \eq{0notinMS}, if one takes $\alpha$ and $\delta_1$ sufficiently small.
\end{pre}

\medskip
On the other hand, if $|\tilde x|\leq \delta'_1$ for small enough $\delta_1'$ and $\tilde\xi\in \R^n\backslash V_\xi(\delta_1)$ (where we set $V_\xi(\delta_1)= \{\tilde \xi;\,\,(Nk)^{\frac 16}|\tilde \xi'|+|\tilde \xi_n|\leq \delta_1\}$), then $\tilde p_N(\tilde x,\tilde \xi )\geq c|\tilde\xi|^2$ for a positive constant $c$, and again, standard techniques of microlocal analytic singularities (see, e.g., \cite{ma1} Theorem 4.2.2) show, for any $m\geq 0$,  the existence of some $\varepsilon_m>1$ (still independent of $N$),
such that,
\begin{equation}
\label{estellip}
\Vert \langle\xi\rangle^mT_1\tilde v_N\Vert_{L^2(\{|\tilde x|\leq \delta_1\}\times V_\xi(\delta_1)^C)}={\mathcal O}(e^{-\varepsilon_m/\tilde h}).
\end{equation}
Gathering (\ref{0notinMS}) and (\ref{estellip}), we obtain,
$$
\Vert \langle\xi\rangle^mT_1\tilde v_N\Vert_{L^2(\{|\tilde x|\leq \delta_1\}\times \R^n)} ={\mathcal O}(e^{-\delta_1'\delta_L/\tilde h}).
$$
 In particular, using the fact that,
\begin{eqnarray*}
&&\Vert T_1\tilde v_N\Vert^2_{L^2(\{|\tilde x|\leq \delta_1\}\times \R^n)}\\
&&\hskip 2cm = (2\pi \tilde h)^nc_1^2\int_{\{|\tilde x|\leq \delta_1\}\times\R^n}e^{-(\tilde x-\tilde y)^2/\tilde h}|\tilde v_N (\tilde y)|^2d\tilde yd\tilde x\\
&&\hskip 2cm \geq (\pi \tilde h)^{-n/2}\int_{|\tilde x -\tilde y|\leq \sqrt{\tilde h}\,,\, |\tilde y|\leq \frac12\delta'_1}e^{-(\tilde x-\tilde y)^2/\tilde h}|\tilde v_N (\tilde y)|^2d\tilde yd\tilde x\\
&&\hskip 2cm \geq \frac{b_n}e\pi ^{-n/2}\Vert\tilde v_N\Vert^2_{L^2(\{|\tilde x|\leq \frac12\delta'_1\})},
\end{eqnarray*}
(where $b_n$ stands for the volume of the unit ball of $\R^n$), and turning back to the previous coordinates, we obtain,
$$
\Vert v_N\Vert_{L^2(\{| x'|\leq \frac12\delta'_1(Nk)^{1/2}, | x_n|\leq \frac12\delta'_1(Nk)^{2/3}\})}={\mathcal O}(h^{\delta'_1L}).
$$
In the same way, working with $\langle\xi\rangle^mT_1\tilde v_N$ instead of $T_1\tilde v_N$, we also find,
$$
\Vert v_N\Vert_{H^m(\{| x'|\leq \frac12\delta'_1(Nk)^{1/2}, | x_n|\leq \frac12\delta'_1(Nk)^{2/3}\})}={\mathcal O}(h^{\delta'_1L}),
$$
for large enough $L$. Since $L$ is arbitrarily large, \eq{vhl} holds uniformly in  ${\mathcal W}(t_N,h)$ by standard Sobolev estimates, and since $\hat x\in\p\open$ was taken arbitrarily,  Proposition \ref{comparison} follows.
\vskip 0.2cm

\section{Asymptotics of the Width}

We calculate the asymptotic formula of $\im \rho(h)$ using the formula \eq{Green} and the results of the preceding sections.

Let $W_\sigma\subset\R^n$ be an $N,h$-dependent open domain containing $\open$ defined by
$$
W_\sigma=\{x;{\rm dist}(x,\open)<\sigma (Nk)^{2/3}\}
$$
for  $1<\sigma<2$.

The boundary $\partial W_\sigma$ is in $U_N$ for $1<\sigma<2$. Hence, replacing $u$ by $w_{CN}$ in the formula \eq{Green} by using Proposition \ref{prop7}, and noticing that $||u||_{L^2(W_N(h))}-1$ is exponentially small, we have
\begin{equation}
\label{width}
\im \rho(h)=-h^2\im\int_{W_\sigma}\frac{\partial w_{CN}}{\partial n}
\overline {w_{CN}}dS+{\mathcal O}(h^{2L} e^{-2S/h}).
\end{equation}

Moreover, the domain of integration $\partial W_\sigma$ can be replaced by $\partial W_\sigma\cap U_{N,1}$ using the facts \eq{2Nk} and Proposition \ref{summary} (i).

Then we can substitute the asymptotic formula \eq{summarywkb}
into the integrand of \eq{width}; for any $L$ large enough, there exist
$\delta_L,c_L(=1/C'_L)>0$ such that for all $N>L/\delta_L$, one has
$$
\begin{array}{l}
h^2\displaystyle{\frac{\p w_{CN}}{\p n}}\overline{ w_{CN}}=h^{1-n/2}
e^{-2(S+\re\tilde\phi)/h}\times 
\\[8pt]
\hspace{2.5cm}\left\{\displaystyle{\sum_{j,k=0}^{L+c_LN|\ln h|}}\left (\frac{\p\tilde\phi}{\p n}\tilde a_j\overline{\tilde a_k}+h\frac{\p \tilde a_j}{\p n}\overline{\tilde a_k}\right )h^{j+k}+\ord(h^L)\right \}.
\end{array}
$$

The vector field $\frac{\p\im\tilde\phi}{\p x}\cdot\frac\p{\p x}$ is transversal to the
caustics $\mathcal C$ where $\im\tilde\phi=\im\phi=0$. Let $\iota$ be the one-to-one map which
associates to a point $x$ in $\partial W_\sigma\cap U_{N,1}$ the point $y=\iota(x)$ on
$\mathcal C$ such that the integral curve of $\frac{\p\im\tilde\phi}{\p
x}\cdot\frac\p{\p x}$ emanating from $y$ passes by $x$.

\begin{lem}
On $\partial W_\sigma\cap U_{N,1}$, the function $\re \tilde\phi(x)$ reaches its (transversally non-degenerate) minimum
$S$  at $\iota^{-1}(\Gamma)$ modulo ${\mathcal O}(h^\infty)$. More precisely, one has,
$$
\re\tilde\phi(x)|_{\partial W_\sigma\cap U_{N,1}}=\phi(\iota(x))+{\mathcal O}(h^\infty),
$$
\end{lem}

\begin{pre}
This is a direct consequence of \eq{lemma} and \eq{imeiko}.
\end{pre}

\begin{lem}
Let $x\in\partial W_\sigma\cap U_{N,1}$, and $y=\iota (x)\in{\mathcal C}$.
There exists a family of smooth functions $\{\beta'_m(y, {\rm dist} (x,{\mathcal C}))\}_{m=0}^\infty$ defined in ${\mathcal C}\times [0,2(Nk)^{2/3})$, with $\beta'_0(y,0)>0$, such that, for any large $L$, there exist $\delta_L,c_L>0$ such that for all $N>L/\delta_L$, one has,
$$
\begin{array}{l}
-h^2\im\displaystyle{\frac{\p w_{CN}}{\p n}}\overline{w_{CN}}=h^{1-n/2}
e^{-2(S+\phi(y))/h}\times 
\\[8pt]
\hspace{2.5cm}\left\{\displaystyle{\sum_{m=0}^{2L+2c_LN|\ln h|+1}}
\beta'_m(y,\delta(Nk)^{2/3})\left(N\ln \frac 1h\right )^{-m}+\ord(h^L)\right \}.
\end{array}
$$
\end{lem}

\begin{pre}
We know by \eq{imphi} and \eq{taj} that
$$
\frac{\p\im\tilde\phi}{\p n}=\ord\left({\rm dist}(x,{\mathcal C})^{1/2}\right ),
\quad
\tilde a_j=\ord\left({\rm dist}(x,{\mathcal C})^{-3j/2-1/4}\right ).
$$
It follows that, for $j+k=m$,
$$
\frac{\p\im\tilde\phi}{\p n}\tilde a_j\overline{\tilde a_k}h^m\sim
{\rm dist}(x,{\mathcal C})^{-3m/2}h^m=
\delta^{-3m/2}\left (\frac h{Nk}\right )^m,
$$
$$
\frac{\p\tilde a_j}{\p n}\overline{\tilde a_k}h^{m+1}\sim
{\rm dist}(x,{\mathcal C})^{-3(m+1)/2}h^{m+1}=
\delta^{-3(m+1)/2}\left (\frac h{Nk}\right )^{m+1}.
$$

In particular, the principal term $\beta'_0(y,0)$ is positive. In fact,
$$
\beta'_0(y,\delta(Nk)^{2/3})=\frac{\p\tilde\phi}{\p n}|\tilde a_0|^2.
$$
In local coordinates as in \S 4, $\partial/\partial n=(\ord (|x|)\p/\p x', (1+\ord (|x|))\p/\p x_n)$ by \eq{odex'2}, and 
$$
-\frac{\p\im\tilde\phi}{\p
x'}|\tilde a_0|^2=\ord (|x'|)+\ord\left(x_n+b(x')\right),
$$
$$
-\frac{\p\im\tilde\phi}{\p
x_n}|\tilde a_0|^2=\frac{\pi
c_0(x',\xi_n^c(x'))^2}{\nu_1(x',\xi_n^c(x'))}+\ord\left(\sqrt{x_n+b(x')}\right),
$$
as $x_n+b(x')$ tends to 0, by \eq{a0} and \eq{imphi}.
\end{pre}

Now expanding $\beta'_m$ in Taylor series with respect to ${\rm dist} (x,{\mathcal C})$, we obtain
$$
-\im\rho=h^{1-n/2}e^{-2S/h}
\displaystyle{\sum_{m=0}^{2L+2c_LN|\ln h|+1}}\left(N\ln \frac 1h\right )^{-m}\sum_{j=0}^{[3L/2]+1}
\delta^{j-3m/2}(Nk)^{2j/3} 
$$
$$
\times\int_{{\mathcal C}\cap U_{N,1}}e^{-2\phi(y)/h}\beta'_{m,j}(y)dy
\,\,+\ord (h^{L+1-n/2}e^{-2S/h})\int_{{\mathcal C}\cap U_{N,1}}e^{-2\phi(y)/h}dy.
$$
To the integrals of the RHS, we apply the stationary phase method using the assumption (A4), which means that the phase $\phi(y)$ attains its transversally non-degenerate minimum on the whole submanifold $\Gamma$. For any large $L$, we obtain,
$$
\int_{{\mathcal C}\cap U_{N,1}}e^{-2\phi(y)/h}\beta'_{m,j}(y)dy
=h^{(n-1-n_\Gamma)/2}\left\{\sum_{l=0}^{L-1}d_{l,j,m}h^l+\ord(h^L)\right\},
$$
where $\{d_{l,j,m}\}$ is a family of real numbers with $d_{0,0,0}>0$.
Here $n_\Gamma={\rm dim}\Gamma$.
Hence we have
$$
-h^{-(1-n_\Gamma)/2}e^{2S/h}\im\rho(h)=\hspace{5cm}
$$
\be
\label{last}
\sum_{(l,j,m)\in{\mathcal N}}
d_{l,j,m}h^lk^{2j/3} 
\left(\ln \frac 1h\right )^{-m}
(N^{2/3}\delta)^{j-3m/2}+\ord(h^L),
\ee
where 
$$
{\mathcal N}=\left\{(l,j,m)\in \N^3;l\leq L-1,\,j\leq \left [\frac{3L}2\right ]+1,\,
m\leq 2(L+c_LN|\ln h|)+1\right\}.
$$
\begin{lem}
\label{lastlemma}
Let $(l,j,m)\in{\mathcal N}$.
The real number $d_{l,j,m}$ vanishes if $j-3m/2\neq 0$.
\end{lem}
\begin{pre}
Observing that $h<<k^{2/3}<<|\ln h|^{-1}$, we introduce an order relation in the set ${\mathcal N}$.
We  write $(l,j,m)<(l',j',m')$ if one of the following three conditions holds:
$$
({\rm i}) \,\,l<l', \,\,({\rm ii})\,\, l=l' \,\,{\rm and}\,\, j<j', \,\,({\rm iii}) \,\,l=l',\,\, j=j', \,\,{\rm and}\,\, m<m'.
$$
Suppose there exists $d_{l,j,m}\neq 0$ with $j-3m/2\neq 0$ and
let $(l_0,j_0,m_0)$ be the smallest among such $(l,m,j)$'s.
Then the RHS of \eq{last} becomes
$$
\sum_{(l,j,m)<(l_0,j_0,m_0)}
d_{l,j,m}h^{l+m}
+d_{l_0,j_0,m_0}
h^{l_0}k^{\frac{2j_0}3} 
\left(\ln \frac 1h\right )^{-m_0}
(N^{2/3}\delta)^{j_0-\frac{3m_0}2}
$$
$$
+\ord \left (
h^{l_0}k^{\frac{2j_0}3} 
\left(\ln \frac 1h\right )^{-m_0}
\right ).
$$
Here, $\delta$ is an arbitrary number varying between 1 and 2 and $d_{l,j,m}$ are independent of $\delta$.
On the other hand, the LHS of \eq{last}  is independent of $\delta$.
This is a contradiction and we have proved Lemma \ref{lastlemma}.
\end{pre}

Then the  proof of Theorem \ref{main} follows from \eq{last} and this Lemma \ref{lastlemma}, since for $j-3m/2=0$, one has
$$
\left(\ln \frac 1h\right )^{-m}k^{2j/3}=h^m .
$$

\section{Appendix}
\subsection{Holomorphic $\delta$-Approximation}
Let $f=f(x)$ be a smooth function on $\R^n$ uniformly bounded together with
all its derivatives. A function 
$\tilde f(x,y)$ on $\R^{2n}$ is said to be an {\it almost-analytic
extension} of $f$ if,
$$
\tilde f(x,0)=f(x),
$$
and, for all $\alpha\in\Z_+^{2n}$,
\begin{equation}
\label{dbar}
\partial^\alpha\left (\frac{\partial}{\partial x}+i
\frac{\partial}{\partial y}\right )\tilde
f(x,y)={\mathcal O}(|y|^\infty),
\end{equation}
as $|y|\to 0_+$, uniformly with respect to $x$.
We can construct an almost-analytic extension (see, e.g., \cite{ms}) by setting,
\begin{equation}
\label{expansion}
\tilde f(x,y)=\sum_{\alpha\in\N^n}
\frac{(iy)^\alpha}{\alpha !}\partial^\alpha f(x)\left
(1-\chi\left(\frac{\epsilon_\alpha}{|y|}\right)\right),
\end{equation}
where $\chi\in C_0^\infty(\R)$ is a fixed cutoff
function that is equal to 1 near 0, and
$(\epsilon_\alpha)_{\alpha\in\N^n}$ is a decreasing sequence of
positive numbers converging to 0 sufficiently rapidly. More precisely, we choose $\varepsilon_\alpha$ such that, for any $\beta\leq\alpha$, one has,
$$
|y|\sup |(1-\chi(\varepsilon_\alpha /|y|))\partial^{\alpha +\beta}f|\leq \alpha!.
$$
Then, the corresponding almost analytic extension have the following elementary properties:

\begin{lem}
\label{almost}
Let $f$ be as above.
\begin{description}
\item[(i)] If $\tilde f(x,y)$ and $\hat f(x,y)$ are both almost-analytic extensions of $f(x)$, then, for any $\delta >0$, one has
\begin{eqnarray*}
&& \tilde f(x,y)-\hat f(x,y)={\mathcal O}(|y|^\infty);\\
&& \sup_{|y|\leq\delta}|\tilde f(x,y)|\leq \min\left\{ \sup|f|+2, \sup|f|+\delta(1+\sup|\nabla f|)\right\}.
\end{eqnarray*}
\item[(ii)] Let $\tilde f$ be an almost-analytic extension of $f$ and let $I_1,\dots,I_n\subset \R$ be bounded open intervals. Then, for any $\delta >0$ there exists a function $f_\delta$, holomorphic in $\Gamma_\delta:=\{z\in\C^n\, ;\, {\rm dist}(z_j,I_j )<\delta,\, j=1,\dots,n\}$, such that, for all $\alpha\in\Z_+^{2n}$, $\beta\in\Z_+^n$, and $N\geq 1$, there exists $C(\alpha, N)>0$, such that,
\begin{eqnarray}
\label{estpresqan1}
&&  \sup_{x+iy\in\Gamma_\delta}\left| \partial^\alpha\left( f_\delta (x+iy) - \tilde f(x,y)\right) \right|\leq C(\alpha, N)\delta^N;\\
\label{estpresqan2}
&& \sup_{z\in\Gamma_\delta}|\partial_z^\beta f_\delta | \leq \sup_{x+iy\in\Gamma_{2\delta}}|\tilde f (x,y)|\delta^{-|\beta|}\beta!,
\end{eqnarray}
uniformly as $\delta\rightarrow 0_+$. (Such a function $f_\delta$ will be called a holomorphic $\delta$-approximation of $f$ on $I_1\times\dots\times I_n$.)
\item[(iii)] Suppose $n=1$ and $f(x)$ is real valued. If $f'(x_0)\ne 0$, then any almost-analytic extension $\tilde f(x,y)$ is one to one in a neighborhood of $(x,y)=(x_0,0)$ and the inverse $\tilde f^{-1}(u,v)$ defined in a neighborhood of $(u,v)=(f(x_0),0)$ is an almost-analytic extension of $f^{-1}(u)$.
\end{description}
\end{lem}

\begin{pre}
The proof of (i) is easy and we proceed with that of (ii). We denote by $\gamma(\delta)$ the (positively oriented) $n$-contour,
$$
\gamma(\delta):= \{ \zeta\in\C^n\, ;\, {\rm dist} (\zeta_j , I)=2\delta,\, j=1,\dots, n\},
$$
and, for  $z\in\gamma(\delta)$, we set,
$$
f_\delta (z) = \frac1{(2i\pi)^n}\int_{\gamma(\delta)}\frac{\tilde f(\re\zeta, \im\zeta)}{(z_1-\zeta_1)\dots (z_n-\zeta_n) }d\zeta_1\dots d\zeta_n.
$$
Then, $f_\delta$ is clearly holomorphic in $\Gamma_\delta$, and since $|z_j-\zeta_j|\geq\delta$ for $\zeta\in\gamma(\delta)$ and $z\in\Gamma_\delta$, (\ref{estpresqan2}) is obtained in a standard way by differentiating under the integral-sign. Moreover, for $z=x+iy\in\Gamma_\delta$, we have,
$$
f_\delta (z) - \tilde f(x,y) =\frac1{(2i\pi)^n}\int_{\gamma(\delta)}\frac{\tilde f(\re\zeta, \im\zeta)-\tilde f(x,y)}{(z_1-\zeta_1)\dots (z_n-\zeta_n) }d\zeta_1\dots d\zeta_n,
$$
and, using the notations $\partial_z=\frac12(\partial_x-i\partial_y)$ and $\overline\partial_z=\frac12(\partial_x+i\partial_y)$, we see that,
\begin{eqnarray*}
&&\tilde f(\re\zeta, \im\zeta)-\tilde f(x,y)\\
&& =(\zeta -z)\int_0^1(\partial_z\tilde f)(t\zeta +(1-t)z)dt +(\overline{\zeta -z})\int_0^1(\overline\partial_z\tilde f)(t\zeta +(1-t)z)dt,
\end{eqnarray*}
where $\partial_z\tilde f (z)$ stands for $\partial_z\tilde f (x,y)$, and similarly for $\overline\partial_z\tilde f (z)$. Therefore, since $\tilde f$ is almost-analytic, and $|\im z| +| \im\zeta|={\mathcal O}(\delta)$, we obtain,
$$
f_\delta (z) - \tilde f(x,y) =\sum_{j=1}^n\int_{\gamma(\delta)}\frac{F_j(z,\zeta)}{\prod_{\ell\not= j}(z_\ell-\zeta_\ell) }d\zeta_1\dots d\zeta_n + r(x,y),
$$
with $F_j(z,\zeta) := \int_0^1(\partial_{z_j}\tilde f)(t\zeta +(1-t)z)dt$, and $\partial^\alpha r ={\mathcal O}(\delta^\infty)$ uniformly. Thus, since $F_j(z,\zeta)\prod_{\ell\not= j}(z_\ell-\zeta_\ell)^{-1}$ depends smoothly on $\zeta_j$ in the domain $A_j:=\{\zeta_j\, ;\, {\rm dist}(\zeta_j ,I)\leq 2\delta\}$, by the Stokes formula, we obtain,
$$
f_\delta (z) - \tilde f(x,y) =-i\sum_{j=1}^n\int_{\gamma_j(\delta)}\frac{\overline\partial_{\zeta_j}F_j(z,\zeta)}{\prod_{\ell\not= j}(z_\ell-\zeta_\ell) }d\zeta_1\dots d\zeta_n\wedge d\overline{\zeta}_j + r(x,y),
$$
with $\gamma_j(\delta):= \{ \zeta_j\in A_j,\, {\rm dist} (\zeta_\ell, I)=2\delta,\, \ell\not= j\}$. Then,  (\ref{estpresqan1}) follows from the fact that $\overline\partial_{\zeta_j}F_j={\mathcal O}(\delta^\infty)$ together with all its derivatives.
\vskip 0.2cm
Now, we prove (iii).
Let us use the coordinates  $(z,\bar
z)=(x+iy,x-iy)$, $(\zeta,\bar
\zeta)=(u+iv,u-iv)$ and regard $\tilde f$ and $g\equiv\tilde f^{-1}$ as functions of
$(z,\bar z)$ and $(\zeta,\bar\zeta)$ respectively. Then
$$
\tilde f(g(\zeta ,\bar \zeta ),\overline{g(\zeta ,\bar \zeta )})=\zeta .
$$
Differentiating by $\bar \zeta $, one gets
$$
\partial_z\tilde f\bar\partial_\zeta  g+\bar\partial_z\tilde f\bar\partial_\zeta  \bar
g=0,
$$
where $\partial_z =\frac 12(\frac{\partial}{\partial x}-i\frac{\partial}{\partial y})$,
$\bar\partial_z =\frac 12(\frac{\partial}{\partial x}+i\frac{\partial}{\partial y})$
and
$\partial_\zeta =\frac 12(\frac{\partial}{\partial u}-i\frac{\partial}{\partial v})$,
$\bar\partial_\zeta =\frac 12(\frac{\partial}{\partial u}+i\frac{\partial}{\partial
v})$.

Since $\partial_z\tilde f$ does not vanish near $x_0$ by assumption, we can conclude that
$\bar\partial_\zeta g={\mathcal O}(|v|^\infty)$, i.e. $g$ is an almost-analytic extension of
$f^{-1}(u)$ if $\bar\partial_z \tilde f$ is, as function of $(u,v)$, of
${\mathcal O}(|v|^\infty)$ as
$v\to 0$.

First, $\bar\partial_z\tilde f={\mathcal O}(|y|^\infty)$ since $\tilde f$ is almost-analytic. On the
other hand, since
$f(x)$ is real-valued, we see from \eq{expansion} that
$v=f'(x)y+{\mathcal O}(y^3)$ as $y\to 0$, and since $f'(x_0)\ne 0$, we also see that $y={\mathcal O}(v)$ as
$v\to 0$. Hence $\bar\partial_z \tilde f={\mathcal O}(|v|^\infty)$.
\end{pre}

\subsection{A Priori Estimates}
We recall some {\it a priori} estimates. The first is
the so-called {\it Agmon estimate} (see for example \cite{hs2}, \cite{ma1}):

\begin{lem}
\label{agmon}
For any $h>0$, $V\in L^\infty(\R^n)$ real-valued, $E\in\R$, $f\in
H^1(\R^n)$, and $\phi$ real-valued and Lipshitz on $\R^n$, one has
$$
\begin{array}{l}
\re \left <e^{\varphi/h}(-h^2\Delta+V-E)f,e^{\varphi/h}f\right >\\[8pt]
=||h\nabla (e^{\varphi/h}f)||^2+\left
<(V-E-|\nabla\varphi(x)|^2)e^{\varphi/h}f,e^{\varphi/h}f\right >
\end{array}
$$
\end{lem}

The second is a microlocal estimate originated by one of the authors
(\cite {ma2}). For $u(x,h)\in S'(\R^n)$ and $\mu >0$, we define the so-called {\it FBI-Bargmann transform} $T_\mu$ by the formula,
\begin{equation}
\label{defT}
T_\mu u(x,\xi,h)=c_\mu\int_{\R^n}
e^{i(x-y)\cdot\xi/h- \mu(x-y)^2/2h}u(y,h)dy,
\end{equation}
where $c_\mu=\mu^{n/4}2^{-n/2}(\pi h)^{-3n/4}$.
The operator $T_\mu$ is unitary from $L^2(\R^n)$ to $L^2(\R^{2n})$,
and $e^{\xi^2/2\mu h}T_\mu u (x,\xi)$ is an entire function of
$z_\mu:=x-i\xi /\mu$.

\begin{prop}
\label{exp}
Let $m\in S_{2n} (1)$, $d\geq 0$,  $p\in S_{2n}(\langle\xi\rangle^{2d})$, and denote by $p_a(x,\xi;h)$ an
almost-analytic extension of
$p$. Let also $k:=h\ln\frac1{h}$, $\rho\in [-1/3,1/3]$ and 
$\psi\in C^\infty(\R^{2n}; \R)$, possibly $h$-dependent, and verifying,
\begin{equation}
\label{estpsi}
\partial_x^\alpha\partial_\xi^\beta\psi ={\mathcal O}\left(k^{-(a|\alpha|+(1-a)|\beta|)}\right),
\end{equation}
for any $\alpha,\beta \in\Z_+^{n}$, with,
\be
\label{condsura}
\frac13 -\min (\rho, 0)\leq a\leq \frac23 -\max (\rho,0).
\ee
Then,  taking $\mu =Ck^{\rho}$ (with $C>0$ constant arbitrary), for $u,v\in L^2(\R^n)$, one has,
\begin{eqnarray}
\label{micro}
\langle mh^{-\psi} T_\mu Pu,h^{-\psi} T_\mu v\rangle
&=&\langle\tilde p(x,\xi;h)h^{-\psi} T_{\mu}u,h^{-\psi}
T_{\mu}v\rangle\\
 &&+ {\mathcal O}\left(\frac{h}{\ln \frac1{h}}\Vert \langle\xi\rangle^{d}h^{-\psi}T_\mu u\Vert\cdot \Vert \langle\xi\rangle^{d}h^{-\psi}T_\mu v\Vert\right),\nonumber
\end{eqnarray}
with,
\begin{eqnarray*}
\tilde p(x,\xi;h)= m(x,\xi )p_a(x-2k\mu^{-1}\partial_{z_\mu}\psi,
\xi +2ik\partial_{z_\mu}\psi)\\
+h\partial_{z_\mu}\left[m(x,\xi )\left(\frac1{\mu}\frac{\partial p_a}{\partial \re x}-i\frac{\partial p_a}{\partial \re \xi}\right)(x-2 k\mu^{-1}\partial_{z_\mu}\psi,
\xi +2ik\partial_{z_\mu}\psi)\right],
\end{eqnarray*}
 where we have set
$\partial_{z_\mu}=(\partial_x+i\mu \partial_\xi)/2$.\\
In particular, if $p$ is real-valued, one obtains,
\begin{eqnarray}
\label{estHp}
&&\label{im}\im\langle h^{-\psi} T_\mu  Pu,h^{-\psi} T_\mu  u\rangle\\
&&=
k\langle  (H_{p}\psi + q_\psi (x, \xi ;h))h^{-\psi}T_\mu  u,
h^{-\psi} T_\mu  u\rangle 
+{\mathcal O}(h)||\langle\xi\rangle^{d}h^{-\psi} T_\mu  u||^2, \nonumber
\end{eqnarray}
with,
\begin{eqnarray*}
q_\psi (x, \xi ;h):&=&
h\mu\sum_{j,\ell =1}^n(\partial_{\xi_j}\partial_{\xi_\ell}p)
 \partial_{\xi_j}\partial_{x_\ell}\psi\\
&& +\sum_{\genfrac{}{}{0cm}{}{\alpha\in\N^{2n}}{ 2\leq|\alpha| \leq3}}\frac{2^{|\alpha|}k^{|\alpha|-1}}{\alpha !}\partial^\alpha p(x,\xi )\im [(-\mu^{-1}\partial_{z_\mu}\psi ,i\partial_{z_\mu}\psi)^\alpha].
\end{eqnarray*}
\end{prop}
\begin{pre} We follow the proof of \cite{ma2} Proposition 3.1 (see also \cite{BoMi} Theorem 3), and we do it for $d=0$ only (the general case $d\geq 0$ can be done along the same lines and is left to the reader).
We have,
\begin{eqnarray}
I:= \langle  mh^{-\psi}T_\mu Pu,h^{-\psi}T_\mu v\rangle = \frac{ c_\mu^2}{(2\pi h)^n}\int e^{\Phi /h}m(x,\xi )p\left(\frac{y+x'}2,\eta \right)\times \nonumber\\
\label{defI}
\times
u(x')\overline{v(y')}d(x',\eta ,y,y',x,\xi), 
\end{eqnarray}
where the integral runs over $\R^{6n}$,
and where we have set,
\begin{equation}
\label{phi}
\Phi = 2k\psi (x,\xi)+i(y'-y)\xi 
-\frac{\mu}2(x -y)^2-\frac{\mu}2({x} -y')^2+i(y-x')\eta .
\end{equation}
Then we observe that, by construction of $p_a$, for all $Y=(y,\eta )\in
\R^{2n}$ and $X =(x-2k\mu^{-1}\partial_{z_\mu}\psi (x,\xi), \xi +2ik\partial_{z_\mu}\psi (x,\xi)  )\in\C^{2n}$ (and setting
$X_s:=sY+(1-s)X$, $0\leq s\leq 1$), we have,
\begin{eqnarray}
 p(Y )-p_a(X )
 &=&
\int_0^1 \left( (Y-\re X )\frac{\partial p_a}{\partial{\re X}}(X_s)
-\im X\frac{\partial p_a}{\partial{\im X}}(X_s)\right) ds\nonumber\\
&=&
\int_0^1 \left( (Y-X )\frac{\partial p_a}{\partial{\re X}}(X_s)
+2i\im X\frac{\partial p_a}{\partial{\overline X}}(X_s)\right) ds \nonumber\\
\label{diff}
&=&
(X -Y)\cdot b_1(x,\xi ,Y ) +r_1(x,\xi ,Y),
\end{eqnarray}
where $b_1$ and $r_1$ are $C^\infty$ on $\R^{4n}$ and verify,
\begin{eqnarray}
\partial_x^\alpha\partial_{\xi}^\beta\partial_{Y}^\gamma b_1 &=&{\mathcal O}(1 +k^{\tau -a|\alpha|-(1-a)|\beta|});\nonumber\\
\label{estr}
\partial_x^\alpha\partial_{\xi}^\beta\partial_{Y}^\gamma r_1 &=&{\mathcal O}((1+k^{\tau -a|\alpha|-(1-a)|\beta|})|\im X|^\infty) ={\mathcal O}(h^\infty),
\end{eqnarray}
 for all
$\alpha,\beta  \in (\Z_+)^{n}$, $\gamma\in (\Z_+)^{2n}$, 
uniformly on $\R^{4n}$, and with $\tau:=\min (1-a,1-a-\rho, a,a+\rho)$. (The last estimate comes from the fact that $|\im X| ={\mathcal O}(k^{1/3})$.)
 
 In the same way, one also has,
\begin{equation}
\label{diff'}
 b_1(x,\xi ,Y ) = b_1(x,\xi,X) + B_2(x,\xi, Y)(X-Y) + r_2(x,\xi,Y),
\end{equation}
 with,
 \begin{eqnarray}
 b_1(x,\xi,X) &=& \frac{\partial p_a}{\partial{\re X}}(X);\nonumber\\
 \partial_x^\alpha\partial_{\xi}^\beta\partial_{Y}^\gamma B_2 &=&{\mathcal O}(1 +k^ {\tau -a|\alpha|-(1-a)|\beta|});
\label{estr2}\\
\partial_x^\alpha\partial_{\xi}^\beta\partial_{Y}^\gamma r_2 &=&{\mathcal O}(h^\infty).\nonumber
 \end{eqnarray}

Inserting (\ref{diff}) and (\ref{diff'}) into (\ref{defI}), we obtain,
\begin{equation}
 \label{Fuj3}
I=\langle mp_a(X) h^{-\psi}T_\mu u , h^{-\psi}T_\mu v\rangle +R_1+R_2+R_3
\end{equation}
with,
 \begin{eqnarray}
\label{R1}
R_1=\frac{ c_\mu^2}{(2\pi h)^n}\int e^{\Phi /h}
\left(X -(\frac{y+x'}2 , 
\eta) \right)\frac{\partial p_a}{\partial{\re X}}(X)\hskip 1pt u(x')\overline{v(y')}\nonumber\\
 \times m(x,\xi )d(x',\eta
,y,y',x,\xi ),
 \end{eqnarray}
\begin{eqnarray}
\label{R2}
R_2=\frac{ c_\mu^2}{(2\pi h)^n}\int e^{\Phi /h}
\langle B\cdot (X -(\frac{y+x'}2 , 
\eta )),X -(\frac{y+x'}2 , 
\eta) \rangle\hskip 1pt u(x')\nonumber\\
 \times \overline{v(y')}m(x,\xi )d(x',\eta
,y,y',x,\xi ),
 \end{eqnarray}
 where we have set, 
 $$
 B:=B_2\left(x,\xi,\frac{y+x'}2,\eta \right),
 $$
 and,
 \begin{eqnarray}
R_3=\frac{ c_\mu^2}{(2\pi h)^n}\int e^{\Phi /h}
r\left(x,\xi , \frac{y+x'}2 , \eta \right) u(x')\overline{v(y')}\nonumber\\
 \times m(x,\xi )d(x',\eta
,y,y',x,\xi ),
\end{eqnarray}
 where,
 $$
 r(x,\xi ,Y):=(X-Y)\cdot r_2(x,\xi,Y) + r_1(x,\xi,Y).
 $$
Then, as in \cite{ma2, BoMi}, we observe that $\Phi$ verifies,
$$
L(\Phi )=X-(\frac{y+x'}2 ,\eta ),
$$
with $L:= \frac12(-\mu^{-1}\partial_x-i\partial_\xi -i\partial_\eta , i\partial_x-\mu\partial_\xi +2i\partial_y)$.
As a consequence, 
$$
e^{\Phi /h}
\left(X -(\frac{y+x'}2 , 
\eta) \right) = hL(e^{\Phi /h}),
$$
and
$$
\langle B\cdot (X -(\frac{y+x'}2 , 
\eta)),X -(\frac{y+x'}2 , 
\eta) \rangle e^{\Phi /h} = h^2\langle B\cdot L , L \rangle e^{\Phi /h}=:h^2Ae^{\Phi /h}.
$$
Thus, making an integration  by parts in  (\ref{R1}), we obtain,
\begin{equation}
\label{Fuj4}
R_1=h\langle p_1(x,\xi) h^{-\psi}T_\mu u , h^{-\psi}T_\mu v\rangle,
\end{equation}
with 
\begin{eqnarray*}
p_1(x,\xi ;h):&=& {}^tL\left(m(x,\xi)\frac{\partial p_a}{\partial{\re X}}(X)\right)\\
 &=&\partial_{z_\mu}\left[m(x,\xi)\left(\frac1{\mu}\frac{\partial p_a}{\partial \re x}(X)-i\frac{\partial p_a}{\partial \re \xi}(X)\right)\right],
\end{eqnarray*}
 (here ${}^tL$ stands for the transposed operator of $L$). In the same way,
making two integrations by parts in  (\ref{R2}), we obtain,
$$
R_2= h^2\langle h^{-\psi}T_{\mu ,f }u, h^{-\psi}T_\mu v\rangle,
$$
where we have set,
\begin{equation}
f\left(x,\xi ,\frac{y+x'}2,\eta \right) = {}^tA \cdot (m(x,\xi)),
\end{equation}
and,
$$
T_{\mu ,f } u (x,\xi ):=c_\mu \int e^{i(x -y)\xi /h - \mu (x -y)^2/2h}
{\rm Op}_h^W (f(x,\xi ,\cdot ))u(y)dy .
$$
In particular, by (\ref{estr2}) and (\ref{condsura}), we see that,
\begin{equation}
\label{estimationf}
 \partial_x^\alpha\partial_{\xi}^\beta\partial_{Y}^\gamma f={\mathcal O}(k^{-1-a|\alpha|-(1-a)|\beta|}).
\end{equation}
With the same notations, we also have,
$$
R_3= \langle h^{-\psi}T_{\mu, m r} u, h^{-\psi}T_\mu v\rangle.
$$
Now, for $g\in\{ f, mr\}$, we observe,
$$
T_{\mu, g}u(x,\xi ) = \left[ T_\mu {\rm Op}_h^W (g(x',\xi',\cdot ))u(x,\xi)\right] \left\vert_{\genfrac{}{}{0cm}{}{x'=x}{\xi'=x}}\right.,
$$
and thus, applying a slight generalization of \cite{ma1} Proposition 3.3.1 to the case $\mu\not=1$, we easily obtain,
\begin{equation}
\label{Tg=gT}
T_{\mu, g}u (x,\xi )={\rm Op}_h (\tilde g)T_\mu u(x,\xi ),
\end{equation}
where ${\rm Op}_h(\tilde g)$ stands for the semiclassical pseudodifferential operator with symbol,
$$
\tilde g(x,x',\xi,\xi', x^*,\xi^*):=g(x,\xi, \frac{x+x'}2 -\xi^*, x^*).
$$
Here, $x^*$ and $\xi^*$ stand for the dual variables of $x$ and $\xi$, respectively.  Then, writing,
$$
\psi(x,\xi )-\psi(x',\xi') = (x-x')\phi_1 + (\xi-\xi')\phi_2,
$$
(with $\phi_j=\phi_j(x,x',\xi,\xi';h)$ smooth),  applying Stokes formula and, in the expression of $h^{-\psi}{\rm Op}_h (\tilde g)h^{\psi}$, performing the change of contour,
$$
\R^{2n}\ni (x^*,\xi^*)\mapsto (x^*,\xi^*) + ik(\phi_1,\phi_2),
$$
we see that $h^{-\psi}{\rm Op}_h (\tilde g)h^{\psi}$ is an $h$-admissible operator with symbol,
$$
g_\psi (x,x',\xi,\xi', x^*,\xi^*) := g_a(x,\xi, \frac{x+x'}2 -\xi^*-ik\phi_2, x^*+ik\phi_1),
$$
where $g_a$ is an almost-analytic extension of $g$. Moreover, by (\ref{estpsi}), we see that $\phi_1$ and $\phi_2$ verify,
\begin{eqnarray*}
&& \partial_{x,x'}^\alpha \partial_{\xi,\xi'}^\beta \phi_1 ={\mathcal O}(k^{-(a+a|\alpha|+(1-a)|\beta|)});\\
&& \partial_{x,x'}^\alpha \partial_{\xi,\xi'}^\beta \phi_2 ={\mathcal O}(k^{-(1-a+a|\alpha|+(1-a)|\beta|)}),
\end{eqnarray*}
and thus, using (\ref{estr2}), (\ref{estimationf}), the fact that $k\geq h$, and the Calder\'on-Vaillancourt theorem (see, e.g., \cite{ma1} Chapter 2, Exercise 15), we obtain,
\begin{eqnarray*}
\Vert h^{-\psi}{\rm Op}_h (\tilde f)h^{\psi} \Vert ={\mathcal O}(k^{-1});\\
\Vert h^{-\psi}{\rm Op}_h (\widetilde {mr})h^{\psi}\Vert ={\mathcal O}(h^\infty ),
\end{eqnarray*}
and therefore,
\begin{eqnarray*}
R_2={\mathcal O}(h^2k^{-1}\Vert h^{-\psi}T_\mu u\Vert\cdot\Vert h^{-\psi}T_\mu v\Vert );\\
R_3={\mathcal O}(h^\infty \Vert h^{-\psi}T_\mu u\Vert\cdot\Vert h^{-\psi}T_\mu v\Vert ),
\end{eqnarray*}
so that, by (\ref{Fuj3}) and (\ref{Fuj4}), (\ref{micro}) follows.

To prove (\ref{estHp}), we first observe, that, by a Taylor expansion, we have,
\begin{eqnarray*}
&& p_a(x-2 k\mu^{-1}\partial_{z_\mu}\psi,
\xi+2ik\partial_{z_\mu}\psi) \\
&& \hskip 1cm = (p -k\mu^{-1}\nabla_x p\nabla_x\psi  -k\mu \nabla_\xi p\nabla_\xi\psi) (x,\xi )+ ikH_p\psi (x,\xi)\\
&&\hskip 1.5cm  +\sum_{\genfrac{}{}{0cm}{}{\alpha\in\N^{2n}}{ 2\leq|\alpha| \leq3}}\frac{(2k)^{|\alpha|}}{\alpha !}\partial^\alpha p(x,\xi )(-\mu^{-1}\partial_{z_\mu}\psi ,i\partial_{z_\mu}\psi)^\alpha +{\mathcal O}(k^{4/3}),
\end{eqnarray*}
and thus, in particular (since $k^{4/3}={\mathcal O}(h)$),
\begin{eqnarray}
\label{im1}
&& \hskip 1cm\im p_a(x-2 k\mu^{-1}\partial_{z_\mu}\psi,
\xi+2ik\partial_{z_\mu}\psi)\\
&&  =  kH_p\psi (x,\xi)+\sum_{\genfrac{}{}{0cm}{}{\alpha\in\N^{2n}}{ 2\leq|\alpha| \leq3}}\frac{(2k)^{|\alpha|}}{\alpha !}\partial^\alpha p(x,\xi )\im [(-\mu^{-1}\partial_{z_\mu}\psi ,i\partial_{z_\mu}\psi)^\alpha]  +{\mathcal O}(h).\nonumber
\end{eqnarray}
Moreover, using (\ref{estpsi}), we also see that,
\begin{eqnarray*}
&&\left(\frac1{\mu}\frac{\partial p_a}{\partial \re x}-i\frac{\partial p_a}{\partial \re \xi}\right)(x-2 k\mu^{-1}\partial_{z_\mu}\psi,
\xi+2ik\partial_{z_\mu}\psi)\\
 &&\hskip 1.5cm= \mu^{-1}\partial_xp(x,\xi )-i\partial_\xi p(x,\xi) +2kM(x,\xi)\partial_{z_\mu}\psi(x,\xi) +{\mathcal O}(k^{2/3}),
\end{eqnarray*}
where $M$ is the  $n\times n$-matrix-valued function,
\begin{eqnarray*}
M &=& -\mu^{-2}(\nabla_x\cdot\nabla_x)p +i\mu^{-1}(\nabla_\xi\cdot\nabla_x+\nabla_x\cdot\nabla_\xi) p +(\nabla_\xi\cdot\nabla_\xi) p\\
&=& (\nabla_\xi\cdot\nabla_\xi) p +{\mathcal O}(k^{1/3}).
\end{eqnarray*}
(Here, we have used the notation $\nabla_x\cdot\nabla_\xi = (\partial_{x_j}\partial_{\xi_\ell})_{1\leq j,\ell\leq n}$.)

Since applying $\partial_{z_\mu}$ makes lose at most $k^{-2/3}$, one also easily obtains,
\begin{eqnarray*}
&&\partial_{z_\mu}\left[\left(\frac1{\mu}\frac{\partial p_a}{\partial \re x}-i\frac{\partial p_a}{\partial \re \xi}\right)(x-2 k\mu^{-1}\partial_{z_\mu}\psi,
\xi+2ik\partial_{z_\mu}\psi)\right]\\
 &&\hskip 2cm=  \frac{\mu}2\Delta_\xi p + 2k\sum_{j,\ell =1}^n(\partial_{\xi_j}\partial_{\xi_\ell}p)\partial_{z_\mu^j}\partial_{z_\mu^\ell}\psi(x,\xi) +{\mathcal O}(1),
\end{eqnarray*}
and thus,  if $p$ is real-valued, one finds,
\begin{eqnarray}
&&\im\partial_{z_\mu}\left[\left(\frac1\mu\frac{\partial p_a}{\partial \re x}-i\frac{\partial p_a}{\partial \re \xi}\right)(x-2 k\mu^{-1}\partial_{z_\mu}\psi,
\xi+2ik\partial_{z_\mu}\psi)\right]\nonumber\\
 && =\frac{k}2\sum_{j,\ell =1}^n\mu(\partial_{\xi_j}\partial_{\xi_\ell}p)(\partial_{\xi_j}\partial_{x_\ell}\psi+\partial_{x_j}\partial_{\xi_\ell}\psi)
  +{\mathcal O}(1)\nonumber\\
 && = k\mu\sum_{j,\ell =1}^n(\partial_{\xi_j}\partial_{\xi_\ell}p)
 (\partial_{x_j}\partial_{\xi_\ell}\psi)
 \label{im2}
 +{\mathcal O}(1).
\end{eqnarray}
Then, (\ref{estHp}) immediately follows from (\ref{im1})-(\ref{im2}).
\end{pre}

\end{document}